\documentclass{ieeetj}
\usepackage{cite}
\usepackage{amsmath,amssymb,amsfonts}
\usepackage{mathtools}
\usepackage{bm}
\usepackage{multirow}
\usepackage{arydshln}
\usepackage{setspace}
\usepackage{algorithmic}
\usepackage{graphicx,color}
\usepackage{textcomp}
\usepackage{xcolor}
\usepackage{hyperref}
\hypersetup{hidelinks=true}
\usepackage{algorithm,algorithmic}
\def\BibTeX{{\rm B\kern-.05em{\sc i\kern-.025em b}\kern-.08em
    T\kern-.1667em\lower.7ex\hbox{E}\kern-.125emX}}
\AtBeginDocument{\definecolor{tmlcncolor}{cmyk}{0.93,0.59,0.15,0.02}\definecolor{NavyBlue}{RGB}{0,86,125}}

\def\OJlogo{\vspace{-4pt}$<$Society logo(s) and publication title will appear here.$>$}
\def\seclogo{\vspace{10pt}$<$Society logo(s) and publication title will appear here.$>$}

\def\authorrefmark#1{\ensuremath{^{\textbf{#1}}}}

\newcommand{\IndFreqbin}{i}
\newcommand{\IndFrame}{j}
\newcommand{\IndSrc}{n}
\newcommand{\IndMic}{m}
\newcommand{\NumFreqbin}{I}
\newcommand{\NumFrame}{J}
\newcommand{\NumSrc}{N}
\newcommand{\NumMic}{M}

\newcommand{\SrcSignal}{s}
\newcommand{\ObsSignal}{x}
\newcommand{\SepSignal}{y}
\newcommand{\abs}[1]{\left| {#1} \right|}
\newcommand{\Hermite}{\mathsf{H}}
\newcommand{\Transpose}{\mathsf{T}}
\newcommand{\const}{\mathrm{const.}}
\newcommand{\ConstWithParam}[1]{\mathrm{const}({#1})}
\newcommand{\Identity}{\bm{E}}
\newcommand{\UnitVec}{\bm{e}}
\newcommand{\Prior}[1]{\hat{#1}}
\newcommand{\AuxCost}[1]{\bar{\mathcal{L}}_{#1}}
\newcommand{\Cost}[1]{\mathcal{L}_{#1}}
\newcommand{\Est}[1]{\tilde{#1}}
\newcommand{\LatestFrame}{\tau}

\newcommand{\MixMat}{\bm{A}}
\newcommand{\MixVec}{\bm{a}}
\newcommand{\DemixMat}{\bm{W}}
\newcommand{\DemixVec}{\bm{w}}
\newcommand{\Basis}{t}
\newcommand{\Activation}{v}
\newcommand{\SrcModel}{r}
\newcommand{\IndBasis}{k}
\newcommand{\NumBasis}{K}
\newcommand{\RegWeight}{\mu}
\newcommand{\ILRMA}{\mathrm{ILRMA}}
\newcommand{\NSR}{\mathrm{NSR}}
\newcommand{\IndTargetSrc}{n^{\upt}}
\newcommand{\CovRegNSR}{\bm{D}}

\newcommand{\upt}{(\mathrm{t})}
\newcommand{\upx}{(\mathrm{x})}
\newcommand{\upn}{(\mathrm{n})}

\newcommand{\Var}{r}
\newcommand{\VarT}{\Var^{\upt}}
\newcommand{\VarN}{\Var^{\upn}}
\newcommand{\SCM}{\bm{R}}
\newcommand{\SCMX}{\SCM^{\upx}}
\newcommand{\SCMN}{\SCM^{\upn}}
\newcommand{\SCMNd}{\SCM'^{\upn}}
\newcommand{\SCMNh}{\SCM'^{\upn}}
\newcommand{\SCMNb}{\breve{\SCM}^{\upn}}
\newcommand{\XMat}{\SCM^{(\mathrm{emp})}}
\newcommand{\XMatInv}{\mathcal{R}^{(\mathrm{emp})}}

\newcommand{\SV}{\bm{a}^{\upt}}
\newcommand{\ShapeIG}{\varrho}
\newcommand{\ScaleIG}{\varsigma}
\newcommand{\WeightSCMN}{\lambda}
\newcommand{\NoiseSV}{\bm{b}}
\newcommand{\NoiseSrcImage}{\bm{z}}
\newcommand{\RCSCME}{\mathrm{RCSCME}}
\newcommand{\trace}{\mathrm{tr}}
\newcommand{\TargetSrcImage}{\hat{\bm{\SrcSignal}}}
\newcommand{\InitLambdaScaling}{\iota}

\newcommand{\IVA}{\mathrm{IVA}}
\newcommand{\CovIVA}{\bm{F}}
\newcommand{\AuxVarIVA}{\theta}
\newcommand{\ForgetIVA}{\alpha}
\newcommand{\InvCovIVA}{\mathcal{F}}
\newcommand{\diffDemixVecIVA}{\bar{\DemixVec}}
\newcommand{\VarISS}{h}

\newcommand{\ModelParameter}{\vartheta}
\newcommand{\WeightMWLE}{\rho}

\newcommand{\ModelMWLE}{p}

\newcommand{\WeightOILRMA}{\WeightMWLE^{(\mathrm{I})}}
\newcommand{\ForgetILRMA}{\beta}
\newcommand{\AuxVarNMFJensen}{l}
\newcommand{\AuxVarNMFTangent}{q}

\newcommand{\EstNumerBasis}{\Est{t}^{(\mathrm{num})}}
\newcommand{\EstDenomBasis}{\Est{t}^{(\mathrm{den})}}
\newcommand{\FramewiseCov}{\bm{G}}
\newcommand{\FramewiseCovReg}{\bm{H}}
\newcommand{\InvFramewiseCov}{\mathcal{G}}
\newcommand{\InvFramewiseCovReg}{\mathcal{H}}

\newcommand{\VecIPInvLemma}{\bar{\DemixVec}}

\newcommand{\WeightORCSCME}{\WeightMWLE^{(\mathrm{R})}}
\newcommand{\ForgetRCSCME}{\gamma}
\newcommand{\ApproxFrameORCSCME}{C}
\newcommand{\AuxMatT}{\bm{\Phi}^{\upt}}
\newcommand{\AuxMatN}{\bm{\Phi}^{\upn}}
\newcommand{\AuxMatSCMX}{\bm{\Psi}}
\newcommand{\AuxScaVarT}{\chi}
\newcommand{\NoiseImageMat}{\bm{P}}
\newcommand{\DiagVec}{\bm{u}}
\newcommand{\NumerWeightSCMN}{\lambda^{(\mathrm{num})}}
\newcommand{\DenomWeightSCMN}{\lambda^{(\mathrm{den})}}
\newcommand{\EstNumerWeightSCMN}{\Est{\lambda}^{(\mathrm{num})}}
\newcommand{\EstDenomWeightSCMN}{\Est{\lambda}^{(\mathrm{den})}}
\newcommand{\EstAuxMatN}{\Est{\bm{\Phi}}^{(\mathrm{n})}}
\newcommand{\EstAuxMatT}{\Est{\bm{\Phi}}^{(\mathrm{t})}}
\newcommand{\EstVarT}{\Est{r}^{\upt}}
\newcommand{\EstVarN}{\Est{r}^{\upn}}
\newcommand{\EstAuxMatSCMX}{\Est{\AuxMatSCMX}}
\newcommand{\EstSCMX}{\tilde{\bm{R}}^{\upx}}
\newcommand{\EstAuxScaVarT}{\Est{\chi}}

\newcommand{\ScalarFirst}{\nu}
\newcommand{\ScalarSecond}{\kappa}
\newcommand{\ScalarComb}{\varphi}
\newcommand{\upaRa}{(\mathrm{aRa})}
\newcommand{\upaRx}{(\mathrm{aRx})}
\newcommand{\upxRx}{(\mathrm{xRx})}

\newcommand{\upbRx}{(\mathrm{bRx})}
\newcommand{\upaRb}{(\mathrm{aRb})}
\newcommand{\upbRa}{(\mathrm{bRa})}
\newcommand{\upbRb}{(\mathrm{bRb})}

\newcommand{\upua}{(\mathrm{ua})}
\newcommand{\upux}{(\mathrm{ux})}

\newcommand{\MPinv}{\bm{\Xi}}

\allowdisplaybreaks

\begin{document}

\receiveddate{XX Month, XXXX}
\reviseddate{XX Month, XXXX}
\accepteddate{XX Month, XXXX}
\publisheddate{XX Month, XXXX}
\currentdate{XX Month, XXXX}
\doiinfo{XXXX.2022.1234567}

\markboth{Online Algorithms for ILRMA and RCSCME Based on MWLE}{Y. Ishikawa {et al.}}

\title{Online Algorithms for Independent Low-Rank Matrix Analysis and Rank-Constrained Spatial Covariance Matrix Estimation Based on Maximum Weighted Likelihood Estimation}

\author{Yuto Ishikawa\authorrefmark{1}, Graduate Student Member, IEEE,\\
Norihiro Takamune\authorrefmark{1},
Tomohiko Nakamura\authorrefmark{1}, Member, IEEE,\\
Daichi Kitamura\authorrefmark{2}, Senior Member, IEEE, 
Hiroshi Saruwatari\authorrefmark{1}, Member, IEEE,\\
Yu Takahashi\authorrefmark{3}, Member, IEEE, and 
Kazunobu Kondo\authorrefmark{3}, Member, IEEE}
\affil{Graduate School of Information Science and Technology, The University of Tokyo, Tokyo 113-8656, Japan}
\affil{National Institute of Technology, Kagawa College, Takamatsu, Kagawa 761-8058, Japan}
\affil{Yamaha Corporation, Hamamatsu-shi, Shizuoka 430-8650, Japan}
\corresp{Corresponding author: Yuto Ishikawa (email: yuto\_ishikawa.jp@ieee.org).}
\authornote{
    This work was supported in part by the JST Moonshot Research and Development (for algorithm development) under Grant JPMJMS2011, in part by the Tateisi Science and Technology Foundation (for numerical experiment), in part by the Kajima Foundation's Support Program for International Joint Research Activities (for practical experiment) under Grant 2024-kyodoshin-05, and in part by the JSPS KAKENHI (for ablation study analysis) under Grant 21H05054.
}

\begin{abstract}
    Real-time multichannel speech extraction (MSE) under diffuse noise conditions is an important task with a wide range of applications, such as speech recognition and hearing aids.
    In this paper, we propose online algorithms for independent low-rank matrix analysis (ILRMA) and rank-constrained spatial covariance matrix estimation (RCSCME).
    Previously, we proposed a real-time extension of the RCSCME-based method: an MSE method based on ILRMA and RCSCME using the blockwise batch algorithm.
    However, it assumes that the spatial characteristics are stationary within a single batch, and thus, in dynamic situations where the target speaker moves, its performance may degrade.
    To address this problem, we derive the online algorithms for ILRMA and RCSCME in the following three steps.
    First, we formulate framewise cost functions for ILRMA and RCSCME on the basis of maximum weighted likelihood estimation.
    Second, we derive the update rules for the framewise cost functions on the basis of auxiliary-function techniques.
    These naive update rules are computationally costly for real-time execution on a practical machine.
    Thus, we finally derive the online algorithms by approximating some intermediate parameters with their estimates.
    Furthermore, we propose stabilization and further acceleration techniques for these online algorithms.
    In experiments, we simulate situations where a target speaker is stationary or moves and show that the proposed method achieves superior speech extraction performance compared with conventional methods.
    In addition, using real-world recorded signals, we demonstrate the effectiveness of the proposed method in practical scenarios.
\end{abstract}



\maketitle

\section{INTRODUCTION}
\label{sec:introduction}

\IEEEPARstart{M}{ultichannel} speech extraction (MSE) is a technique to extract a specific target speech signal from noisy mixtures recorded by a microphone array~\cite{Cardoso1993IET,Takahashi2009TASLP,Miyazaki2012IEICED,Miyazaki2014SigPro,Aprilyanti2015AST,Koldovsky2019IEEE-TSP,Scheibler2019WASPAA,Kubo2020TASLP,Brendel2020IEEE-TSP,Brendel2021EUSIPCO,Ikeshita2021IEEE-TSP,Ikeshita2021IEEE-SPL,Jansky2021EURASIP-JAMSP,Ueda2024EURASIP-JASMP,Nishida2023EUSIPCO,Zmolikova2023IEEE-SPM}.
MSE has been widely applied in various acoustic fields, such as preprocessing of speech recognition for robots~\cite{Nakadai2009ICASSP,Kawahara2024CA} and hearing aids~\cite{Sunohara2017ICASSP,Haruta2021EUSIPCO,Une2019APSIPA}. 
In practical situations, that is, noisy environments, MSE helps improve the speech recognition performance of robots and reduce listening effort for hearing-aid users, thereby enabling smoother communication.
Furthermore, real-time applicability of MSE is often required in such situations.
If the processing delay of MSE becomes too large, it can delay subsequent processes in dialogue systems or cause discomfort for hearing-aid users, which hinders smooth communication.
Therefore, for real-world applications, the entire process from the observation of noisy signals to the output of extracted target speech signal must be performed in real time.
For these reasons, in this paper, we address the real-time MSE problem for a single target speaker in a diffuse noise environment.
Furthermore, for simplicity, we consider a situation where the approximate location of the target speaker and the geometry of the microphone array are known beforehand.
Methods that require large amounts of training data, such as deep neural network-based methods, are beyond the scope of this research.

\begin{figure*}[t]
    \centering
    \includegraphics[width=0.85\linewidth]{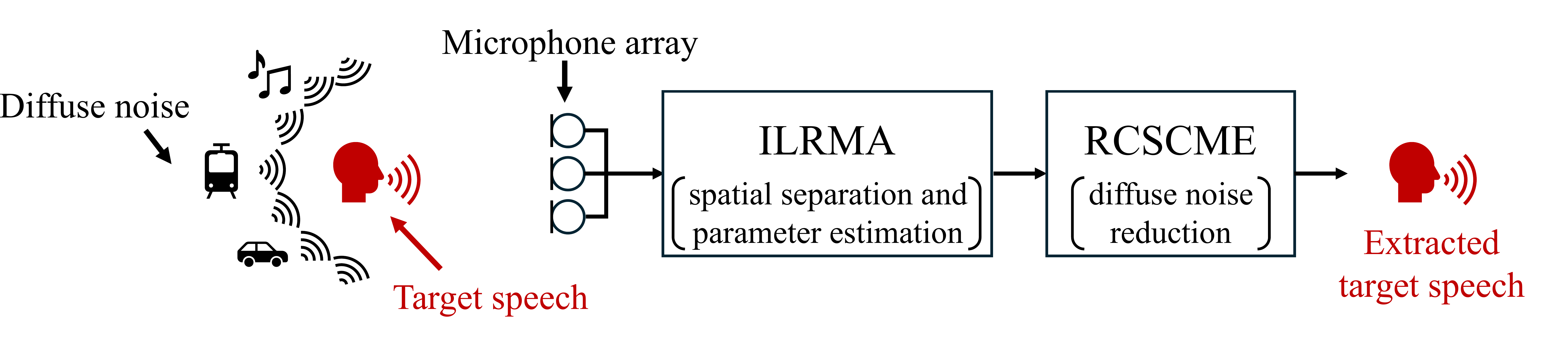}
    \caption{Flow of RCSCME-based speech extraction method under diffuse noise condition.}
    \label{fig:schematic_offline_SE}
\end{figure*}

In an offline scenario, we previously proposed a state-of-the-art MSE method based on independent low-rank matrix analysis (ILRMA)~\cite{Kitamura2016TASLP,Kitamura2018ASS_book} and rank-constrained spatial covariance matrix estimation (RCSCME)~\cite{Kubo2020TASLP}.
ILRMA is a state-of-the-art blind source separation (BSS) method~\cite{Pal2013COCOSDA,Naik2014BSS,Sawada2019APSIPA}, which is a technique for separating each source signal from observed mixtures without any prior information.
Under (over-)determined conditions, that is, the number of microphones is greater than or equal to that of sources, ILRMA estimates the time-invariant demixing matrix in the time-frequency domain using the short-time Fourier transform (STFT) to separate the observed signals into each source signal.
In addition, if prior information about the spatial characteristics is available, it can be incorporated as a spatial regularizer for ILRMA to improve the source separation performance~\cite{Mitsui2018ICASSP}.
For example, we have proposed null-based spatially regularized ILRMA (NSR-ILRMA)~\cite{Ishikawa2025Access} as a spatially regularized ILRMA that uses only the steering vector of a single target speech.
However, when we apply ILRMA or its spatially regularized extensions to mixtures recorded in a diffuse noise environment, the separated signal corresponding to the target speech contains residual diffuse noise components in principle~\cite{Araki2003EURASIP,Araki2003IEEETSAP}.
On the other hand, the other separated signals cancel the target speech signal accurately~\cite{Takahashi2009TASLP}.
Owing to this property, RCSCME fixes some spatial parameters estimated from the outputs of ILRMA or its spatially regularized extensions, resulting in efficient estimation for remaining parameters and accurate extraction of the target speech signal.
In \cite{Kubo2020TASLP}, it is demonstrated that RCSCME outperforms conventional MSE methods, such as multichannel nonnegative matrix factorization (MNMF)~\cite{Ozerov2010IEEE-TASLP,Sawada2013IEEE-TASLP} and independent vector extraction (IVE)~\cite{Koldovsky2019IEEE-TSP}.
Therefore, RCSCME can be regarded as a state-of-the-art MSE method, and we aim to develop a real-time MSE method based on RCSCME.
Hereafter, we refer to the MSE method based on ILRMA (or its spatially regularized extensions) and RCSCME as the RCSCME-based method.
Fig.~\ref{fig:schematic_offline_SE} shows a flow of the RCSCME-based method under a diffuse noise condition.

When considering the real-time MSE problem, the most straightforward approach is to execute an offline MSE method every time a new frame of an observed signal is obtained, that is, to execute it within the STFT shift length.
However, in a real-time scenario, new frames of the observed signal arrive continuously, and thus, we must consider very long observed signals.
Therefore, it is impractical to directly process such long observed signals within the STFT shift length even on extremely high-performance computational resources.
To address this, more efficient approaches have been proposed: a blockwise batch algorithm~\cite{Mukai2004IEICE,Mori2006EURASIP,Hiekata2009AST} and online algorithm~\cite{Taniguchi2014HSCMA,Nakashima2022APSIPA,Ueda2021EUSIPCO,Ueda2024IEEE-TASLP}.
In the blockwise batch algorithm, the observed signals are divided into small batches, and an offline method is applied to each batch.
An advantage of the blockwise batch algorithm is the high applicability of existing methods owing to its simplicity.
We previously proposed a real-time extension of the RCSCME-based method based on the blockwise batch algorithm (\textit{B-RCSCME})~\cite{Ishikawa2025Access}.
In the RCSCME-based method, ILRMA requires sufficiently long observed signals to achieve high source separation performance and involves numerous matrix operations.
Thus, ILRMA is computationally costlier than RCSCME, making it difficult to execute the entire process of the RCSCME-based method within the STFT shift length.
In B-RCSCME, we focus on the facts that RCSCME only requires the time-invariant parameters obtained from the ILRMA outputs and is sufficiently fast to be executed within the STFT shift length.
Then, we divide the entire process into two parts: ILRMA and RCSCME parts.
The ILRMA part estimates time-invariant parameters (e.g., demixing matrix) using ILRMA or its spatially regularized extensions, while the RCSCME part utilizes the outputs of the ILRMA part and extracts the target speech signal using RCSCME.
By applying the blockwise batch algorithm to these two parts and executing them in parallel, we can perform the RCSCME-based method in real time.
The ILRMA part estimates the time-invariant spatial parameters at a relatively long interval, and the RCSCME part uses the latest outputs of the ILRMA part and extracts the target speech signal at the STFT shift length interval.
However, when the spatial characteristics change significantly, such as when the target speaker moves, the spatial information becomes nonstationary within the interval of the ILRMA part.
Thus, the spatial parameters for the inputs of the RCSCME part may become inappropriate, and the speech extraction performance may degrade.
On the other hand, in the online algorithm, some intermediate parameters that involve a summation over time frames are approximated so that they can be updated sequentially without recalculating the entire summation, resulting in reduction of computational costs and execution in a frame-by-frame manner.
Owing to the frame-by-frame processing, methods based on the online algorithm can track changes in spatial characteristics and are expected to operate robustly under dynamic conditions.
Note that in the field of offline MSE methods, several methods based on IVE have been proposed to handle a moving target speaker~\cite{Koldovsky2019IEEE-TSP,Jansky2021EURASIP-JAMSP}.
However, these methods estimate a demixing filter that broadly enhances the direction where the target speaker is present during the observation, and thus, their performance is limited in scenarios where the target speaker moves over a wide spatial region.
In such cases, real-time methods based on online algorithms are expected to provide more robust performance.
As a method based on the online algorithm, online extensions of independent vector analysis (IVA)~\cite{Kim2007TASLP,Hiroe2006ICA,Ono2011WASPAA} and IVE have been proposed~\cite{Taniguchi2014HSCMA,Nakashima2022APSIPA,Ueda2021EUSIPCO,Ueda2024IEEE-TASLP} and are referred to as online IVA (O-IVA) and online IVE (O-IVE), respectively.
IVA is one of the major offline BSS methods.
In both IVA and IVE, it is necessary to compute the weighted spatial covariance matrix, and sufficiently long signals are required to estimate it accurately.
A high computational cost for calculating the weighted spatial covariance matrix makes it difficult to simply perform IVA and IVE at the STFT shift length interval.
To perform IVA and IVE in real time, O-IVA and O-IVE approximate the weighted spatial covariance matrix using a sequential update based on an autoregressive model.
However, since O-IVA and O-IVE are based on their corresponding offline methods, their ability to capture the spectral structures of sources is limited.
As demonstrated in \cite{Kitamura2016TASLP}, ILRMA achieved superior performance to IVA in the offline setting by introducing a more expressive source model, and therefore an online algorithm for ILRMA is expected to outperform O-IVA and O-IVE.
Note that since (O-)IVA and (O-)IVE are methods that estimate the linear demixing filters as ILRMA, there is also a limitation for diffuse noise conditions.

Our previously proposed B-RCSCME has two major limitations.
First, in the ILRMA part, the demixing matrix is estimated at relatively long intervals, resulting in performance degradation when the spatial characteristics change abruptly.
Second, in the RCSCME part, although the iterative updates in RCSCME consist only of scalar operations and can be computed efficiently, we have experimentally found that the initialization of RCSCME is computationally costly because of numerous matrix operations including the computation of a Moore--Penrose (MP) inverse.
In this study, to achieve high real-time speech extraction performance even in situations where the target speaker moves, we propose an online extension of the RCSCME-based method: online algorithms for NSR-ILRMA and RCSCME.
We derive the proposed online algorithms in the following three steps similar to \cite{Ishikawa2025EUSIPCO}.
First, we formulate framewise cost functions for NSR-ILRMA and RCSCME.
The cost functions for offline NSR-ILRMA and RCSCME are derived on the basis of maximum likelihood estimation (MLE), and thus, the likelihoods of all time frames are equally considered.
However, when we consider situations where the spatial characteristics may change, past frames may affect speech extraction performance at the current frame.
To address this issue, we introduce the concept of maximum weighted likelihood estimation (MWLE)~\cite{Ahmed2005IEEETR,Fung2022InsuranceME} into NSR-ILRMA and RCSCME.
MWLE is an extension of MLE that reduces the influence of outliers by applying small weights to the likelihoods of data points identified as outliers.
We assume that the signals observed in the past frames can originate from a spatial system different from those observed in the recent frames.
Under this assumption, we derive the cost functions for NSR-ILRMA and RCSCME on the basis of MWLE by assigning smaller weights to the likelihoods of past frames at each frame estimation.
In addition, an appropriate design of the weight parameters enable us to derive sequential updates for intermediate parameters.
In the second step, we derive naive update rules for the framewise cost function on the basis of the majorization-minimization (MM) or the maximization-equalization (ME) algorithms~\cite{Hunter2000JCGS,Fevotte2009NC}.
Unfortunately, the naive update rules have high computational costs; thus, they are unsuitable for real-time processing.
In the third step, to reduce the computational cost, we approximate some intermediate parameters that involve a summation over time frames using estimates obtained at the previous frame estimation and derive the efficient online algorithms for NSR-ILRMA and RCSCME by minimizing the approximate cost functions.
Furthermore, we also propose stabilization and further acceleration techniques for the online algorithms.
In experiments, we simulate two scenarios where the target speaker is either stationary or moving, and demonstrate that the proposed method outperforms the conventional methods in terms of real-time speech extraction performance and operates robustly against a moving speaker.
Furthermore, using real-world recorded signals, we confirm that the proposed method also achieves higher performance than the conventional methods in a practical acoustic environment.

This paper includes our earlier work~\cite{Ishikawa2025EUSIPCO}.
In offline IVA, the separated signals are assumed to follow the same probabilistic generative model over all time frames.
However, in a real-time scenario, the assumption that the spatial characteristics are stationary becomes unrealistic.
Therefore, in \cite{Ishikawa2025EUSIPCO}, we introduce a framewise probabilistic generative model for the separated signal and derive a framewise cost function.
Using this framewise cost function, we then derive a naive update rule.
Finally, by approximating the weighted spatial covariance matrix using estimates obtained at the past frame, we obtain an online algorithm for IVA.
However, this approach is applicable only to methods that assume a time-invariant generative model (e.g., IVA) and cannot be applied to methods that assume a time-varying generative model (e.g., ILRMA and RCSCME).
In this paper, we address this limitation by introducing the concept of MWLE.
The contributions of this paper are as follows.
\begin{itemize}
    \item We derived online algorithms for NSR-ILRMA and RCSCME on the basis of MWLE to achieve real-time MSE under diffuse noise conditions. 
    \item We proposed several stabilization and further acceleration techniques for online algorithms.
    \item We demonstrated the effectiveness of the proposed method through real-time speech extraction experiments in both simulated and real-world diffuse noise environments. 
\end{itemize}

The rest of this paper is organized as follows.
In Section~\ref{sec:related_methods}, we review offline methods (ILRMA, NSR-ILRMA, and RCSCME) and the conventional real-time methods (B-RCSCME and O-IVA).
In Section~\ref{sec:proposed_method}, we propose online algorithms for NSR-ILRMA and RCSCME based on MWLE.
We also propose several stabilization techniques for online NSR-ILRMA and acceleration techniques for online RCSCME to achieve stable and efficient real-time processing.
In Section~\ref{sec:experiments}, we simulate situations where the target speaker is either stationary or moving under diffuse noise conditions and demonstrate that the proposed method achieves higher real-time speech extraction performance than conventional methods.
Furthermore, we use real-world-recorded signals and confirm the effectiveness of the proposed method in practical situations.
Finally, in Section~\ref{sec:conclusion}, we conclude and summarize the paper.

\section{RELATED WORKS}
\label{sec:related_methods}

In this section, we explain the related offline methods, i.e., ILRMA~\cite{Kitamura2016TASLP,Kitamura2018ASS_book}, NSR-ILRMA~\cite{Ishikawa2025Access}, and RCSCME~\cite{Kubo2020TASLP}.
Then, we describe the conventional real-time methods, B-RCSCME~\cite{Ishikawa2025Access} and O-IVA~\cite{Taniguchi2014HSCMA,Nakashima2022APSIPA}.

\subsection{ILRMA AND NSR-ILRMA}
\label{ssec:ILRMAand_NSR-ILRMA}

Let $\bm{\ObsSignal}_{\IndFreqbin\IndFrame} = (\ObsSignal_{\IndFreqbin \IndFrame 1}, ..., \ObsSignal_{\IndFreqbin \IndFrame \NumMic})^{\Transpose} \in \mathbb{C}^{\NumMic}$, $\bm{\SrcSignal}_{\IndFreqbin \IndFrame} = (\SrcSignal_{\IndFreqbin \IndFrame 1}, ..., \SrcSignal_{\IndFreqbin \IndFrame \NumSrc})^{\Transpose} \in \mathbb{C}^{\NumSrc}$, and $\bm{\SepSignal}_{\IndFreqbin \IndFrame} = (\SepSignal_{\IndFreqbin \IndFrame 1}, ..., \SepSignal_{\IndFreqbin \IndFrame \NumSrc})^{\Transpose} \in \mathbb{C}^{\NumSrc}$ be the STFTs of the observed, source, and separated signals, respectively.
Here, $\IndFreqbin \in \{1, ..., \NumFreqbin\}$, $\IndFrame \in \{1, ..., \NumFrame\}$, $\IndMic \in \{1, ..., \NumMic\}$, and $\IndSrc \in \{1, ..., \NumSrc\}$ are the indices of the frequency bins, time frames, microphones, and sources, respectively, and $^{\Transpose}$ represents the transpose.
In ILRMA~\cite{Kitamura2016TASLP}, it is assumed that each source is a point source and the window length of an STFT is sufficiently larger than the room reverberation time.
Under these assumptions, instantaneous mixing in the time-frequency domain approximately holds and the observed signals can be modeled as
\begin{align}
    \label{eq:def:ObsSignal}
    \bm{\ObsSignal}_{\IndFreqbin \IndFrame} = \MixMat_{\IndFreqbin} \bm{\SrcSignal}_{\IndFreqbin \IndFrame},
\end{align}
where $\MixMat_{\IndFreqbin} = (\MixVec_{\IndFreqbin 1}, ..., \MixVec_{\IndFreqbin \NumSrc}) \in \mathbb{C}^{\NumMic \times \NumSrc}$ is the mixing matrix, which represents time-invariant spatial characteristics of the mixing system, and $\MixVec_{\IndFreqbin \IndSrc}$ is the steering vector of the $\IndSrc$th source.
If $\NumMic = \NumSrc$ and $\MixMat_{\IndFreqbin}$ is regular, the separated signal $\bm{\SepSignal}_{\IndFreqbin\IndFrame}$ can be obtained as
\begin{align}
    \label{eq:def:SepSignal}
    \bm{\SepSignal}_{\IndFreqbin \IndFrame} = \DemixMat_{\IndFreqbin} \bm{\ObsSignal}_{\IndFreqbin \IndFrame},
\end{align}
where $\DemixMat_{\IndFreqbin} = (\bm{\DemixVec}_{\IndFreqbin 1}, ..., \bm{\DemixVec}_{\IndFreqbin \NumSrc})^{\Hermite} =: \MixMat_{\IndFreqbin}^{-1}$ is called the demixing matrix and $^{\Hermite}$ denotes the Hermitian transpose.

In ILRMA, each time-frequency bin of the separated signal $\SepSignal_{\IndFreqbin \IndFrame \IndSrc}$ is assumed to follow a univariate complex Gaussian distribution whose mean and time-varying variance are $0$ and $\SrcModel_{\IndFreqbin\IndFrame\IndSrc} > 0$, respectively, as
\begin{align}
    p_{\mathrm{UG}}(\SepSignal_{\IndFreqbin\IndFrame\IndSrc}; 0, \SrcModel_{\IndFreqbin\IndFrame\IndSrc}) \propto \frac{1}{\SrcModel_{\IndFreqbin\IndFrame\IndSrc}} \exp{\biggl( - \frac{\abs{\SepSignal_{\IndFreqbin\IndFrame\IndSrc}}^{2}}{\SrcModel_{\IndFreqbin\IndFrame\IndSrc}}\biggr)}.
\end{align}
Here, the time-varying variance $\SrcModel_{\IndFreqbin \IndFrame \IndSrc}$ is modeled by nonnegative matrix factorization (NMF)~\cite{Lee1999Nature} as
\begin{align}
    \label{eq:def:SrcModel}
    \SrcModel_{\IndFreqbin \IndFrame \IndSrc} = \sum_{\IndBasis} \Basis_{\IndFreqbin \IndBasis \IndSrc} \Activation_{\IndBasis \IndFrame \IndSrc},
\end{align}
where $\Basis_{\IndFreqbin \IndBasis \IndSrc} \geq 0$ and $\Activation_{\IndBasis \IndFrame \IndSrc} \geq 0$ are the NMF variables and are called basis and activation variables, respectively, and $\IndBasis \in \{1, ..., \NumBasis\}$ is the index of the NMF bases.
The cost function $\Cost{\ILRMA}$ is given as a negative log-likelihood of the observed signals, excluding constant terms independent of the objective variables $\DemixMat_{\IndFreqbin}$, $\Basis_{\IndFreqbin\IndBasis\IndSrc}$, and $\Activation_{\IndBasis\IndFrame\IndSrc}$, as follows:
\begin{align}
    \label{eq:Cost:NaiveILRMA}
    \Cost{\ILRMA} &= \sum_{\IndFreqbin, \IndFrame, \IndSrc} \Biggl(\frac{\abs{\DemixVec_{\IndFreqbin \IndSrc}^{\Hermite} \bm{\ObsSignal}_{\IndFreqbin \IndFrame}}^{2}}{\sum_{\IndBasis} \Basis_{\IndFreqbin \IndBasis \IndSrc} \Activation_{\IndBasis \IndFrame \IndSrc}} + \log{\sum_{\IndBasis} \Basis_{\IndFreqbin \IndBasis \IndSrc} \Activation_{\IndBasis \IndFrame \IndSrc}}\Biggr) 
    \nonumber\\
    &\phantom{=}- \NumFrame \sum_{\IndFreqbin} \log{\abs{\det{\DemixMat_{\IndFreqbin}}}^{2}}.
\end{align}

To improve the performance of ILRMA, its extensions utilizing spatial regularization have been proposed~\cite{Mitsui2018ICASSP}.
In the supposed situation where the approximate direction of the target speaker and the microphone array geometry are known, we can obtain the approximate steering vector of the target speech in advance.
Note that in practice, we cannot obtain the precise steering vector owing to factors such as reverberation and slight differences in the position of the target speaker.
As a spatially regularized ILRMA that only uses the approximate steering vector of the target speech, we previously proposed NSR-ILRMA~\cite{Ishikawa2025Access}.
In NSR-ILRMA, we introduce the following regularizer into ILRMA:
\begin{align}
    \label{eq:Regularizer:NSR-ILRMA}
    \NumFrame \sum_{\IndFreqbin, \IndSrc} \RegWeight_{\IndFreqbin\IndSrc} (1 - \delta_{\IndSrc \IndTargetSrc}) \abs{\DemixVec_{\IndFreqbin\IndSrc}^{\Hermite} \Prior{\MixVec}_{\IndFreqbin}^{\upt}}^{2},
\end{align}
where $\RegWeight_{\IndFreqbin \IndSrc} \geq 0$, $\delta_{\IndSrc \IndTargetSrc}$, $\IndTargetSrc$, and $\Prior{\MixVec}_{\IndFreqbin}^{\upt} \in \mathbb{C}^{\NumSrc}$ denote a weight parameter for the regularizer, Kronecker's delta, the source index corresponding to the target speech, and the prior steering vector of the target speech, respectively.
This regularizer induces the demixing filters for noise, $\DemixVec_{\IndFreqbin\IndSrc}\ (\IndSrc \ne \IndTargetSrc)$, to form the null of the prior steering vector of the target speech, $\Prior{\MixVec}_{\IndFreqbin}^{\upt}$, and thus, it is expected that the $\IndTargetSrc$th separated signal, $\SepSignal_{\IndFreqbin\IndFrame\IndTargetSrc} = \DemixVec_{\IndFreqbin\IndTargetSrc}^{\Hermite} \bm{\ObsSignal}_{\IndFreqbin\IndFrame}$, will contain the target speech signal.
By introducing this regularizer (\ref{eq:Regularizer:NSR-ILRMA}) into the cost function of ILRMA (\ref{eq:Cost:NaiveILRMA}), we obtain the cost function of NSR-ILRMA, $\Cost{\NSR}$, as
\begin{align}
    \label{eq:Cost:NSR-ILRMA}
    \Cost{\NSR} = \Cost{\ILRMA} + \NumFrame \sum_{\IndFreqbin, \IndSrc} \RegWeight_{\IndFreqbin \IndSrc} (1 - \delta_{\IndSrc \IndTargetSrc}) \abs{\DemixVec_{\IndFreqbin \IndSrc}^{\Hermite} \Prior{\MixVec}_{\IndFreqbin}^{\upt}}^{2}.
\end{align}
$\Cost{\NSR}$ is minimized by iteratively updating the demixing matrix $\DemixMat_{\IndFreqbin}$ and the NMF variables $\Basis_{\IndFreqbin\IndBasis\IndSrc}$ and $\Activation_{\IndBasis\IndFrame\IndSrc}$.
For the NMF variables $\Basis_{\IndFreqbin\IndBasis\IndSrc}$ and $\Activation_{\IndBasis\IndFrame\IndSrc}$, the update rules based on the MM algorithm are given in \cite{Kitamura2016TASLP} as 
\begin{align}
    \label{eq:NSR-ILRMA:update:basis}
    \Basis_{\IndFreqbin\IndBasis\IndSrc} &\leftarrow \Basis_{\IndFreqbin\IndBasis\IndSrc} \sqrt{\frac{\sum_{\IndFrame} \abs{\SepSignal_{\IndFreqbin\IndFrame\IndSrc}}^{2} \Activation_{\IndBasis\IndFrame\IndSrc} \bigl(\sum_{\IndBasis'} \Basis_{\IndFreqbin\IndBasis'\IndSrc} \Activation_{\IndBasis'\IndFrame\IndSrc} \bigr)^{-2}}{\sum_{\IndFrame} \Activation_{\IndBasis\IndFrame\IndSrc} \bigl(\sum_{\IndBasis'} \Basis_{\IndFreqbin\IndBasis'\IndSrc} \Activation_{\IndBasis'\IndFrame\IndSrc} \bigr)^{-1}}},
    \\
    \label{eq:NSR-ILRMA:update:activation}
    \Activation_{\IndBasis\IndFrame\IndSrc} &\leftarrow \Activation_{\IndBasis\IndFrame\IndSrc} \sqrt{\frac{\sum_{\IndFreqbin} \abs{\SepSignal_{\IndFreqbin\IndFrame\IndSrc}}^{2} \Basis_{\IndFreqbin\IndBasis\IndSrc} \bigl(\sum_{\IndBasis'} \Basis_{\IndFreqbin\IndBasis'\IndSrc} \Activation_{\IndBasis'\IndFrame\IndSrc} \bigr)^{-2}}{\sum_{\IndFreqbin} \Basis_{\IndFreqbin\IndBasis\IndSrc} \bigl(\sum_{\IndBasis'} \Basis_{\IndFreqbin\IndBasis'\IndSrc} \Activation_{\IndBasis'\IndFrame\IndSrc} \bigr)^{-1}}}.
\end{align}
Details of the MM algorithm are provided in Appendix~A.
The demixing matrix $\DemixMat_{\IndFreqbin}$ is updated by iterative projection (IP)~\cite{Ono2011WASPAA} as follows:
\begin{align}
    \label{eq:IP:NSR-ILRMA:CovMat_Reg}
    \CovRegNSR_{\IndFreqbin \IndSrc} &\leftarrow \frac{1}{\NumFrame} \sum_{\IndFrame=1}^{\NumFrame} \frac{\bm{\ObsSignal}_{\IndFreqbin \IndFrame} \bm{\ObsSignal}_{\IndFreqbin \IndFrame}^{\Hermite}}{\SrcModel_{\IndFreqbin \IndFrame \IndSrc}} + \RegWeight_{\IndFreqbin \IndSrc} (1 - \delta_{\IndSrc \IndTargetSrc}) \Prior{\MixVec}_{\IndFreqbin}^{\upt} \bigl( \Prior{\MixVec}_{\IndFreqbin}^{\upt} \bigr)^{\Hermite},
    \\
    \label{eq:IP:NSR-ILRMA:vec}
    \DemixVec_{\IndFreqbin\IndSrc} &\leftarrow \bigl(\DemixMat_{\IndFreqbin} \CovRegNSR_{\IndFreqbin \IndSrc}\bigr)^{-1} \UnitVec_{\IndSrc},
    \\
    \label{eq:IP:NSR-ILRMA:w}
    \DemixVec_{\IndFreqbin\IndSrc} &\leftarrow \DemixVec_{\IndFreqbin\IndSrc} / \sqrt{\DemixVec_{\IndFreqbin\IndSrc}^{\Hermite} \CovRegNSR_{\IndFreqbin\IndSrc} \DemixVec_{\IndFreqbin\IndSrc}},
\end{align}
where $\UnitVec_{\IndSrc} \in \mathbb{R}^{\NumSrc}$ is a one-hot vector whose $\IndSrc$th element is one and the others are zero.
The update of each variable guarantees the monotonically nonincreasing property of the cost function $\Cost{\NSR}$.

In NSR-ILRMA, the scale of $\SepSignal_{\IndFreqbin\IndFrame\IndSrc}$ can vary across the frequency bins~\cite{Ishikawa2025Access}.
To fix the scales of $\SepSignal_{\IndFreqbin\IndFrame\IndSrc}$ among the frequency bins, the projection back method~\cite{Murata2001Neurocomputing} is applied to $\SepSignal_{\IndFreqbin\IndFrame\IndSrc}$ after the parameter estimation.

\subsection{RCSCME}
\label{ssec:RCSCME}

RCSCME assumes a situation where a single (point-source) target speaker exists in a diffuse noise environment.
More specifically, it relies on the following spatial and statistical assumptions: the target speech source is a point source and its power spectrogram is sparse, whereas the noise is diffuse and not a point source.
When ILRMA, NSR-ILRMA, or other source separation methods based on linear demixing filters are applied to mixtures recorded in this situation, a residual diffuse noise component remains in the separated signal corresponding to the target speech in principle~\cite{Araki2003EURASIP,Araki2003IEEETSAP}.
On the other hand, the other $\NumMic-1$ separated signals contain only the diffuse noise component and cancel the target speech signal accurately~\cite{Takahashi2009TASLP}.
Utilizing this property, we can estimate the rank-$(\NumMic-1)$ component of the spatial covariance matrix (SCM) for the diffuse noise and the steering vector of the target speech with high accuracy.
In RCSCME, these parameters, calculated from the outputs of ILRMA or NSR-ILRMA, are fixed, and then the deficient rank-$1$ component of the diffuse noise SCM and the time-varying variances of the target speech and diffuse noise are estimated.
Finally, using the estimated parameters, a multichannel Wiener filter is applied to the observed signal to accurately extract the target speech signal.

In RCSCME, the observed signal $\bm{\ObsSignal}_{\IndFreqbin\IndFrame}$ is assumed to follow a multivariate complex Gaussian distribution\footnote{
    Note that in \cite{Kubo2020TASLP}, a multivariate complex generalized Gaussian distribution was used for the generative model of the observed signal.
    However, it was also reported in \cite{Kubo2020TASLP} that if we use a multivariate complex Gaussian distribution as the generative model, the speech extraction performance is sufficiently high and the number of iterations needed to achieve the peak performance is small.
    In this research, it is desirable to reduce the computational cost for real-time applicability, and thus, we focus on the assumption that the observed signal follows a multivariate complex Gaussian distribution.
} as
\begin{align}
    \label{eq:RCSCME:generative_model_obs_signal}
    p_{\mathrm{MG}}(\bm{\ObsSignal}_{\IndFreqbin\IndFrame}; \bm{0}_{\NumMic}, \SCMX_{\IndFreqbin\IndFrame}) \propto \frac{1}{\det{\SCMX_{\IndFreqbin\IndFrame}}} \exp{\Bigl( - \bm{\ObsSignal}_{\IndFreqbin\IndFrame}^{\Hermite} \bigl( \SCMX_{\IndFreqbin\IndFrame} \bigr)^{-1} \bm{\ObsSignal}_{\IndFreqbin\IndFrame} \Bigr)}, 
\end{align}
where $\bm{0}_{\NumMic} \in \mathbb{R}^{\NumMic}$ is the $\NumMic$-dimensional zero vector and $\SCMX_{\IndFreqbin\IndFrame} \in \mathbb{C}^{\NumMic \times \NumMic}$ is the covariance matrix of the observed signal.
$\SCMX_{\IndFreqbin \IndFrame}$ is modeled by the time-varying weighted summation of the SCMs of the directional target speech and diffuse noise as
\begin{align}
    \label{eq:RCSCME:SCMX_model}
    \SCMX_{\IndFreqbin \IndFrame} = \VarT_{\IndFreqbin \IndFrame} \SV_{\IndFreqbin} \bigl( \SV_{\IndFreqbin} \bigr)^{\Hermite} + \VarN_{\IndFreqbin \IndFrame} \SCMN_{\IndFreqbin},
\end{align}
where $\VarT_{\IndFreqbin\IndFrame} > 0$ and $\VarN_{\IndFreqbin \IndFrame} > 0$ are the time-varying variances of the target speech and diffuse noise, respectively, $\SV_{\IndFreqbin} \in \mathbb{C}^{\NumMic}$ is the time-invariant steering vector of the target speech, and $\SCMN_{\IndFreqbin} \in \mathbb{C}^{\NumMic \times \NumMic}$ is the time-invariant full-rank SCM of the diffuse noise.
Here, $\SV_{\IndFreqbin} \bigl( \SV_{\IndFreqbin} \bigr)^{\Hermite}$ represents the time-invariant rank-$1$ SCM of the directional target speech.
To induce sparsity, we introduce the inverse gamma distribution for a prior distribution of $\VarT_{\IndFreqbin\IndFrame}$ as
\begin{align}
    \label{eq:RCSCME:prior_VarT}
    p_{\mathrm{IG}}\bigl( \VarT_{\IndFreqbin\IndFrame}; \ShapeIG, \ScaleIG \bigr) \propto \bigl( \VarT_{\IndFreqbin\IndFrame} \bigr)^{-\ShapeIG-1} \exp{\biggl( - \frac{\ScaleIG}{\VarT_{\IndFreqbin\IndFrame}} \biggr)},
\end{align}
where $\ShapeIG > 0$ and $\ScaleIG > 0$ are the shape and scale parameters of the inverse gamma distributions, respectively.
For the time-invariant parameters $\SV_{\IndFreqbin}$ and $\SCMN_{\IndFreqbin}$, RCSCME utilizes the demixing matrix estimated by (NSR-)ILRMA, $\DemixMat_{\IndFreqbin}$, and the channel index corresponding to the target speech, $\IndTargetSrc$. 
That is, $\SV_{\IndFreqbin}$ can be calculated as the $\IndTargetSrc$th column vector of $\DemixMat_{\IndFreqbin}^{-1}$.
For $\SCMN_{\IndFreqbin}$, since the $\NumMic - 1$ separated signals estimated by (NSR-)ILRMA can accurately estimate the diffuse noise, $\SCMN_{\IndFreqbin}$ is modeled in \cite{Kubo2020TASLP} by a full-rank matrix as
\begin{align}
    \label{eq:RCSCME:model_SCMN}
    \SCMN_{\IndFreqbin} &= \SCMNd_{\IndFreqbin} + \WeightSCMN_{\IndFreqbin} \NoiseSV_{\IndFreqbin} \NoiseSV_{\IndFreqbin}^{\Hermite},
    \\
    \label{eq:RCSCME:def:SCMNd}
    \SCMNd_{\IndFreqbin} &= \frac{1}{\NumFrame} \sum_{\IndFrame = 1}^{\NumFrame} \NoiseSrcImage_{\IndFreqbin\IndFrame} \NoiseSrcImage_{\IndFreqbin\IndFrame}^{\Hermite},
    \\
    \label{eq:RCSCME:def:NoiseSrcImage}
    \NoiseSrcImage_{\IndFreqbin\IndFrame} 
    &= \DemixMat_{\IndFreqbin}^{-1} \bigl( \DemixVec_{\IndFreqbin1}^{\Hermite} \bm{\ObsSignal}_{\IndFreqbin\IndFrame}, ..., \DemixVec_{\IndFreqbin(\IndTargetSrc-1)}^{\Hermite} \bm{\ObsSignal}_{\IndFreqbin\IndFrame}, 0,
    \nonumber\\
    & \phantom{= \DemixMat_{\IndFreqbin}^{-1} = } \DemixVec_{\IndFreqbin(\IndTargetSrc+1)}^{\Hermite} \bm{\ObsSignal}_{\IndFreqbin\IndFrame}, ..., \DemixVec_{\IndFreqbin\NumMic}^{\Hermite} \bm{\ObsSignal}_{\IndFreqbin\IndFrame}
    \bigr)^{\Transpose}
    \nonumber\\
    &= \bigl( \Identity_{\NumMic} - \SV_{\IndFreqbin} \DemixVec_{\IndFreqbin\IndTargetSrc}^{\Hermite} \bigr) \bm{\ObsSignal}_{\IndFreqbin\IndFrame},
\end{align}
where $\SCMNd_{\IndFreqbin} \in \mathbb{C}^{\NumMic \times \NumMic}$ is the rank-$(\NumMic-1)$ component of the diffuse noise SCM, $\NoiseSV_{\IndFreqbin} \in \mathbb{C}^{\NumMic}$ is a vector that is linearly independent of all the column vectors of $\SCMNd_{\IndFreqbin}$, $\WeightSCMN_{\IndFreqbin}$ is a scalar variable, $\NoiseSrcImage_{\IndFreqbin\IndFrame} \in \mathbb{C}^{\NumMic}$ is the source image of diffuse noise excluding the $\IndTargetSrc$th channel, and $\Identity_{\NumMic} \in \mathbb{C}^{\NumMic \times \NumMic}$ is the $\NumMic$-dimensional identity matrix.
Here, $\WeightSCMN_{\IndFreqbin} \NoiseSV_{\IndFreqbin} \NoiseSV_{\IndFreqbin}^{\Hermite}$ represents a rank-$1$ matrix that complements the deficient rank-$1$ component of $\SCMN_{\IndFreqbin}$.
$\NoiseSV_{\IndFreqbin}$ is set to, for example, $\SV_{\IndFreqbin}$ or a unit eigenvector of $\SCMNd_{\IndFreqbin}$ corresponding to a zero eigenvalue.
For $\IndTargetSrc$, if we use ILRMA as a preprocessing of RCSCME, since ILRMA is a fully blind method and it is not determined which of the demixing filters estimated by ILRMA corresponds to the target speech, we should determine the channel index corresponding to the target speech, $\IndTargetSrc$, by some means.
In \cite{Kubo2020TASLP}, for example, the maximum kurtosis criterion~\cite{Fujihara2008IWAENC} is used and works well in offline scenarios.
On the other hand, if we use NSR-ILRMA as a preprocessing of RCSCME, the spatial regularizer directly controls the target speech channel index.

In RCSCME, the parameters calculated from the estimates of ILRMA (i.e., $\SV_{\IndFreqbin}$, $\SCMNd_{\IndFreqbin}$, and $\NoiseSV_{\IndFreqbin}$) are fixed, and the to-be-estimated parameters are $\VarT_{\IndFreqbin\IndFrame}$, $\VarN_{\IndFreqbin\IndFrame}$, and $\WeightSCMN_{\IndFreqbin}$.
To estimate these parameters on the basis of maximum a posteriori, the cost function $\Cost{\RCSCME}$ is given as the negative log-posterior of the observed signals with the prior distribution of $\VarT_{\IndFreqbin\IndFrame}$, excluding constant terms independent of the objective variables $\VarT_{\IndFreqbin\IndFrame}$, $\VarN_{\IndFreqbin\IndFrame}$, and $\WeightSCMN_{\IndFreqbin}$, as follows:
\begin{align}
    \label{eq:def:Cost_RCSCME}
    \Cost{\RCSCME} = \sum_{\IndFreqbin, \IndFrame} \Biggl(& \bm{\ObsSignal}_{\IndFreqbin\IndFrame}^{\Hermite} \bigl( \SCMX_{\IndFreqbin\IndFrame} \bigr)^{-1} \bm{\ObsSignal}_{\IndFreqbin\IndFrame} + \log{\det{\SCMX_{\IndFreqbin\IndFrame}}} 
    \nonumber\\
    &+ (\ShapeIG + 1) \log{\VarT_{\IndFreqbin\IndFrame}} + \frac{\ScaleIG}{\VarT_{\IndFreqbin\IndFrame}} \Biggr).
\end{align}

The update rule of the objective variables is derived on the basis of the ME algorithm~\cite{Fevotte2011NC} in \cite{Kubo2020TASLP} as follows\footnote{
Note that in \cite{Kubo2020TASLP}, the update rule of the objective variables based on the MM algorithm is also derived.
However, it was also reported in \cite{Kubo2020TASLP} that the number of iterations to achieve high speech extraction performance when using the ME algorithm is smaller than that when using the MM algorithm.
For real-time implementation, it is desirable to reduce the number of iterations, and thus, we adopt the ME-algorithm-based update rule.
}:
\begin{align}
    \label{eq:RCSCME:update:VarT}
    \VarT_{\IndFreqbin\IndFrame} &\leftarrow \VarT_{\IndFreqbin\IndFrame} \left(
        \frac{
            \abs{\bm{\ObsSignal}_{\IndFreqbin\IndFrame}^{\Hermite} \bigl(\SCMX_{\IndFreqbin\IndFrame}\bigr)^{-1} \SV_{\IndFreqbin}}^{2}
            + \ScaleIG / \bigl(\VarT_{\IndFreqbin\IndFrame}\bigr)^{2} 
        }{
            \bigl(\SV_{\IndFreqbin}\bigr)^{\Hermite} \bigl( \SCMX_{\IndFreqbin\IndFrame} \bigr)^{-1} \SV_{\IndFreqbin} 
            + (\ShapeIG + 1) / \VarT_{\IndFreqbin\IndFrame}
        }
    \right),
    \\
    \label{eq:RCSCME:update:VarN}
    \VarN_{\IndFreqbin\IndFrame} &\leftarrow \VarN_{\IndFreqbin\IndFrame} \left(
        \frac{
            \bm{\ObsSignal}_{\IndFreqbin\IndFrame}^{\Hermite} \bigl( \SCMX_{\IndFreqbin\IndFrame} \bigr)^{-1} \SCMN_{\IndFreqbin} \bigl( \SCMX_{\IndFreqbin\IndFrame} \bigr)^{-1} \bm{\ObsSignal}_{\IndFreqbin\IndFrame}
        }{
            \trace \Bigl( \bigl( \SCMX_{\IndFreqbin\IndFrame} \bigr)^{-1} \SCMN_{\IndFreqbin} \Bigr)
        }
    \right),
    \\
    \label{eq:RCSCME:update:WeightSCMN}
    \WeightSCMN_{\IndFreqbin} &\leftarrow \WeightSCMN_{\IndFreqbin} \left(
        \frac{
            \sum_{\IndFrame=1}^{\NumFrame} \VarN_{\IndFreqbin\IndFrame} \abs{\NoiseSV_{\IndFreqbin}^{\Hermite} \bigl( \SCMX_{\IndFreqbin\IndFrame} \bigr)^{-1} \bm{\ObsSignal}_{\IndFreqbin\IndFrame}}^{2}
        }{
            \sum_{\IndFrame=1}^{\NumFrame} \VarN_{\IndFreqbin\IndFrame} \NoiseSV_{\IndFreqbin}^{\Hermite} \bigl( \SCMX_{\IndFreqbin\IndFrame} \bigr)^{-1} \NoiseSV_{\IndFreqbin}
        }
    \right).
\end{align} 
Note that $\SCMX_{\IndFreqbin\IndFrame}$ is updated only after the update of $\VarN_{\IndFreqbin\IndFrame}$ and $\WeightSCMN_{\IndFreqbin}$ using (\ref{eq:RCSCME:SCMX_model}).
Details of the ME algorithm are provided in Appendix~A.
To stabilize the performance, we initialize the objective variables $\VarT_{\IndFreqbin\IndFrame}$ and $\VarN_{\IndFreqbin\IndFrame}$ as
\begin{align}
    \label{eq:RCSCME:initialize:VarT}
    \VarT_{\IndFreqbin\IndFrame} &= \abs{\DemixVec_{\IndFreqbin\IndTargetSrc}^{\Hermite} \bm{\ObsSignal}_{\IndFreqbin\IndFrame}}^{2},
    \\
    \label{eq:RCSCME:initialize:VarN}
    \VarN_{\IndFreqbin\IndFrame} &= \frac{1}{\NumMic} \NoiseSrcImage_{\IndFreqbin\IndFrame}^{\Hermite} \bigl( \SCMNd_{\IndFreqbin} \bigr)^{+} \NoiseSrcImage_{\IndFreqbin\IndFrame},
\end{align}
where the operator $^{+}$ for matrices denotes the MP inverse.
In \cite{Kubo2020TASLP}, $\WeightSCMN_{\IndFreqbin}$ is initialized as the minimum nonzero eigenvalue of $\SCMNd_{\IndFreqbin}$.
However, it is computationally costly to calculate the minimum eigenvalue, which can hinder real-time implementation.
To reduce the computational cost, in \cite{Ishikawa2025Access}, the average of the eigenvalues is used and $\WeightSCMN_{\IndFreqbin}$ is initialized as 
\begin{align}
    \label{eq:RCSCME:initialization:WeightSCMN}
    \WeightSCMN_{\IndFreqbin} = \InitLambdaScaling \frac{\trace \bigl( \SCMNd_{\IndFreqbin} \bigr)}{\NumMic},
\end{align}
where $\InitLambdaScaling$ is a hyperparameter to improve the speech extraction performance.

After the above parameter estimation, the source image of the target speech signal is extracted using a multichannel Wiener filter as
\begin{align}
    \label{eq:RCSCME:MWF}
    \TargetSrcImage_{\IndFreqbin\IndFrame} 
    = \VarT_{\IndFreqbin\IndFrame} \SV_{\IndFreqbin} \bigl( \SV_{\IndFreqbin} \bigr)^{\Hermite} \bigl( \SCMX_{\IndFreqbin\IndFrame} \bigr)^{-1} \bm{\ObsSignal}_{\IndFreqbin\IndFrame}.
\end{align}

The update rules (\ref{eq:RCSCME:update:VarT})--(\ref{eq:RCSCME:update:WeightSCMN}) include computationally expensive matrix operations, such as the computation of $\bigl( \SCMX_{\IndFreqbin\IndFrame} \bigr)^{-1}$, which have become a bottleneck for efficient updates.
In \cite{Kubo2020TASLP}, an acceleration technique that enables the iterative updates to consist solely of scalar operations by pre-computing the matrix operations collectively in the initialization step has also been proposed.
This results in a short computation time per iteration.
Additionally, as reported in \cite{Kubo2020TASLP}, the number of iterations needed to achieve sufficiently high speech extraction performance is small.
Consequently, the entire process of RCSCME is fast.
It has also been reported in \cite{Kubo2020TASLP} that RCSCME achieves a high speech extraction performance.

\subsection{B-RCSCME}
\label{ssec:blockwise-RCSCME}

\begin{figure*}[t]
    \centering
    \includegraphics[width=0.9\linewidth]{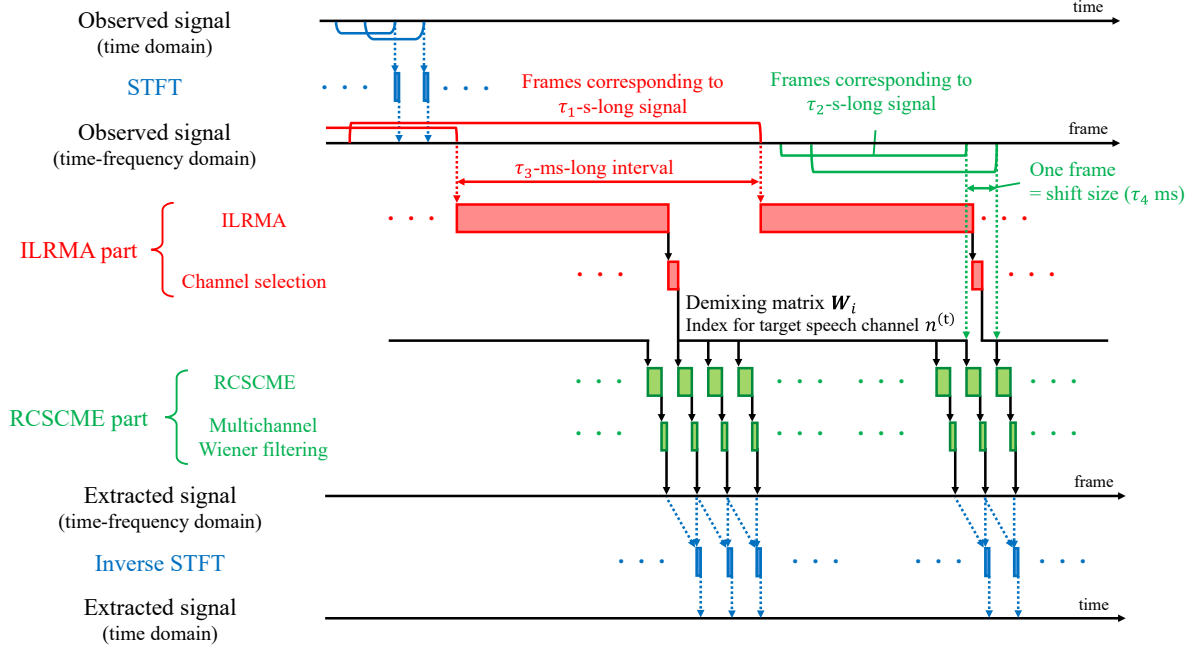}
    \caption{Schematic of parallel processing of B-RCSCME. This figure is cited from \cite{Ishikawa2025Access}.}
    \label{fig:schematic_B-RCSCME}
\end{figure*}

As discussed in Section~\ref{sec:related_methods}-\ref{ssec:RCSCME}, the processing of RCSCME is computationally efficient enough for real-time processing on modern computers.
However, (NSR-)ILRMA involves numerous matrix operations and requires a sufficient number of iterations to converge.
Thus, performing the entire procedure consisting of (NSR-)ILRMA and RCSCME within the shift length would require extremely high-performance computational resources, which is impractical for wide-range applications.

To address this issue, in \cite{Ishikawa2025Access}, we focused on the fact that the output of ILRMA required in RCSCME is only the demixing matrix $\DemixMat_{\IndFreqbin}$, which is a time-invariant parameter.
Consequently, we divide the entire process into the ILRMA and RCSCME parts.
If the spatial characteristics do not change abruptly (e.g., by a sudden large movement of the target speaker), the estimated $\DemixMat_{\IndFreqbin}$ is likely to remain similar across frames.
Considering this, it is expected that using $\DemixMat_{\IndFreqbin}$ estimated from slightly earlier frames does not considerably degrade the separation performance.
Thus, B-RCSCME executes the ILRMA and RCSCME parts in parallel by introducing the blockwise batch algorithm~\cite{Mukai2004IEICE,Mori2006EURASIP}.
Fig.~\ref{fig:schematic_B-RCSCME} shows the schematic of B-RCSCME.
In the ILRMA part, (NSR-)ILRMA estimates the demixing matrix $\DemixMat_{\IndFreqbin}$ and the target speech channel index $\IndTargetSrc$ at a relatively long interval.
In parallel, the RCSCME part is executed at every shift length interval using the latest $\DemixMat_{\IndFreqbin}$ and $\IndTargetSrc$ estimated by the ILRMA part.
Note that $\DemixMat_{\IndFreqbin}$ and $\IndTargetSrc$ used in the RCSCME part can be the estimates obtained one execution interval of the ILRMA part earlier.

Owing to this parallel implementation, the ILRMA part is not required to be computed within the shift length, which enables real-time execution of RCSCME under practical computational resources.

\subsection{O-IVA}
\label{ssec:online-iva}

O-IVA is an online extension of offline IVA and performs parameter estimation efficiently in a frame-by-frame manner~\cite{Taniguchi2014HSCMA,Nakashima2022APSIPA}.
In offline IVA, the separated signals are assumed to follow the following multivariate Laplace distribution with a mean of $\bm{0}_{\NumFreqbin}$ and a scale matrix of $\Identity_{\NumFreqbin}$~\cite{Kim2007TASLP,Hiroe2006ICA}: 
\begin{align}
    \label{eq:IVA:generative_model_of_separated_signal}
    p_{\mathrm{ML}}\bigl(\{\SepSignal_{\IndFreqbin\IndFrame\IndSrc}\}_{\IndFreqbin}^{\NumFreqbin}; \bm{0}_{\NumFreqbin}, \Identity_{\NumFreqbin}\bigr) \propto \exp{\Biggl( \sqrt{\sum_{i} \abs{\SepSignal_{\IndFreqbin\IndFrame\IndSrc}}^{2}} \Biggr)}.
\end{align}
The cost function $\Cost{\IVA}$ is given as a negative log-likelihood of the observed signals, excluding the terms independent of the objective variable $\DemixMat_{\IndFreqbin}$, as follows:
\begin{align}
    \label{eq:IVA:Cost}
    \Cost{\IVA} = \sum_{\IndSrc} \sum_{\IndFrame=1}^{\NumFrame} \sqrt{\sum_{\IndFreqbin} \abs{\DemixVec_{\IndFreqbin\IndSrc}^{\Hermite} \bm{\ObsSignal}_{\IndFreqbin\IndFrame}}^{2}} - \NumFrame \sum_{\IndFreqbin} \log{\abs{\det{\DemixMat_{\IndFreqbin}}}^{2}}.
\end{align}

It is difficult to directly minimize $\Cost{\IVA}$ with respect to $\DemixMat_{\IndFreqbin}$.
Thus, in auxiliary-function-based IVA (AuxIVA)~\cite{Ono2011WASPAA}, an iterative update rule is derived on the basis of the MM algorithm as follows.
First, we consider the following auxiliary function of $\Cost{\IVA}$:
\begin{align}
    \label{eq:IVA:AuxFunction}
    \AuxCost{\IVA} &= \sum_{\IndSrc} \sum_{\IndFrame=1}^{\NumFrame} \biggl( \frac{\sum_{\IndFreqbin} \abs{\DemixVec_{\IndFreqbin\IndSrc}^{\Hermite} \bm{\ObsSignal}_{\IndFreqbin\IndFrame}}^{2}}{2 \AuxVarIVA_{\IndFrame\IndSrc}} + \frac{\AuxVarIVA_{\IndFrame\IndSrc}}{2} \biggr) 
    \nonumber\\
    &\phantom{=} - \NumFrame \sum_{\IndFreqbin} \log{\abs{\det{\DemixMat_{\IndFreqbin}}}^{2}}
    \\
    &= \NumFrame \biggl( \sum_{\IndFreqbin, \IndSrc} \DemixVec_{\IndFreqbin\IndSrc}^{\Hermite} \CovIVA_{\IndFreqbin\IndSrc} \DemixVec_{\IndFreqbin\IndSrc} - \sum_{\IndFreqbin} \log{\abs{\det{\DemixMat_{\IndFreqbin}}}^{2}} \biggr) + \const,
\end{align}
where $\const$ denotes the term independent of $\DemixVec_{\IndFreqbin\IndSrc}$, $\AuxVarIVA_{\IndFrame\IndSrc} \geq 0$ is an auxiliary variable, and $\CovIVA_{\IndFreqbin\IndSrc}$ is a weighted spatial covariance matrix and is defined as
\begin{align}
    \label{eq:IVA:def:CovMatrix_AuxIVA}
    \CovIVA_{\IndFreqbin\IndSrc} = \frac{1}{\NumFrame} \sum_{\IndFrame=1}^{\NumFrame} \frac{\bm{\ObsSignal}_{\IndFreqbin\IndFrame}\bm{\ObsSignal}_{\IndFreqbin\IndFrame}^{\Hermite}}{2 \AuxVarIVA_{\IndFrame\IndSrc}}.
\end{align}
Here, $\Cost{\IVA} = \AuxCost{\IVA}$ holds if and only if
\begin{align}
    \label{eq:IVA:AuxVarCondition}
    \AuxVarIVA_{\IndFrame\IndSrc} = \sqrt{\sum_{\IndFreqbin} \abs{\DemixVec_{\IndFreqbin\IndSrc}^{\Hermite} \bm{\ObsSignal}_{\IndFreqbin\IndFrame}}^{2}}
\end{align}
holds.
Then, the cost function $\Cost{\IVA}$ is minimized by iteratively repeating the following two processes.
\begin{itemize}
    \item Update $\AuxVarIVA_{\IndFrame\IndSrc}$ and $\CovIVA_{\IndFreqbin\IndSrc}$ using (\ref{eq:IVA:AuxVarCondition}) and (\ref{eq:IVA:def:CovMatrix_AuxIVA}), respectively
    \item Minimize $\AuxCost{\IVA}$ with respect to $\DemixMat_{\IndFreqbin}$
\end{itemize}
For minimizing the auxiliary function $\AuxCost{\IVA}$ with respect to the demixing matrix $\DemixMat_{\IndFreqbin}$, iterative update rules such as IP and iterative source steering (ISS)~\cite{Scheibler2020ICASSP} are often used.

When we consider online execution of IVA, the computational cost to update the weighted covariance matrix $\CovIVA_{\IndFreqbin\IndSrc}$ poses a significant problem.
In offline AuxIVA, the auxiliary variables are updated each time the demixing matrix is updated, and thus, $\CovIVA_{\IndFreqbin\IndSrc}$ also needs to be recalculated using (\ref{eq:IVA:def:CovMatrix_AuxIVA}).
Consequently, when a sufficiently long batch is used to ensure high separation performance, the computational cost for calculating $\CovIVA_{\IndFreqbin\IndSrc}$ becomes high, which hinders real-time processing.
In O-IVA~\cite{Taniguchi2014HSCMA,Nakashima2022APSIPA}, to reduce the computational cost, when the $\LatestFrame$th frame of the observed signal is obtained, the approximate weighted spatial covariance matrix $\Est{\CovIVA}_{\IndFreqbin\IndSrc, \LatestFrame}$ is updated in an autoregressive manner as 
\begin{align}
    \label{eq:O-IVA_update_CovMat}
    \Est{\CovIVA}_{\IndFreqbin\IndSrc, \LatestFrame} \leftarrow (1 - \ForgetIVA) \frac{\bm{\ObsSignal}_{\IndFreqbin\LatestFrame} \bm{\ObsSignal}_{\IndFreqbin\LatestFrame}^{\Hermite}}{2 \AuxVarIVA_{\LatestFrame\IndSrc}} + \ForgetIVA \Est{\CovIVA}_{\IndFreqbin\IndSrc, \LatestFrame-1},
\end{align}
where $\ForgetIVA \in (0,1)$ denotes the forgetting factor, and the variable with a tilde and the subscript $\LatestFrame$, that is, $\Est{\cdot}_{,\LatestFrame}$, indicates the estimated parameter at the $\LatestFrame$th frame.
Here, $\Est{\CovIVA}_{\IndFreqbin\IndSrc, 0}$ is initialized as $\epsilon_{\IVA} \Identity_{\NumSrc}$ with the stability parameter $\epsilon_{\IVA}$.
The autoregressive update of $\Est{\CovIVA}_{\IndFreqbin\IndSrc,\LatestFrame}$ corresponds to assigning exponentially decaying weights to the observed signals, thereby allowing the estimation of $\Est{\CovIVA}_{\IndFreqbin\IndSrc,\LatestFrame}$ to emphasize more recent observations.
Therefore, it is expected to operate robustly even under conditions where the spatial characteristics change.
O-IVA iteratively repeats the following processes at the $\LatestFrame$th frame.
\begin{itemize}
    \item Update $\AuxVarIVA_{\LatestFrame\IndSrc}$ and $\Est{\CovIVA}_{\IndFreqbin\IndSrc, \LatestFrame}$ using (\ref{eq:IVA:AuxVarCondition}) and (\ref{eq:O-IVA_update_CovMat}), respectively
    \item Update $\DemixMat_{\IndFreqbin}$ by IP or ISS using $\Est{\CovIVA}_{\IndFreqbin\IndSrc, \LatestFrame}$ 
\end{itemize}
In addition, since spatial characteristics are not expected to change abruptly in the neighboring frames, we utilize the estimate of $\DemixMat_{\IndFreqbin}$ from the $(\LatestFrame-1)$th frame as the initial value for $\DemixMat_{\IndFreqbin}$ in the $\LatestFrame$th frame, reducing the number of iterations.
The detailed explanations for the update rules based on IP and ISS in O-IVA are provided in Appendix~B.
Similar to NSR-ILRMA, since the scale of $\SepSignal_{\IndFreqbin\LatestFrame\IndSrc}$ can vary across the frequency bins, the projection back method is applied to $\SepSignal_{\IndFreqbin\LatestFrame\IndSrc}$ after the estimation at the $\LatestFrame$th frame.

Furthermore, O-IVE, an online extension of IVE, has also been proposed~\cite{Ueda2021EUSIPCO,Ueda2024IEEE-TASLP}.
IVE extends IVA for (multiple) target source(s) extraction by explicitly assigning the same source model to the non-target (noise) sources.
In O-IVE, the weighted SCMs of the target and noise sources are updated in an autoregressive manner, as in O-IVA, and the demixing filter for each source is updated at each frame.

\section{PROPOSED METHODS}
\label{sec:proposed_method}

In this section, we describe the proposed online extension of the RCSCME-based method, that is, online algorithms for NSR-ILRMA and RCSCME.
In Section~\ref{sec:proposed_method}-\ref{ssec:motivation_and_strategy}, we present the motivation of the proposed method and the common strategy for deriving the online algorithms.
In Sections~\ref{sec:proposed_method}-\ref{ssec:online_NSR-ILRMA} and \ref{sec:proposed_method}-\ref{ssec:online_RCSCME}, we derive the proposed online algorithms for NSR-ILRMA and RCSCME, respectively.

\subsection{MOTIVATION AND STRATEGY}
\label{ssec:motivation_and_strategy}

\begin{figure*}[t]
    \centering
    \includegraphics[width=0.9\linewidth]{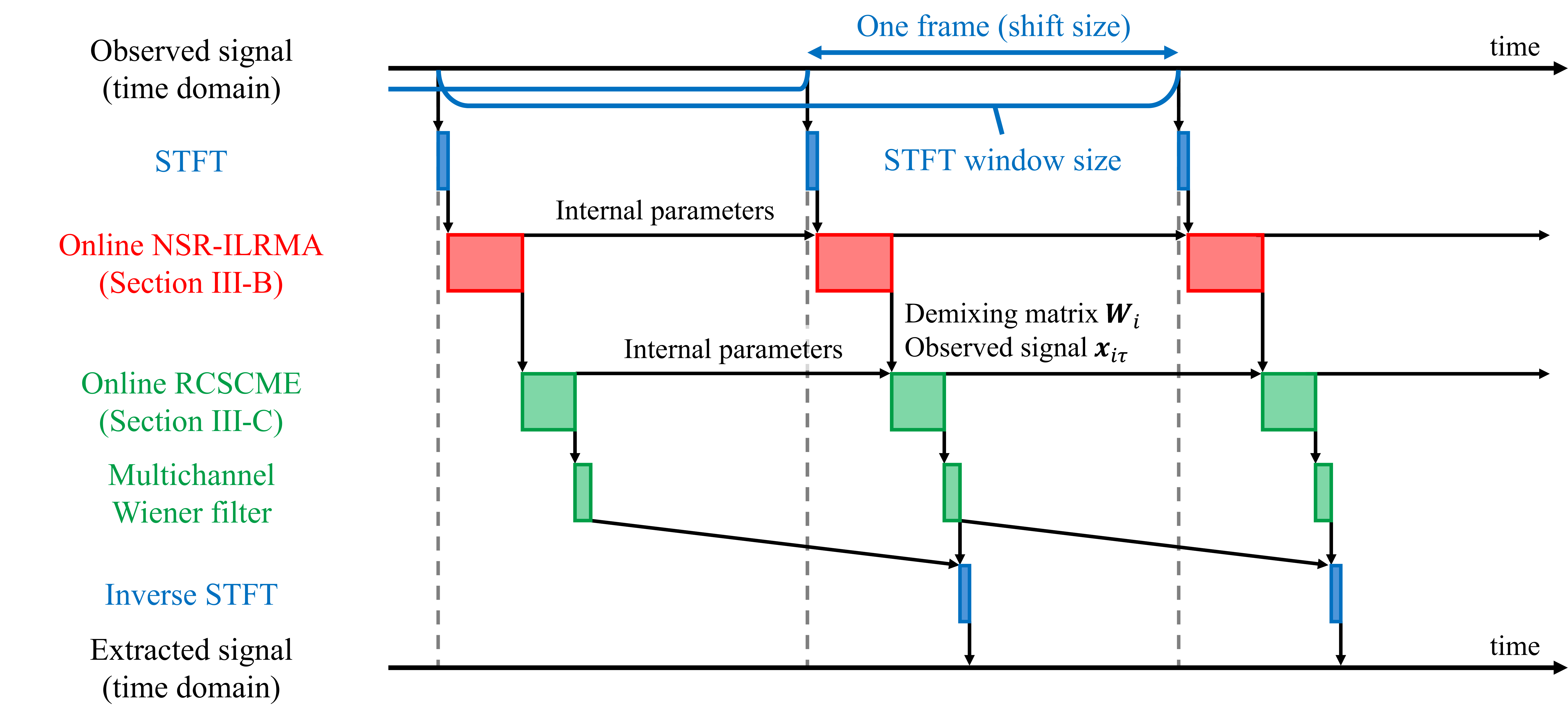}
    \caption{Schematic of online processing for RCSCME-based speech extraction method.}
    \label{fig:scheme_proposed_method}
\end{figure*}

In this paper, we address the real-time MSE problem in a scenario where a single directional and potentially moving target speaker exists under diffuse noise conditions.
As discussed in Section~\ref{sec:related_methods}-\ref{ssec:blockwise-RCSCME}, we previously proposed B-RCSCME for real-time MSE.
B-RCSCME relies on the blockwise batch algorithm that assumes the spatial characteristics do not change abruptly and estimates the time-invariant spatial parameters at relatively long intervals compared with the STFT shift length.
However, in scenarios where the spatial characteristics change (e.g., the target speaker moves), long interval for estimating the spatial parameters in B-RCSCME may affect the speech extraction performance.
On the other hand, as discussed in Section~\ref{sec:related_methods}-\ref{ssec:online-iva}, O-IVA updates the spatial parameters every time a new frame is observed, and it has been reported that O-IVA performs robustly when the source position changes by exponentially forgetting older observed signals.
Therefore, in this paper, we first propose an online algorithm for NSR-ILRMA.
In addition to online NSR-ILRMA, we incorporate the idea of O-IVA into RCSCME and derive an online algorithm for RCSCME to improve the robustness against a moving target speaker and reduce the computational cost of RCSCME.

The proposed online algorithms are derived in the following three steps as in \cite{Ishikawa2025EUSIPCO}.
In the first step, we formulate framewise cost functions for NSR-ILRMA and RCSCME.
In B-RCSCME, it is assumed that the spatial characteristics do not change abruptly within a single batch.
Thus, in practical scenarios where the spatial characteristics may change (e.g., when the target speaker moves), past frames in the batch may affect the speech extraction performance.
To address this problem, we introduce the concept of MWLE~\cite{Ahmed2005IEEETR,Fung2022InsuranceME} into NSR-ILRMA and RCSCME.
MWLE is an extension of MLE that reduces the influence of outliers by applying small weights to the likelihoods of data points identified as outliers by some means.
In this paper, we derive cost functions for NSR-ILRMA and RCSCME on the basis of MWLE by assigning smaller weights to the likelihoods of past frames to emphasize the recent frames of the observed signals.
Note that this formulation can be reasonably interpreted from the perspective that older frames of the observed signals can originate from different spatial characteristics compared with the current frames.
When the observed signals up to the $\LatestFrame$th frame are obtained, we consider the following cost function:
\begin{align}
    \label{eq:Cost_MWLE}
    \Cost{\mathrm{MWLE},\LatestFrame} = \sum_{\IndFrame=1}^{\LatestFrame} \WeightMWLE_{\IndFrame,\LatestFrame} \log{\ModelMWLE(\{\bm{\ObsSignal}_{\IndFreqbin\IndFrame}\}_{\IndFreqbin}; \ModelParameter_{\IndFrame})},
\end{align}
where $\WeightMWLE_{\IndFrame,\LatestFrame} \in [0, 1]$ is a weight parameter for the $\IndFrame$th-frame likelihood, $\ModelMWLE(\cdot)$ represents some generative model, and $\ModelParameter_{\IndFrame}$ is a set of model parameters.
In the second step, we derive the naive update rules for the framewise cost functions obtained in the first step.
Unfortunately, the naive update rules have the high computational costs, which hinders real-time execution.
Therefore, in the third step, to reduce the computational cost while securing a sufficient number of frames to achieve high performance, we approximate some time-varying parameters by those estimated at the previous frame and derive efficient online algorithms.
Table~1 illustrates the position of the proposed methods among the conventional offline BSS and MSE methods and their corresponding online extensions.
Furthermore, we propose techniques for online NSR-ILRMA to update the NMF bases stably and for online RCSCME to efficiently calculate the MP inverse by reconsidering the diffuse noise SCM model inspired by O-IVA.
We also incorporate the acceleration technique proposed for offline RCSCME in \cite{Kubo2020TASLP} into our online RCSCME to achieve efficient parameter updates.

\begin{table}
    \label{table:relationship}
    \caption{
        Position of proposed methods among conventional offline BSS and MSE methods and their corresponding online extensions
    }
    \includegraphics[width=\columnwidth]{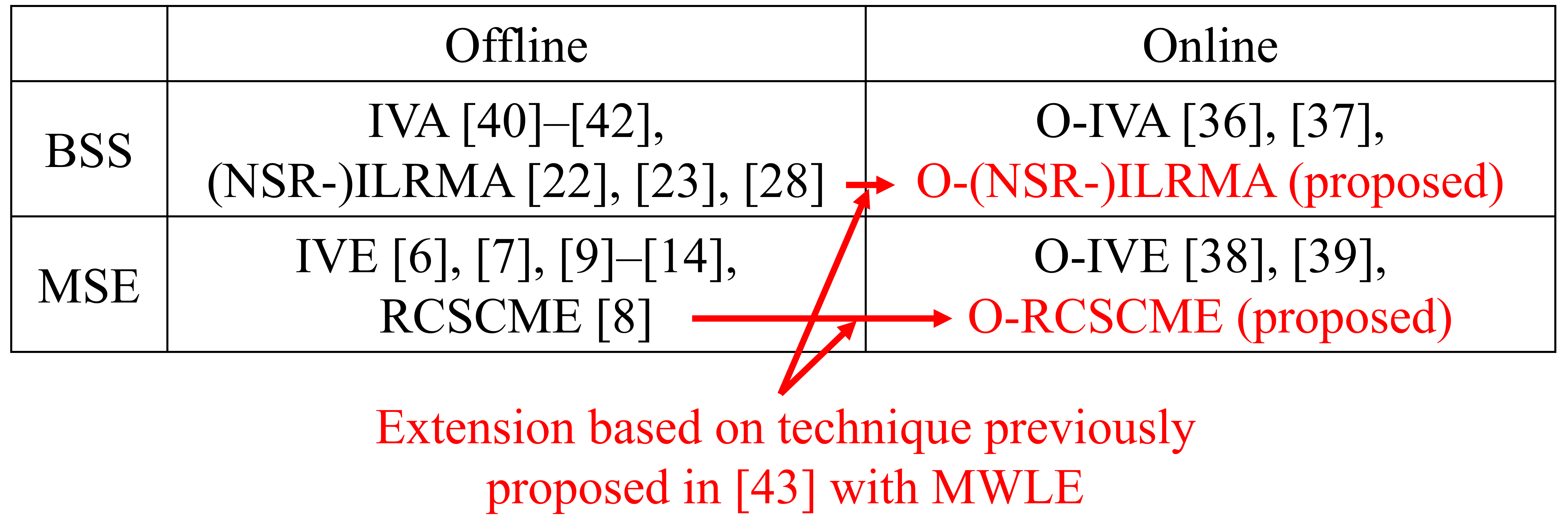}
\end{table}

Fig.~\ref{fig:scheme_proposed_method} shows a schematic of the proposed online RCSCME-based speech extraction method.
The online algorithms for NSR-ILRMA and RCSCME (referred to as \textit{online NSR-ILRMA} and \textit{online RCSCME}, respectively) are executed sequentially every time a new frame of the observed signal is obtained.
Hereafter, we refer to the proposed online RCSCME-based speech extraction method as \textit{O-RCSCME}.

\subsection{ONLINE NSR-ILRMA}
\label{ssec:online_NSR-ILRMA}

\subsubsection{Framewise Formulation}
\label{sssec:O-NSR-ILRMA:framewise_formulation}

When the observed signal up to the $\LatestFrame$th frame is obtained, we weight the likelihood of each frame $\IndFrame$ in (\ref{eq:Cost:NSR-ILRMA}) with a weight parameter $\WeightOILRMA_{\IndFrame,\LatestFrame} > 0$ for NSR-ILRMA and derive the framewise cost function of NSR-ILRMA $\Cost{\NSR,\LatestFrame}$ as
\begin{align}
    \label{eq:O-NSR-ILRMA:def:cost}
    &\Cost{\NSR,\LatestFrame}
    \nonumber\\
    &= \sum_{\IndFreqbin} \sum_{\IndFrame=1}^{\LatestFrame} \WeightOILRMA_{\IndFrame,\LatestFrame} \Biggl[ \sum_{\IndSrc} \left( \frac{\abs{\DemixVec_{\IndFreqbin\IndSrc}^{\Hermite} \bm{\ObsSignal}_{\IndFreqbin\IndFrame}}^{2}}{\sum_{\IndBasis} \Basis_{\IndFreqbin\IndBasis\IndSrc} \Activation_{\IndBasis\IndFrame\IndSrc}} + \log{\sum_{\IndBasis} \Basis_{\IndFreqbin\IndBasis\IndSrc} \Activation_{\IndBasis\IndFrame\IndSrc}}\right)
    \nonumber\\
    &\phantom{=========} - \log{\abs{\det{\DemixMat_{\IndFreqbin}}}^{2}} \Biggr]
    \nonumber\\
    &\phantom{=} + \sum_{\IndFreqbin,\IndSrc} \RegWeight_{\IndFreqbin\IndSrc,\LatestFrame} (1 - \delta_{\IndSrc\IndTargetSrc}) \abs{\DemixVec_{\IndFreqbin\IndSrc}^{\Hermite} \Prior{\MixVec}_{\IndFreqbin,\LatestFrame}}^{2},
\end{align}
where $\RegWeight_{\IndFreqbin\IndSrc,\LatestFrame} \geq 0$ and $\Prior{\MixVec}_{\IndFreqbin,\LatestFrame} \in \mathbb{C}^{\NumMic}$ are a weight parameter for the regularizer and a prior steering vector of the target speech at the $\LatestFrame$th frame, respectively.
Here, we can set $\RegWeight_{\IndFreqbin\IndSrc,\LatestFrame}$ independently at each frame.
For example, if the movement of the target speaker can be detected, assigning a larger weight to the regularizer is expected to improve tracking performance for the moving speaker.
In this paper, we simply set $\RegWeight_{\IndFreqbin\IndSrc,\LatestFrame}$ to balance the scales of the weighted likelihood and the prior distribution as 
\begin{align}
    \label{eq:O-NSR-ILRMA:def_reg_weight}
    \RegWeight_{\IndFreqbin\IndSrc,\LatestFrame} = \Biggl( \sum_{\IndFrame=1}^{\LatestFrame} \WeightOILRMA_{\IndFrame,\LatestFrame} \Biggr) \bar{\RegWeight}_{\IndFreqbin\IndSrc}
\end{align}
with a frame-independent parameter $\bar{\RegWeight}_{\IndFreqbin\IndSrc} \geq 0$.
For simplicity, we impose the following scaling constraint for $\WeightOILRMA_{\IndFrame,\LatestFrame}$: 
\begin{align}
    \label{eq:O-NSR-ILRMA:constraint_weight_OILRMA}
    \sum_{\IndFrame=1}^{\LatestFrame} \WeightOILRMA_{\IndFrame,\LatestFrame} = 1.
\end{align}

\subsubsection{Naive Update Rule}
\label{sssec:O-NSR-ILRMA:naive_update_rule}

To obtain the online algorithm for NSR-ILRMA, we first derive a naive update rule for $\Cost{\NSR,\LatestFrame}$ on the basis of the MM algorithm.
For the demixing matrix $\DemixMat_{\IndFreqbin}$, the update rule can be derived using IP as follows:
\begin{align}
    \label{eq:O-NSR-ILRMA:naive_update_framewise_cov}
    \FramewiseCov_{\IndFreqbin\IndSrc} &\leftarrow \sum_{\IndFrame=1}^{\LatestFrame} \WeightOILRMA_{\IndFrame,\LatestFrame} \frac{\bm{\ObsSignal}_{\IndFreqbin\IndFrame} \bm{\ObsSignal}_{\IndFreqbin\IndFrame}^{\Hermite}}{\SrcModel_{\IndFreqbin\IndFrame\IndSrc}},
    \\
    \label{eq:O-NSR-ILRMA:naive_update_cov_and_reg}
    \FramewiseCovReg_{\IndFreqbin\IndSrc} &\leftarrow \FramewiseCov_{\IndFreqbin\IndSrc} + \bar{\RegWeight}_{\IndFreqbin\IndSrc} (1 - \delta_{\IndSrc\IndTargetSrc}) \Prior{\MixVec}_{\IndFreqbin,\LatestFrame} \Prior{\MixVec}_{\IndFreqbin,\LatestFrame}^{\Hermite},
    \\
    \label{eq:O-NSR-ILRMA:naive_update_intermediate_vec}
    \DemixVec_{\IndFreqbin\IndSrc} &\leftarrow \bigl( \DemixMat_{\IndFreqbin} \FramewiseCovReg_{\IndFreqbin\IndSrc} \bigr)^{-1} \UnitVec_{\IndSrc},
    \\
    \label{eq:O-NSR-ILRMA:naive_update_demix_vec}
    \DemixVec_{\IndFreqbin\IndSrc} &\leftarrow \DemixVec_{\IndFreqbin\IndSrc} / \sqrt{\DemixVec_{\IndFreqbin\IndSrc}^{\Hermite} \FramewiseCovReg_{\IndFreqbin\IndSrc} \DemixVec_{\IndFreqbin\IndSrc}},
\end{align}
where $\SrcModel_{\IndFreqbin\IndFrame\IndSrc} = \sum_{\IndBasis} \Basis_{\IndFreqbin\IndBasis\IndSrc} \Activation_{\IndBasis\IndFrame\IndSrc}$.
For the NMF variables $\Basis_{\IndFreqbin\IndBasis\IndSrc}$ and $\Activation_{\IndBasis\IndFrame\IndSrc}$, we design an auxiliary function of $\Cost{\NSR,\LatestFrame}$ in the same way as offline ILRMA~\cite{Kitamura2016TASLP} as follows.
Since the first term of (\ref{eq:O-NSR-ILRMA:def:cost}) is a reciprocal function for the summation of the NMF variables, we apply Jensen's inequality to this term with an auxiliary variable $\AuxVarNMFJensen_{\IndFreqbin\IndFrame\IndBasis\IndSrc} \geq 0$ that satisfies $\sum_{\IndBasis} \AuxVarNMFJensen_{\IndFreqbin\IndFrame\IndBasis\IndSrc} = 1$ and obtain the following inequality:
\begin{align}
    \label{eq:O-NSR-ILRMA:Jensens_inequality}
    \frac{1}{\sum_{\IndBasis} \Basis_{\IndFreqbin\IndBasis\IndSrc} \Activation_{\IndBasis\IndFrame\IndSrc}} \leq \sum_{\IndBasis} \frac{\AuxVarNMFJensen_{\IndFreqbin\IndFrame\IndBasis\IndSrc}^{2}}{\Basis_{\IndFreqbin\IndBasis\IndSrc} \Activation_{\IndBasis\IndFrame\IndSrc}}.
\end{align}
Since the second term in (\ref{eq:O-NSR-ILRMA:def:cost}) is a logarithm function, we utilize the relationship between a concave function and its tangent line and obtain an inequality with an auxiliary variable $\AuxVarNMFTangent_{\IndFreqbin\IndFrame\IndSrc} \geq 0$ as
\begin{align}
    \label{eq:O-NSR-ILRMA:Tangent_inequality}
    \log{\sum_{\IndBasis} \Basis_{\IndFreqbin\IndBasis\IndSrc} \Activation_{\IndBasis\IndFrame\IndSrc}} \leq \frac{1}{\AuxVarNMFTangent_{\IndFreqbin\IndFrame\IndSrc}} \Biggl( \sum_{\IndBasis} \Basis_{\IndFreqbin\IndBasis\IndSrc} \Activation_{\IndBasis\IndFrame\IndSrc} - \AuxVarNMFTangent_{\IndFreqbin\IndFrame\IndSrc} \Biggr) + \log{\AuxVarNMFTangent_{\IndFreqbin\IndFrame\IndSrc}}.
\end{align}
The equality of (\ref{eq:O-NSR-ILRMA:Jensens_inequality}) and (\ref{eq:O-NSR-ILRMA:Tangent_inequality}) holds if and only if the following conditions hold:
\begin{align}
    \label{eq:O-NSR-ILRMA:AuxVarCondition_Jensen}
    \AuxVarNMFJensen_{\IndFreqbin\IndFrame\IndBasis\IndSrc} &= \frac{\Basis_{\IndFreqbin\IndBasis\IndSrc} \Activation_{\IndBasis\IndFrame\IndSrc}}{\sum_{\IndBasis'} \Basis_{\IndFreqbin\IndBasis'\IndSrc} \Activation_{\IndBasis'\IndFrame\IndSrc}},
    \\
    \label{eq:O-NSR-ILRMA:AuxVarCondition_Tangent}
    \AuxVarNMFTangent_{\IndFreqbin\IndFrame\IndSrc} &= \sum_{\IndBasis} \Basis_{\IndFreqbin\IndBasis\IndSrc} \Activation_{\IndBasis\IndFrame\IndSrc}.
\end{align}
By applying inequalities (\ref{eq:O-NSR-ILRMA:Jensens_inequality}) and (\ref{eq:O-NSR-ILRMA:Tangent_inequality}) to the framewise cost function (\ref{eq:O-NSR-ILRMA:def:cost}), we obtain the following auxiliary function of $\Cost{\NSR,\LatestFrame}$ with respect to the NMF variables:
\begin{align}
    \label{eq:O-NSR-ILRMA:AuxFunction_NMF}
    \AuxCost{\NSR,\LatestFrame} 
    &= \sum_{\IndFreqbin,\IndSrc} \sum_{\IndFrame=1}^{\LatestFrame} \WeightOILRMA_{\IndFrame,\LatestFrame} \Biggl( \sum_{\IndBasis} \frac{\AuxVarNMFJensen_{\IndFreqbin\IndFrame\IndBasis\IndSrc}^{2} \abs{\SepSignal_{\IndFreqbin\IndFrame\IndSrc}}^{2}}{\Basis_{\IndFreqbin\IndBasis\IndSrc} \Activation_{\IndBasis\IndFrame\IndSrc}}  
    \nonumber\\
    &\phantom{=========} + \frac{\sum_{\IndBasis} \Basis_{\IndFreqbin\IndBasis\IndSrc} \Activation_{\IndBasis\IndFrame\IndSrc}}{\AuxVarNMFTangent_{\IndFreqbin\IndFrame\IndSrc}} - 1 + \log{\AuxVarNMFTangent_{\IndFreqbin\IndFrame\IndSrc}} \Biggr) 
    \nonumber\\
    &\phantom{=} - \sum_{\IndFreqbin} \log{\abs{\det{\DemixMat_{\IndFreqbin}}}^{2}}
    \nonumber\\
    &\phantom{=} + \sum_{\IndFreqbin,\IndSrc} \bar{\RegWeight}_{\IndFreqbin\IndSrc} (1 - \delta_{\IndSrc\IndTargetSrc}) \abs{\DemixVec_{\IndFreqbin\IndSrc}^{\Hermite} \Prior{\MixVec}_{\IndFreqbin,\LatestFrame}}^{2}.
\end{align} 
Here, $\AuxCost{\NSR,\LatestFrame} = \Cost{\NSR,\LatestFrame}$ holds if and only if (\ref{eq:O-NSR-ILRMA:AuxVarCondition_Jensen}) and (\ref{eq:O-NSR-ILRMA:AuxVarCondition_Tangent}) hold.
To minimize the auxiliary function $\AuxCost{\NSR,\LatestFrame}$ with respect to the NMF variables $\Basis_{\IndFreqbin\IndBasis\IndSrc}$ and $\Activation_{\IndBasis\IndFrame\IndSrc}$, they are updated as follows:
\begin{align}
    \label{eq:O-NSR-ILRMA:naive_update_activation}
    \Activation_{\IndBasis\IndFrame\IndSrc} &\leftarrow \sqrt{\frac{\sum_{\IndFreqbin} \AuxVarNMFJensen_{\IndFreqbin\IndFrame\IndBasis\IndSrc}^{2} \abs{\SepSignal_{\IndFreqbin\IndFrame\IndSrc}}^{2} \Basis_{\IndFreqbin\IndBasis\IndSrc}^{-1}}{\sum_{\IndFreqbin} \Basis_{\IndFreqbin\IndBasis\IndSrc} \AuxVarNMFTangent_{\IndFreqbin\IndFrame\IndSrc}^{-1}}},
    \\
    \label{eq:O-NSR-ILRMA:naive_update_basis}
    \Basis_{\IndFreqbin\IndBasis\IndSrc} &\leftarrow \sqrt{\frac{\sum_{\IndFrame=1}^{\LatestFrame} \WeightOILRMA_{\IndFrame,\LatestFrame} \AuxVarNMFJensen_{\IndFreqbin\IndFrame\IndBasis\IndSrc}^{2} \abs{\SepSignal_{\IndFreqbin\IndFrame\IndSrc}}^{2} \Activation_{\IndBasis\IndFrame\IndSrc}^{-1}}{\sum_{\IndFrame=1}^{\LatestFrame} \WeightOILRMA_{\IndFrame,\LatestFrame} \Activation_{\IndBasis\IndFrame\IndSrc} \AuxVarNMFTangent_{\IndFreqbin\IndFrame\IndSrc}^{-1}}}.
\end{align}
Here, since the auxiliary function is minimized with respect to the auxiliary variables $\AuxVarNMFJensen_{\IndFreqbin\IndFrame\IndBasis\IndSrc}$ and $\AuxVarNMFTangent_{\IndFreqbin\IndFrame\IndSrc}$ when the equalities in (\ref{eq:O-NSR-ILRMA:Jensens_inequality}) and (\ref{eq:O-NSR-ILRMA:Tangent_inequality}) hold, $\AuxVarNMFJensen_{\IndFreqbin\IndFrame\IndBasis\IndSrc}$ and $\AuxVarNMFTangent_{\IndBasis\IndFrame\IndSrc}$ are updated using (\ref{eq:O-NSR-ILRMA:AuxVarCondition_Jensen}) and (\ref{eq:O-NSR-ILRMA:AuxVarCondition_Tangent}), respectively, after each update of $\Activation_{\IndBasis\IndFrame\IndSrc}$ or $\Basis_{\IndFreqbin\IndBasis\IndSrc}$.
It is guaranteed that the update of each variable monotonically nonincreases the framewise cost function $\Cost{\NSR,\LatestFrame}$.
Note that if we set $\WeightOILRMA_{\IndFrame,\LatestFrame} = 1 / \LatestFrame \  (\forall \IndFrame)$, the entire update rule coincides with that of offline NSR-ILRMA, as described in Section~\ref{sec:related_methods}-\ref{ssec:ILRMAand_NSR-ILRMA}.
In addition, if we set $\WeightOILRMA_{\IndFrame,\LatestFrame}$ to 
\begin{align}
    \WeightOILRMA_{\IndFrame,\LatestFrame} = 
    \begin{cases}
        1 / \NumFrame^{(\mathrm{I})},    &\text{(if $\LatestFrame - \NumFrame^{(\mathrm{I})} + 1 \leq \IndFrame \leq \LatestFrame$)}
        \\
        0,   &\text{(otherwise)}
    \end{cases}
\end{align}
the entire update rule coincides with that of NSR-ILRMA in B-RCSCME using a $\NumFrame^{(\mathrm{I})}$-frame-long batch, as described in Section~\ref{sec:related_methods}-\ref{ssec:blockwise-RCSCME}.

\subsubsection{Online Update Rule}
\label{sssec:O-NSR-ILRMA:online_update_rule}

In the update algorithm derived in Section~\ref{sec:proposed_method}-\ref{ssec:online_NSR-ILRMA}-\ref{sssec:O-NSR-ILRMA:naive_update_rule}, the auxiliary variables and $\FramewiseCov_{\IndFreqbin\IndSrc}$ need to be recalculated each time the NMF variables and the demixing matrix are updated; 
thus, it is difficult to perform the naive update rule within the STFT shift length.
To reduce the computational cost, we approximate some intermediate parameters that involve summation over time frames in cost or auxiliary function using estimates at the ($\LatestFrame-1$)th-frame estimation.
We then minimize the approximate cost or auxiliary function.

To derive an efficient update rule for the demixing matrix $\DemixMat_{\IndFreqbin}$, we approximate $\SrcModel_{\IndFreqbin\IndFrame\IndSrc}$ for $\IndFrame = 1, ..., \LatestFrame-1$ in (\ref{eq:O-NSR-ILRMA:def:cost}) using its estimate $\Est{\SrcModel}_{\IndFreqbin\IndFrame\IndSrc,\LatestFrame-1}$ obtained at the $(\LatestFrame-1)$th-frame estimation.
Then, the approximate cost function with respect to $\DemixMat_{\IndFreqbin}$ is expressed as follows:
\begin{align}
    \label{eq:O-NSR-ILRMA:approx_cost_for_DemixMat}
    \Cost{\NSR,\LatestFrame} \approx &\sum_{\IndFreqbin,\IndSrc} \DemixVec_{\IndFreqbin\IndSrc}^{\Hermite} \Bigl(
        \Est{\FramewiseCov}_{\IndFreqbin\IndSrc,\LatestFrame} + \bar{\RegWeight}_{\IndFreqbin\IndSrc} (1 - \delta_{\IndSrc\IndTargetSrc}) \Prior{\MixVec}_{\IndFreqbin,\LatestFrame} \Prior{\MixVec}_{\IndFreqbin,\LatestFrame}^{\Hermite}
    \Bigr) \DemixVec_{\IndFreqbin\IndSrc}
    \nonumber\\
    &- \sum_{\IndFreqbin} \log{\abs{\det{\DemixMat_{\IndFreqbin}}}^{2}} + \ConstWithParam{\{ \DemixMat_{\IndFreqbin} \}_{\IndFreqbin}},
\end{align}
where $\ConstWithParam{\{ \DemixMat_{\IndFreqbin} \}_{\IndFreqbin}}$ denotes the term independent of $\DemixMat_{\IndFreqbin}\ (\forall \IndFreqbin)$.
Here, $\Est{\FramewiseCov}_{\IndFreqbin\IndSrc,\LatestFrame}$ is an approximation of $\FramewiseCov_{\IndFreqbin\IndSrc}$ by fixing $\SrcModel_{\IndFreqbin\IndFrame\IndSrc}$ for $\IndFrame = 1, ..., \LatestFrame-1$ in (\ref{eq:O-NSR-ILRMA:naive_update_framewise_cov}) to their estimates $\Est{\SrcModel}_{\IndFreqbin\IndFrame\IndSrc,\LatestFrame-1}$, and it is defined as
\begin{align}
    \label{eq:O-NSR-ILRMA:define_Est_Framewise_Cov}
    \Est{\FramewiseCov}_{\IndFreqbin\IndSrc,\LatestFrame} &= 
    \WeightOILRMA_{\LatestFrame,\LatestFrame} \frac{\bm{\ObsSignal}_{\IndFreqbin\LatestFrame} \bm{\ObsSignal}_{\IndFreqbin\LatestFrame}^{\Hermite}}{\SrcModel_{\IndFreqbin\LatestFrame\IndSrc}} + \sum_{\IndFrame=1}^{\LatestFrame-1} \WeightOILRMA_{\IndFrame,\LatestFrame} \frac{\bm{\ObsSignal}_{\IndFreqbin\IndFrame} \bm{\ObsSignal}_{\IndFreqbin\IndFrame}^{\Hermite}}{\Est{\SrcModel}_{\IndFreqbin\IndFrame\IndSrc,\LatestFrame-1}}
    \nonumber\\
    &= \WeightOILRMA_{\LatestFrame,\LatestFrame} \frac{\bm{\ObsSignal}_{\IndFreqbin\LatestFrame} \bm{\ObsSignal}_{\IndFreqbin\LatestFrame}^{\Hermite}}{\SrcModel_{\IndFreqbin\LatestFrame\IndSrc}} + \sum_{\IndFrame = 1}^{\LatestFrame-1} \frac{\WeightOILRMA_{\IndFrame,\LatestFrame}}{\WeightOILRMA_{\IndFrame,\LatestFrame-1}} \WeightOILRMA_{\IndFrame,\LatestFrame-1} \frac{\bm{\ObsSignal}_{\IndFreqbin\IndFrame} \bm{\ObsSignal}_{\IndFreqbin\IndFrame}^{\Hermite}}{\Est{\SrcModel}_{\IndFreqbin\IndFrame\IndSrc,\LatestFrame-1}}.
\end{align}
If $\WeightOILRMA_{\IndFrame,\LatestFrame} / \WeightOILRMA_{\IndFrame,\LatestFrame-1}$ is independent of $\IndFrame$, the second term of the right-hand side of (\ref{eq:O-NSR-ILRMA:define_Est_Framewise_Cov}) can be expressed using the estimate of $\Est{\FramewiseCov}_{\IndFreqbin\IndSrc,\LatestFrame-1}$.
To satisfy this condition, in this paper, we set $\WeightOILRMA_{\IndFrame,\LatestFrame}$ on the basis of the exponential smoothing to
\begin{align}
    \label{eq:O-NSR-ILRMA:exp_smoothing_WeightOILRMA}
    \WeightOILRMA_{\IndFrame,\LatestFrame} \propto \ForgetILRMA^{\LatestFrame-\IndFrame},
\end{align}
where $\ForgetILRMA \in (0, 1)$ denotes the forgetting factor for online NSR-ILRMA.
From (\ref{eq:O-NSR-ILRMA:constraint_weight_OILRMA}) and (\ref{eq:O-NSR-ILRMA:exp_smoothing_WeightOILRMA}), we obtain 
\begin{align}
    \label{eq:O-NSR-ILRMA:WeightOILRMA}
    \WeightOILRMA_{\IndFrame,\LatestFrame} = \frac{1 - \ForgetILRMA}{1 - \ForgetILRMA^{\LatestFrame}} \ForgetILRMA^{\LatestFrame - \IndFrame}.
\end{align}
As a result, the update of $\Est{\FramewiseCov}_{\IndFreqbin\IndSrc,\LatestFrame}$ can be expressed as 
\begin{align}
    \label{eq:O-NSR-ILRMA:update_Est_Framewise_Cov}
    \Est{\FramewiseCov}_{\IndFreqbin\IndSrc,\LatestFrame} = \frac{1 - \ForgetILRMA}{1 - \ForgetILRMA^{\LatestFrame}} \frac{\bm{\ObsSignal}_{\IndFreqbin\LatestFrame} \bm{\ObsSignal}_{\IndFreqbin\LatestFrame}^{\Hermite}}{\SrcModel_{\IndFreqbin\LatestFrame\IndSrc}} + \frac{\ForgetILRMA - \ForgetILRMA^{\LatestFrame}}{1 - \ForgetILRMA^{\LatestFrame}} \Est{\FramewiseCov}_{\IndFreqbin\IndSrc,\LatestFrame-1},
\end{align}
where $\SrcModel_{\IndFreqbin\LatestFrame\IndSrc} = \sum_{\IndBasis} \Basis_{\IndFreqbin\IndBasis\IndSrc} \Activation_{\IndBasis\LatestFrame\IndSrc}$ and $\Est{\FramewiseCov}_{\IndFreqbin\IndSrc,1}$ is set to 
\begin{align}
    \label{eq:O-NSR-ILRMA:initialization_framewise_cov}
    \Est{\FramewiseCov}_{\IndFreqbin\IndSrc,1} = \frac{\bm{\ObsSignal}_{\IndFreqbin1} \bm{\ObsSignal}_{\IndFreqbin1}^{\Hermite}}{\SrcModel_{\IndFreqbin1\IndSrc}} + \epsilon_{\ILRMA} \Identity_{\NumSrc}
\end{align}
with a sufficiently small parameter $\epsilon_{\ILRMA}$ for stability.
Note that this update rule can be regarded as a type of autoregressive update; however, unlike the update rule of the weighted covariance matrix in O-IVA, (\ref{eq:O-IVA_update_CovMat}), its coefficients vary depending the number of observed frames $\LatestFrame$.
Furthermore, since $0 < \ForgetILRMA < 1$, when a sufficiently large number of observed frames are obtained, $\ForgetILRMA^{\LatestFrame}$ converges to zero, and (\ref{eq:O-NSR-ILRMA:update_Est_Framewise_Cov}) can be written as 
\begin{align}
    \label{eq:O-NSR-ILRMA:autoregressive_update_Est_FramewiseCov_limit}
    \Est{\FramewiseCov}_{\IndFreqbin\IndSrc,\LatestFrame} = (1 - \ForgetILRMA) \frac{\bm{\ObsSignal}_{\IndFreqbin\LatestFrame} \bm{\ObsSignal}_{\IndFreqbin\LatestFrame}^{\Hermite}}{\SrcModel_{\IndFreqbin\LatestFrame\IndSrc}} + \ForgetILRMA \Est{\FramewiseCov}_{\IndFreqbin\IndSrc,\LatestFrame-1},
\end{align}
which results in a form similar to that in the autoregressive update in O-IVA.
By minimizing the approximate cost function (\ref{eq:O-NSR-ILRMA:approx_cost_for_DemixMat}) with respect to $\DemixMat_{\IndFreqbin}$, we can derive the update rule for $\DemixMat_{\IndFreqbin}$ using IP as follows:
\begin{align}
    \label{eq:O-NSR-ILRMA:update_Est_FramewiseCovReg}
    \Est{\FramewiseCovReg}_{\IndFreqbin\IndSrc,\LatestFrame} &\leftarrow \Est{\FramewiseCov}_{\IndFreqbin\IndSrc,\LatestFrame} + \bar{\RegWeight}_{\IndFreqbin\IndSrc} (1 - \delta_{\IndSrc\IndTargetSrc}) \Prior{\MixVec}_{\IndFreqbin,\LatestFrame} \Prior{\MixVec}_{\IndFreqbin,\LatestFrame}^{\Hermite},
    \\
    \label{eq:O-NSR-ILRMA:update_Est_intermediate_vec}
    \DemixVec_{\IndFreqbin\IndSrc} &\leftarrow \bigl( \DemixMat_{\IndFreqbin} \Est{\FramewiseCovReg}_{\IndFreqbin\IndSrc,\LatestFrame} \bigr)^{-1} \UnitVec_{\IndSrc},
    \\
    \label{eq:O-NSR-ILRMA:update_DemixVec}
    \DemixVec_{\IndFreqbin\IndSrc} &\leftarrow \DemixVec_{\IndFreqbin\IndSrc} / \sqrt{\DemixVec_{\IndFreqbin\IndSrc}^{\Hermite} \Est{\FramewiseCovReg}_{\IndFreqbin\IndSrc,\LatestFrame} \DemixVec_{\IndFreqbin\IndSrc}}.
\end{align}
Furthermore, we consider performing the matrix inversion in (\ref{eq:O-NSR-ILRMA:update_Est_intermediate_vec}) efficiently in a similar way to \cite{Taniguchi2014HSCMA} (see also Appendix~B for details).
First, we transform (\ref{eq:O-NSR-ILRMA:update_Est_intermediate_vec}) using the relationship between the mixing and demixing matrices (i.e., $\DemixMat_{\IndFreqbin} = \MixMat_{\IndFreqbin}^{-1}$) as 
\begin{align}
    \label{eq:O-NSR-ILRMA:transform_intermediate_vec}
    \DemixVec_{\IndFreqbin\IndSrc} 
    &= \Est{\FramewiseCovReg}_{\IndFreqbin\IndSrc,\LatestFrame}^{-1} \DemixMat_{\IndFreqbin}^{-1} \UnitVec_{\IndSrc}
    \nonumber\\
    &= \Est{\FramewiseCovReg}_{\IndFreqbin\IndSrc,\LatestFrame}^{-1} \MixVec_{\IndFreqbin\IndSrc}.
\end{align}
Then, since $\Est{\FramewiseCovReg}_{\IndFreqbin\IndSrc,\LatestFrame}$ is updated by summing the full-rank matrix $\Est{\FramewiseCov}_{\IndFreqbin\IndSrc,\LatestFrame}$ and the rank-1 matrix in (\ref{eq:O-NSR-ILRMA:update_Est_FramewiseCovReg}), $\Est{\FramewiseCovReg}_{\IndFreqbin\IndSrc,\LatestFrame}^{-1} =: \Est{\InvFramewiseCovReg}_{\IndFreqbin\IndSrc,\LatestFrame}$ can be calculated using the Sherman--Morrison formula as follows:
\begin{align}
    \label{eq:O-NSR-ILRMA:rank-1-update_InvFramewiseCovReg}
    \Est{\InvFramewiseCovReg}_{\IndFreqbin\IndSrc,\LatestFrame} = \Est{\InvFramewiseCov}_{\IndFreqbin\IndSrc,\LatestFrame} - (1 - \delta_{\IndSrc\IndTargetSrc}) \frac{\Est{\InvFramewiseCov}_{\IndFreqbin\IndSrc,\LatestFrame} \Prior{\MixVec}_{\IndFreqbin,\LatestFrame} \Prior{\MixVec}_{\IndFreqbin,\LatestFrame}^{\Hermite} \Est{\InvFramewiseCov}_{\IndFreqbin\IndSrc,\LatestFrame}}{\bar{\RegWeight}_{\IndFreqbin\IndSrc}^{-1} + \Prior{\MixVec}_{\IndFreqbin,\LatestFrame}^{\Hermite} \Est{\InvFramewiseCov}_{\IndFreqbin\IndSrc,\LatestFrame} \Prior{\MixVec}_{\IndFreqbin,\LatestFrame}},
\end{align}
where $\Est{\InvFramewiseCov}_{\IndFreqbin\IndSrc,\LatestFrame} := \Est{\FramewiseCov}_{\IndFreqbin\IndSrc,\LatestFrame}^{-1}$.
By using $\Est{\InvFramewiseCovReg}_{\IndFreqbin\IndSrc,\LatestFrame}$, we can express the update rule for $\DemixMat_{\IndFreqbin}$, (\ref{eq:O-NSR-ILRMA:transform_intermediate_vec}) and (\ref{eq:O-NSR-ILRMA:update_DemixVec}), as 
\begin{align}
    \label{eq:O-NSR-ILRMA:transform_update_DemixVec}
    \DemixVec_{\IndFreqbin\IndSrc} 
    \leftarrow \Est{\InvFramewiseCovReg}_{\IndFreqbin\IndSrc,\LatestFrame} \MixVec_{\IndFreqbin\IndSrc} / \sqrt{\MixVec_{\IndFreqbin\IndSrc}^{\Hermite} \Est{\InvFramewiseCovReg}_{\IndFreqbin\IndSrc,\LatestFrame} \MixVec_{\IndFreqbin\IndSrc}}.
\end{align}
However, computing $\MixVec_{\IndFreqbin\IndSrc} = \DemixMat_{\IndFreqbin}^{-1} \UnitVec_{\IndSrc}$ at each iteration is computationally expensive; thus, we consider computing $\MixMat_{\IndFreqbin}$ efficiently, as in \cite{Taniguchi2014HSCMA}.
The update rule for $\DemixMat_{\IndFreqbin}$ can be written as
\begin{align}
    \label{eq:O-NSR-ILRMA:def_intermediate_vec_nu}
    \VecIPInvLemma_{\IndFreqbin\IndSrc} &\leftarrow \Est{\InvFramewiseCovReg}_{\IndFreqbin\IndSrc,\LatestFrame} \MixVec_{\IndFreqbin\IndSrc} / \sqrt{\MixVec_{\IndFreqbin\IndSrc}^{\Hermite} \Est{\InvFramewiseCovReg}_{\IndFreqbin\IndSrc,\LatestFrame} \MixVec_{\IndFreqbin\IndSrc}},
    \\
    \label{eq:O-NSR-ILRMA:rank-1-update_DemixMat}
    \DemixMat_{\IndFreqbin} &\leftarrow \DemixMat_{\IndFreqbin} + \UnitVec_{\IndSrc} (\VecIPInvLemma_{\IndFreqbin\IndSrc} - \DemixVec_{\IndFreqbin\IndSrc})^{\Hermite}.
\end{align}
By applying the Sherman--Morrison formula, we obtain an efficient update rule for $\MixMat_{\IndFreqbin}$ as
\begin{align}
    \label{eq:O-NSR-ILRMA:rank-1-update_MixMat}
    \MixMat_{\IndFreqbin} \leftarrow \Biggl(
        \Identity_{\NumSrc} - \frac{\MixVec_{\IndFreqbin\IndSrc} (\VecIPInvLemma_{\IndFreqbin\IndSrc} - \DemixVec_{\IndFreqbin\IndSrc})^{\Hermite}}{\VecIPInvLemma_{\IndFreqbin\IndSrc}^{\Hermite} \MixVec_{\IndFreqbin\IndSrc}}
    \Biggr) \MixMat_{\IndFreqbin}.
\end{align}
For the computation of $\Est{\InvFramewiseCov}_{\IndFreqbin\IndSrc,\LatestFrame}$, since $\Est{\FramewiseCov}_{\IndFreqbin\IndSrc,\LatestFrame}$ is updated by summing the full-rank and rank-1 matrices in (\ref{eq:O-NSR-ILRMA:update_Est_Framewise_Cov}), $\Est{\InvFramewiseCov}_{\IndFreqbin\IndSrc,\LatestFrame}$ can also be calculated using the Sherman--Morrison formula as follows:
\begin{align}
    \label{eq:O-NSR-ILRMA:rank-1-update_Est_InvFramewiseCov}
    \Est{\InvFramewiseCov}_{\IndFreqbin\IndSrc,\LatestFrame} = \frac{1 - \ForgetILRMA^{\LatestFrame}}{\ForgetILRMA - \ForgetILRMA^{\LatestFrame}} \Biggl(
        \Est{\InvFramewiseCov}_{\IndFreqbin\IndSrc,\LatestFrame-1} - \frac{
            \Est{\InvFramewiseCov}_{\IndFreqbin\IndSrc,\LatestFrame-1} \bm{\ObsSignal}_{\IndFreqbin\LatestFrame} \bm{\ObsSignal}_{\IndFreqbin\LatestFrame}^{\Hermite} \Est{\InvFramewiseCov}_{\IndFreqbin\IndSrc,\LatestFrame-1}
        }{
            \frac{\ForgetILRMA - \ForgetILRMA^{\LatestFrame}}{1 - \ForgetILRMA} \SrcModel_{\IndFreqbin\LatestFrame\IndSrc} + \bm{\ObsSignal}_{\IndFreqbin\LatestFrame}^{\Hermite} \Est{\InvFramewiseCov}_{\IndFreqbin\IndSrc,\LatestFrame-1} \bm{\ObsSignal}_{\IndFreqbin\LatestFrame}
        }
    \Biggr),
\end{align}
where $\Est{\InvFramewiseCov}_{\IndFreqbin\IndSrc,1}$ is calculated from its definition and (\ref{eq:O-NSR-ILRMA:initialization_framewise_cov}) as 
\begin{align}
    \label{eq:O-NSR-ILRMA:initlaization_inv_framewise_cov}
    \Est{\InvFramewiseCov}_{\IndFreqbin\IndSrc,1} = \epsilon_{\ILRMA}^{-1} \Identity_{\NumSrc} - \frac{\bm{\ObsSignal}_{\IndFreqbin1} \bm{\ObsSignal}_{\IndFreqbin1}^{\Hermite}}{ \epsilon_{\ILRMA}^{2} \SrcModel_{\IndFreqbin1\IndSrc} + \epsilon_{\ILRMA} \| \bm{\ObsSignal}_{\IndFreqbin1} \|^{2}}.
\end{align}
In summary, the overall update rule for $\DemixMat_{\IndFreqbin}$ can be expressed as follows:
\begin{align}
    \label{eq:O-NSR-ILRMA:online_update_cov}
    \Est{\InvFramewiseCov}_{\IndFreqbin\IndSrc,\LatestFrame} &\leftarrow
    \frac{1 - \ForgetILRMA^{\LatestFrame}}{\ForgetILRMA - \ForgetILRMA^{\LatestFrame}} \Biggl( \Est{\InvFramewiseCov}_{\IndFreqbin\IndSrc,\LatestFrame-1} - \frac{\Est{\InvFramewiseCov}_{\IndFreqbin\IndSrc,\LatestFrame-1} \bm{\ObsSignal}_{\IndFreqbin\LatestFrame} \bm{\ObsSignal}_{\IndFreqbin\LatestFrame}^{\Hermite} \Est{\InvFramewiseCov}_{\IndFreqbin\IndSrc,\LatestFrame-1} }{\frac{\ForgetILRMA - \ForgetILRMA^{\LatestFrame}}{1 - \ForgetILRMA} \SrcModel_{\IndFreqbin\LatestFrame\IndSrc} + \bm{\ObsSignal}_{\IndFreqbin\LatestFrame}^{\Hermite} \Est{\InvFramewiseCov}_{\IndFreqbin\IndSrc,\LatestFrame-1} \bm{\ObsSignal}_{\IndFreqbin\LatestFrame}} \Biggr),
    \\
    \label{eq:O-NSR-ILRMA:online_update_cov_and_reg}
    \Est\InvFramewiseCovReg_{\IndFreqbin\IndSrc,\LatestFrame} &\leftarrow \Est{\InvFramewiseCov}_{\IndFreqbin\IndSrc,\LatestFrame} - (1 - \delta_{\IndSrc\IndTargetSrc}) \frac{\Est{\InvFramewiseCov}_{\IndFreqbin\IndSrc,\LatestFrame} \Prior{\MixVec}_{\IndFreqbin,\LatestFrame} \Prior{\MixVec}_{\IndFreqbin,\LatestFrame}^{\Hermite} \Est{\InvFramewiseCov}_{\IndFreqbin\IndSrc,\LatestFrame}}{\bar{\RegWeight}_{\IndFreqbin\IndSrc}^{-1} + \Prior{\MixVec}_{\IndFreqbin,\LatestFrame}^{\Hermite} \Est{\InvFramewiseCov}_{\IndFreqbin\IndSrc,\LatestFrame} \Prior{\MixVec}_{\IndFreqbin,\LatestFrame}}, 
    \\
    \label{eq:O-NSR-ILRMA:online_udpate_intermediate_vec}
    \VecIPInvLemma_{\IndFreqbin\IndSrc} &\leftarrow \Est{\InvFramewiseCovReg}_{\IndFreqbin\IndSrc,\LatestFrame} \MixVec_{\IndFreqbin\IndSrc} / \sqrt{\MixVec_{\IndFreqbin\IndSrc}^{\Hermite} \Est{\InvFramewiseCovReg}_{\IndFreqbin\IndSrc,\LatestFrame} \MixVec_{\IndFreqbin\IndSrc}},
    \\
    \label{eq:O-NSR-ILRMA:online_update_mix_mat}
    \MixMat_{\IndFreqbin} &\leftarrow \Biggl( \Identity_{\NumSrc} - \frac{\MixVec_{\IndFreqbin\IndSrc} ( \VecIPInvLemma_{\IndFreqbin\IndSrc} - \DemixVec_{\IndFreqbin\IndSrc} )^{\Hermite}}{\VecIPInvLemma_{\IndFreqbin\IndSrc}^{\Hermite} \MixVec_{\IndFreqbin\IndSrc}} \Biggr) \MixMat_{\IndFreqbin},
    \\
    \label{eq:O-NSR-ILRMA:online_update_demix_vec}
    \DemixVec_{\IndFreqbin\IndSrc} &\leftarrow \VecIPInvLemma_{\IndFreqbin\IndSrc}.
\end{align}

To derive an efficient update rule for the NMF variables, we approximate $\AuxVarNMFJensen_{\IndFreqbin\IndFrame\IndBasis\IndSrc}$, $\AuxVarNMFTangent_{\IndFreqbin\IndFrame\IndSrc}$, $\Activation_{\IndBasis\IndFrame\IndSrc}$, and $\SepSignal_{\IndFreqbin\IndFrame\IndSrc}$ for $\IndFrame = 1,..., \LatestFrame - 1$ in (\ref{eq:O-NSR-ILRMA:AuxFunction_NMF}) by their estimates $\Est{\AuxVarNMFJensen}_{\IndFreqbin\IndFrame\IndBasis\IndSrc,\LatestFrame-1}$, $\Est{\AuxVarNMFTangent}_{\IndFreqbin\IndFrame\IndSrc,\LatestFrame-1}$, $\Est{\Activation}_{\IndBasis\IndFrame\IndSrc,\LatestFrame-1}$, and $\Est{\SepSignal}_{\IndFreqbin\IndFrame\IndSrc,\LatestFrame-1}$ obtained at the $(\LatestFrame-1)$th-frame estimation, respectively.
Then, the approximate auxiliary function with respect to the NMF variables $\Basis_{\IndFreqbin\IndBasis\IndSrc}$ and $\Activation_{\IndBasis\LatestFrame\IndSrc}$ is expressed as follows:
\begin{align}
    \label{eq:O-NSR-ILRMA:approx_aux_function_for_NMF}
    \Cost{\NSR,\LatestFrame} \approx
    &\sum_{\IndFreqbin,\IndBasis,\IndSrc} \Biggl[
        \WeightOILRMA_{\LatestFrame,\LatestFrame} \Biggl(
            \frac{\AuxVarNMFJensen_{\IndFreqbin\LatestFrame\IndBasis\IndSrc}^{2} \abs{\SepSignal_{\IndFreqbin\LatestFrame\IndSrc}}^{2}}{\Basis_{\IndFreqbin\IndBasis\IndSrc} \Activation_{\IndBasis\LatestFrame\IndSrc}} + \frac{\Basis_{\IndFreqbin\IndBasis\IndSrc} \Activation_{\IndBasis\LatestFrame\IndSrc}}{\AuxVarNMFTangent_{\IndFreqbin\LatestFrame\IndSrc}}
        \Biggr)
        \nonumber\\
        +& \sum_{\IndFrame=1}^{\LatestFrame-1} \WeightOILRMA_{\IndFrame,\LatestFrame} \Biggl( 
            \frac{\Est{\AuxVarNMFJensen}_{\IndFreqbin\IndFrame\IndSrc,\LatestFrame-1}^{2} \abs{\Est{\SepSignal}_{\IndFreqbin\IndFrame\IndSrc,\LatestFrame-1}}^{2}}{\Basis_{\IndFreqbin\IndBasis\IndSrc} \Est{\Activation}_{\IndBasis\IndFrame\IndSrc,\LatestFrame-1}} + \frac{\Basis_{\IndFreqbin\IndBasis\IndSrc} \Est{\Activation}_{\IndBasis\IndFrame\IndSrc,\LatestFrame-1}}{\Est{\AuxVarNMFTangent}_{\IndFreqbin\IndFrame\IndSrc,\LatestFrame-1}}
        \Biggr)
    \Biggr]
    \nonumber\\
    &+ \const,
\end{align}
where $\const$ denotes the term independent of $\Basis_{\IndFreqbin\IndBasis\IndSrc}\ (\forall \IndFreqbin,\IndBasis,\IndSrc)$ and $\Activation_{\IndBasis\LatestFrame\IndSrc}\ (\forall \IndBasis,\IndSrc)$.
By setting the gradients of (\ref{eq:O-NSR-ILRMA:approx_aux_function_for_NMF}) with respect to $\Basis_{\IndFreqbin\IndBasis\IndSrc}$ and $\Activation_{\IndBasis\LatestFrame\IndSrc}$ to zero, we can derive the update rules for $\Basis_{\IndFreqbin\IndBasis\IndSrc}$ and $\Activation_{\IndBasis\LatestFrame\IndSrc}$ as follows:
\begin{align}
    \label{eq:O-NSR-ILRMA:update_activation_1}
    \Activation_{\IndBasis\LatestFrame\IndSrc} &\leftarrow \Activation_{\IndBasis\LatestFrame\IndSrc} \sqrt{\frac{
        \sum_{\IndFreqbin} \AuxVarNMFJensen_{\IndFreqbin\LatestFrame\IndBasis\IndSrc}^{2} \abs{\SepSignal_{\IndFreqbin\LatestFrame\IndSrc}}^{2} \Basis_{\IndFreqbin\IndBasis\IndSrc}^{-1}
    }{
        \sum_{\IndFreqbin} \Basis_{\IndFreqbin\IndBasis\IndSrc} \AuxVarNMFTangent_{\IndFreqbin\LatestFrame\IndSrc}^{-1}
    }},
    \\
    \label{eq:O-NSR-ILRMA:define_numer_basis_1}
    \EstNumerBasis_{\IndFreqbin\IndBasis\IndSrc,\LatestFrame} &= \WeightOILRMA_{\LatestFrame,\LatestFrame} \frac{\AuxVarNMFJensen_{\IndFreqbin\LatestFrame\IndBasis\IndSrc}^{2} \abs{\SepSignal_{\IndFreqbin\LatestFrame\IndSrc}}^{2}}{\Activation_{\IndBasis\LatestFrame\IndSrc}} 
    \nonumber\\
    &\phantom{=}+ \sum_{\IndFrame=1}^{\LatestFrame-1} \frac{\WeightOILRMA_{\IndFrame,\LatestFrame}}{\WeightOILRMA_{\IndFrame,\LatestFrame-1}} \WeightOILRMA_{\IndFrame,\LatestFrame-1} \frac{\Est{\AuxVarNMFJensen}_{\IndFreqbin\IndFrame\IndBasis\IndSrc,\LatestFrame-1}^{2} \abs{\Est{\SepSignal}_{\IndFreqbin\IndFrame\IndSrc,\LatestFrame-1}}^{2}}{\Est{\Activation}_{\IndBasis\IndFrame\IndSrc,\LatestFrame-1}},
    \\
    \label{eq:O-NSR-ILRMA:define_denom_basis_1}
    \EstDenomBasis_{\IndFreqbin\IndBasis\IndSrc,\LatestFrame} &= \WeightOILRMA_{\LatestFrame,\LatestFrame} \frac{\Activation_{\IndBasis\LatestFrame\IndSrc}}{\AuxVarNMFTangent_{\IndFreqbin\LatestFrame\IndSrc}} 
    \nonumber\\
    &\phantom{=} + \sum_{\IndFrame=1}^{\LatestFrame-1} \frac{\WeightOILRMA_{\IndFrame,\LatestFrame}}{\WeightOILRMA_{\IndFrame,\LatestFrame-1}} \WeightOILRMA_{\IndFrame,\LatestFrame-1} \frac{\Est{\Activation}_{\IndBasis\IndFrame\IndSrc,\LatestFrame-1}}{\Est{\AuxVarNMFTangent}_{\IndFreqbin\IndFrame\IndSrc,\LatestFrame-1}},
    \\
    \label{eq:O-NSR-ILRMA:update_basis_with_num_and_den_1}
    \Basis_{\IndFreqbin\IndBasis\IndSrc} &\leftarrow \sqrt{\frac{\EstNumerBasis_{\IndFreqbin\IndBasis\IndSrc,\LatestFrame}}{\EstDenomBasis_{\IndFreqbin\IndBasis\IndSrc,\LatestFrame}}},
\end{align} 
where $\EstNumerBasis_{\IndFreqbin\IndBasis\IndSrc,\LatestFrame}$ and $\EstDenomBasis_{\IndFreqbin\IndBasis\IndSrc,\LatestFrame}$ are nonnegative intermediate variables.
Similarly to the discussion on the update rule for the demixing matrix, if $\WeightOILRMA_{\IndFrame,\LatestFrame} / \WeightOILRMA_{\IndFrame,\LatestFrame-1}$ is independent of $\IndFrame$, the second term of the right-hand sides of (\ref{eq:O-NSR-ILRMA:define_numer_basis_1}) and (\ref{eq:O-NSR-ILRMA:define_denom_basis_1}) can be expressed using the estimates of $\EstNumerBasis_{\IndFreqbin\IndBasis\IndSrc,\LatestFrame-1}$ and $\EstDenomBasis_{\IndFreqbin\IndBasis\IndSrc,\LatestFrame-1}$, respectively.
Therefore, by substituting (\ref{eq:O-NSR-ILRMA:WeightOILRMA}) into (\ref{eq:O-NSR-ILRMA:define_numer_basis_1}) and (\ref{eq:O-NSR-ILRMA:define_denom_basis_1}), we can express the update of $\EstNumerBasis_{\IndFreqbin\IndBasis\IndSrc,\LatestFrame}$ and $\EstDenomBasis_{\IndFreqbin\IndBasis\IndSrc,\LatestFrame}$ as 
\begin{align}
    \label{eq:O-NSR-ILRMA:autoregressive_update_basis_numerator}
    \EstNumerBasis_{\IndFreqbin\IndBasis\IndSrc,\LatestFrame} &= \frac{1 - \ForgetILRMA}{1 - \ForgetILRMA^{\LatestFrame}} \frac{\AuxVarNMFJensen_{\IndFreqbin\LatestFrame\IndBasis\IndSrc}^{2} \abs{\SepSignal_{\IndFreqbin\LatestFrame\IndSrc}}^{2}}{\Activation_{\IndBasis\LatestFrame\IndSrc}} + \frac{\ForgetILRMA - \ForgetILRMA^{\LatestFrame}}{1 - \ForgetILRMA^{\LatestFrame}} \EstNumerBasis_{\IndFreqbin\IndBasis\IndSrc,\LatestFrame-1},
    \\
    \label{eq:O-NSR-ILRMA:autoregressive_update_basis_denomerator}
    \EstDenomBasis_{\IndFreqbin\IndBasis\IndSrc,\LatestFrame} &= \frac{1 - \ForgetILRMA}{1 - \ForgetILRMA^{\LatestFrame}} \frac{\Activation_{\IndBasis\LatestFrame\IndSrc}}{\AuxVarNMFTangent_{\IndFreqbin\LatestFrame\IndSrc}} + \frac{\ForgetILRMA - \ForgetILRMA^{\LatestFrame}}{1 - \ForgetILRMA^{\LatestFrame}} \EstDenomBasis_{\IndFreqbin\IndBasis\IndSrc,\LatestFrame-1},
\end{align} 
where $\EstNumerBasis_{\IndFreqbin\IndBasis\IndSrc,0}$ and $\EstDenomBasis_{\IndFreqbin\IndBasis\IndSrc,0}$ are set to $0$.
Finally, by substituting (\ref{eq:O-NSR-ILRMA:AuxVarCondition_Jensen}) and (\ref{eq:O-NSR-ILRMA:AuxVarCondition_Tangent}) into (\ref{eq:O-NSR-ILRMA:update_activation_1}), (\ref{eq:O-NSR-ILRMA:autoregressive_update_basis_numerator}), (\ref{eq:O-NSR-ILRMA:autoregressive_update_basis_denomerator}), and (\ref{eq:O-NSR-ILRMA:update_basis_with_num_and_den_1}), we can derive the entire update rule for the NMF variables as follows:
\begin{align}
    \label{eq:O-NSR-ILRMA:online_update_activation}
    \Activation_{\IndBasis\LatestFrame\IndSrc} &\leftarrow 
    \Activation_{\IndBasis\LatestFrame\IndSrc} \sqrt{\frac{
        \sum_{\IndFreqbin} \Basis_{\IndFreqbin\IndBasis\IndSrc} \abs{\SepSignal_{\IndFreqbin\LatestFrame\IndSrc}}^{2} \bigl( \sum_{\IndBasis'} \Basis_{\IndFreqbin\IndBasis'\IndSrc} \Activation_{\IndBasis'\LatestFrame\IndSrc} \bigr)^{-2}
    }{
        \sum_{\IndFreqbin} \Basis_{\IndFreqbin\IndBasis\IndSrc} \bigl( \sum_{\IndBasis'} \Basis_{\IndFreqbin\IndBasis'\IndSrc} \Activation_{\IndBasis'\LatestFrame\IndSrc} \bigr)^{-1}
    }},
    \\
    \label{eq:O-NSR-ILRMA:online_update_basis_numerator}
    \EstNumerBasis_{\IndFreqbin\IndBasis\IndSrc,\LatestFrame} &\leftarrow
    \frac{1 - \ForgetILRMA}{1 - \ForgetILRMA^{\LatestFrame}} \frac{\Basis_{\IndFreqbin\IndBasis\IndSrc}^{2} \Activation_{\IndBasis\LatestFrame\IndSrc} \abs{\SepSignal_{\IndFreqbin\LatestFrame\IndSrc}}^{2}}{\bigl( \sum_{\IndBasis'} \Basis_{\IndFreqbin\IndBasis'\IndSrc} \Activation_{\IndBasis'\LatestFrame\IndSrc} \bigr)^{2}} + \frac{\ForgetILRMA - \ForgetILRMA^{\LatestFrame}}{1 - \ForgetILRMA^{\LatestFrame}} \EstNumerBasis_{\IndFreqbin\IndBasis\IndSrc,\LatestFrame-1},
    \\
    \label{eq:O-NSR-ILRMA:online_update_basis_denomerator}
    \EstDenomBasis_{\IndFreqbin\IndBasis\IndSrc,\LatestFrame} &\leftarrow \frac{1 - \ForgetILRMA}{1 - \ForgetILRMA^{\LatestFrame}} \frac{\Activation_{\IndBasis\LatestFrame\IndSrc}}{\sum_{\IndBasis'} \Basis_{\IndFreqbin\IndBasis'\IndSrc} \Activation_{\IndBasis'\LatestFrame\IndSrc}} + \frac{\ForgetILRMA - \ForgetILRMA^{\LatestFrame}}{1 - \ForgetILRMA^{\LatestFrame}} \EstDenomBasis_{\IndFreqbin\IndBasis\IndSrc,\LatestFrame-1},
    \\
    \label{eq:O-NSR-ILRMA:online_update_basis}
    \Basis_{\IndFreqbin\IndBasis\IndSrc} &\leftarrow \sqrt{\frac{\EstNumerBasis_{\IndFreqbin\IndBasis\IndSrc,\LatestFrame}}{\EstDenomBasis_{\IndFreqbin\IndBasis\IndSrc,\LatestFrame}}}.
\end{align}

Note that the scale of $\SepSignal_{\IndFreqbin\IndFrame\IndSrc}$ can vary across the frequency bins, as in offline NSR-ILRMA~\cite{Ishikawa2025Access}.
When the separated signal estimated by online NSR-ILRMA is output, it is necessary to apply the projection back method to $\SepSignal_{\IndFreqbin\IndFrame\IndSrc}$ to fix their scales.

\subsubsection{Techniques for Stability}
\label{sssec:O-NSR-ILRMA:technique_for_stability}

We have experimentally found that the updates of the NMF variables frequently became numerically unstable and output NaN either immediately after the execution started or after a sufficiently long time.
To perform online NSR-ILRMA stably, we introduce two stabilization techniques.

First, the cause of the numerical instability that occurs immediately after the execution started can be explained as follows.
In (NSR-)ILRMA, the $\NumFreqbin \times \LatestFrame$ matrix representing the power spectrogram of the separated signal for each source is modeled using rank-$\NumBasis$ nonnegative matrices by NMF.
However, when the number of observed frames is very small (i.e., $\LatestFrame < \NumBasis$), the rank of the $\NumFreqbin \times \LatestFrame$ matrix for the power spectrogram is less than $\NumBasis$.
This can cause the NMF bases to overfit the data, resulting in a numerical instability.
To mitigate this, we limit the number of to-be-estimated bases to, at most, $\LatestFrame$.
As a result, this technique is expected to improve numerical stability when the number of observed frames is very small.

Second, as a possible cause of the numerical instability that occurs after a long time, we experimentally observed that the power of $\Basis_{\IndFreqbin\IndBasis\IndSrc}$ often concentrated in specific frequency bins, which may cause inappropriate models for real-world signals.
To address this problem, we reset $\Basis_{\IndFreqbin\IndBasis\IndSrc}$ to a random value when the ratio of the maximum value of $\Basis_{\IndFreqbin\IndBasis\IndSrc}$ across frequency bins to its total summation, $\max_{\IndFreqbin} ( \Basis_{\IndFreqbin\IndBasis\IndSrc} ) / \sum_{\IndFreqbin'} \Basis_{\IndFreqbin'\IndBasis\IndSrc}$, exceeds a threshold value $\epsilon_{\mathrm{basis}}$.
This technique is expected to improve numerical stability to some extent.

\subsection{ONLINE RCSCME}
\label{ssec:online_RCSCME}

\subsubsection{Framewise Formulation}
\label{sssec:O-RCSCME:framewise_formulation}

When the observed signal up to the $\LatestFrame$th frame is obtained, we weight the likelihood and prior of each frame $\IndFrame$ in (\ref{eq:def:Cost_RCSCME}) with a weight parameter $\WeightORCSCME_{\IndFrame,\LatestFrame} > 0$ for RCSCME and derive the framewise cost function of RCSCME $\Cost{\RCSCME,\LatestFrame}$ as 
\begin{align}
    \label{eq:O-RCSCME:def:cost}
    &\Cost{\RCSCME,\LatestFrame}
    \nonumber\\
    &= \sum_{\IndFreqbin} \sum_{\IndFrame=1}^{\LatestFrame} \WeightORCSCME_{\IndFrame,\LatestFrame} \biggl( \bm{\ObsSignal}_{\IndFreqbin\IndFrame}^{\Hermite} \bigl( \SCMX_{\IndFreqbin\IndFrame} \bigr)^{-1} \bm{\ObsSignal}_{\IndFreqbin\IndFrame} + \log{\det{\SCMX_{\IndFreqbin\IndFrame}}}\biggr)
    \nonumber\\
    &\phantom{=} + \sum_{\IndFreqbin} \sum_{\IndFrame=1}^{\LatestFrame} \WeightORCSCME_{\IndFrame,\LatestFrame} \biggl( (\bar{\ShapeIG} + 1) \log{\VarT_{\IndFreqbin\IndFrame}} + \frac{\bar{\ScaleIG}}{\VarT_{\IndFreqbin\IndFrame}} \biggr),
\end{align}
where $\bar{\ShapeIG}$ and $\bar{\ScaleIG}$ are shape and scale parameters of inverse gamma distribution for the weighted prior of $\VarT_{\IndFreqbin\IndFrame}$, respectively.
Note that the weight parameter for RCSCME, $\WeightORCSCME_{\IndFrame,\LatestFrame}$, can be set independently of that for (NSR-)ILRMA, $\WeightOILRMA_{\IndFrame,\LatestFrame}$.
For simplicity, we impose the following scaling constraint for $\WeightORCSCME_{\IndFrame,\LatestFrame}$:
\begin{align}
    \label{eq:O-RCSCME:constraint_weight_ORCSCME}
    \sum_{\IndFrame = 1}^{\LatestFrame} \WeightORCSCME_{\IndFrame,\LatestFrame} = 1.
\end{align}

Furthermore, we reconsider the model of the diffuse noise SCM $\SCMN_{\IndFreqbin}$.
In offline RCSCME, $\SCMN_{\IndFreqbin}$ is modeled by (\ref{eq:RCSCME:model_SCMN})--(\ref{eq:RCSCME:def:NoiseSrcImage}) and can be interpreted as assigning equal weights to the noise source image $\NoiseSrcImage_{\IndFreqbin\IndFrame}$ for every frame.
To align with the concept of MWLE, we remodel $\SCMN_{\IndFreqbin}$ using $\WeightORCSCME_{\IndFrame,\LatestFrame}$ as follows: 
\begin{align}
    \label{eq:O-RCSCME:model_noise_SCM}
    \SCMN_{\IndFreqbin} &= \SCMNh_{\IndFreqbin,\LatestFrame} + \WeightSCMN_{\IndFreqbin,\LatestFrame} \NoiseSV_{\IndFreqbin} \NoiseSV_{\IndFreqbin}^{\Hermite},
    \\
    \label{eq:O-RCSCME:model_noise_SCM_rank-M-1}
    \SCMNh_{\IndFreqbin,\LatestFrame} &= \sum_{\IndFrame = 1}^{\LatestFrame} \WeightORCSCME_{\IndFrame,\LatestFrame} \NoiseSrcImage_{\IndFreqbin\IndFrame} \NoiseSrcImage_{\IndFreqbin\IndFrame}^{\Hermite},
    \\
    \label{eq:O-RCSCME:model_noise_src_image}
    \NoiseSrcImage_{\IndFreqbin\IndFrame} &= \bigl( \Identity_{\NumSrc} - \SV_{\IndFreqbin} \DemixVec_{\IndFreqbin\IndTargetSrc}^{\Hermite} \bigr) \bm{\ObsSignal}_{\IndFreqbin\IndFrame}.
\end{align}
For further discussion, we transform (\ref{eq:O-RCSCME:model_noise_SCM_rank-M-1}) and (\ref{eq:O-RCSCME:model_noise_src_image}) as 
\begin{align}
    \label{eq:O-RCSCME:transform_model_noise_SCM_rank-M-1}
    \SCMNh_{\IndFreqbin,\LatestFrame} &= \NoiseImageMat_{\IndFreqbin} \XMat_{\IndFreqbin,\LatestFrame} \NoiseImageMat_{\IndFreqbin}^{\Hermite},
    \\
    \label{eq:O-RCSCME:model_noise_image_mat}
    \NoiseImageMat_{\IndFreqbin} &= \Identity_{\NumMic} - \SV_{\IndFreqbin} \DemixVec_{\IndFreqbin\IndTargetSrc}^{\Hermite},
    \\
    \label{eq:O-RCSCME:model_XMat}
    \XMat_{\IndFreqbin,\LatestFrame} &= \sum_{\IndFrame=1}^{\LatestFrame} \WeightORCSCME_{\IndFrame,\LatestFrame} \bm{\ObsSignal}_{\IndFreqbin\IndFrame} \bm{\ObsSignal}_{\IndFreqbin\IndFrame}^{\Hermite}.
\end{align}
Note that if we set $\WeightORCSCME_{\IndFrame,\LatestFrame} = 1 / \LatestFrame \ (\forall \IndFrame)$, the model of $\SCMN_{\IndFreqbin}$ in online RCSCME, (\ref{eq:O-RCSCME:model_noise_SCM}) and (\ref{eq:O-RCSCME:transform_model_noise_SCM_rank-M-1})--(\ref{eq:O-RCSCME:model_XMat}), coincides with that in offline RCSCME, (\ref{eq:RCSCME:model_SCMN})--(\ref{eq:RCSCME:def:NoiseSrcImage}).
In addition, if we set $\WeightORCSCME_{\IndFrame,\LatestFrame}$ to 
\begin{align}
    \label{eq:O-RCSCME:condition_online_to_blockwise}
    \WeightORCSCME_{\IndFrame,\LatestFrame} = 
    \begin{cases}
        1 / \NumFrame^{(\mathrm{R})},    &\text{(if $\LatestFrame - \NumFrame^{(\mathrm{R})} + 1 \leq \IndFrame \leq \LatestFrame$)},
        \\
        0 ,  &\text{(otherwise)}
    \end{cases}
\end{align}
the model of $\SCMN_{\IndFreqbin}$ coincides with that of RCSCME in B-RCSCME using a $\NumFrame^{(\mathrm{R})}$-frame-long batch.

\subsubsection{Naive Update Rule}
\label{sssec:O-RCSCME:naive_update_rule}

To obtain the online algorithm for RCSCME, we first derive a naive update rule for $\Cost{\RCSCME,\LatestFrame}$ on the basis of the ME algorithm, as in \cite{Kubo2020TASLP}.
We consider the following auxiliary function $\AuxCost{\RCSCME,\LatestFrame}$ using the inequalities derived in \cite{Kubo2020TASLP}:
\begin{align}
    \label{eq:O-RCSCME:auxiliary_cost}
    &\AuxCost{\RCSCME,\LatestFrame}
    \nonumber\\
    &= \sum_{\IndFreqbin} \sum_{\IndFrame=1}^{\LatestFrame} \WeightORCSCME_{\IndFrame,\LatestFrame} \Biggl(
        \frac{\bigl| \bm{\ObsSignal}_{\IndFreqbin\IndFrame}^{\Hermite} \AuxMatT_{\IndFreqbin\IndFrame} \SV_{\IndFreqbin} \bigr|^{2}}{\VarT_{\IndFreqbin\IndFrame}} 
    \nonumber\\
    &\phantom{=}+ \frac{\bm{\ObsSignal}_{\IndFreqbin\IndFrame}^{\Hermite} \bigl( \AuxMatN_{\IndFreqbin\IndFrame} \bigr)^{\Hermite} \SCMNb_{\IndFreqbin,\LatestFrame} \AuxMatN_{\IndFreqbin\IndFrame} \bm{\ObsSignal}_{\IndFreqbin\IndFrame}}{\VarN_{\IndFreqbin\IndFrame}}
    + \frac{\bigl| \DiagVec_{\IndFreqbin}^{\Hermite} \AuxMatN_{\IndFreqbin\IndFrame} \bm{\ObsSignal}_{\IndFreqbin\IndFrame} \bigr|^{2}}{\VarN_{\IndFreqbin\IndFrame} \WeightSCMN_{\IndFreqbin,\LatestFrame}}
    \nonumber\\
    &\phantom{=}  + \log{\det{\AuxMatSCMX_{\IndFreqbin\IndFrame}}} + \VarT_{\IndFreqbin\IndFrame} \bigl( \SV_{\IndFreqbin} \bigr)^{\Hermite} \AuxMatSCMX_{\IndFreqbin\IndFrame}^{-1} \SV_{\IndFreqbin}
    \nonumber\\
    &\phantom{=} + \VarN_{\IndFreqbin\IndFrame} \Bigl( \WeightSCMN_{\IndFreqbin,\LatestFrame} \NoiseSV_{\IndFreqbin}^{\Hermite} \AuxMatSCMX_{\IndFreqbin\IndFrame}^{-1} \NoiseSV_{\IndFreqbin} + \trace \bigl( \AuxMatSCMX_{\IndFreqbin\IndFrame}^{-1} \SCMNh_{\IndFreqbin,\LatestFrame} \bigr) \Bigr) - \NumMic
    \nonumber\\
    &\phantom{=} + (\bar{\ShapeIG} + 1) \biggl( \frac{\VarT_{\IndFreqbin\IndFrame}}{\AuxScaVarT_{\IndFreqbin\IndFrame}} - 1 + \log{\AuxScaVarT_{\IndFreqbin\IndFrame}} \biggr) + \frac{\bar{\ScaleIG}}{\VarT_{\IndFreqbin\IndFrame}}
    \Biggr),
\end{align}
where $\AuxMatT_{\IndFreqbin\IndFrame} \in \mathbb{C}^{\NumMic \times \NumMic}$, $\AuxMatN_{\IndFreqbin\IndFrame} \in \mathbb{C}^{\NumMic \times \NumMic}$, $\AuxMatSCMX_{\IndFreqbin\IndFrame} \in \mathbb{C}^{\NumMic \times \NumMic}$, and $\AuxScaVarT_{\IndFreqbin\IndFrame} > 0$ are auxiliary variables satisfying that $\AuxMatT_{\IndFreqbin\IndFrame} + \AuxMatN_{\IndFreqbin\IndFrame} = \bm{\ObsSignal}_{\IndFreqbin\IndFrame} \bm{\ObsSignal}_{\IndFreqbin\IndFrame}^{\Hermite} / \| \bm{\ObsSignal}_{\IndFreqbin\IndFrame} \|^{2}$ holds and $\AuxMatSCMX_{\IndFreqbin\IndFrame}$ is a positive definite matrix, $\DiagVec_{\IndFreqbin} \in \mathbb{C}^{\NumMic}$ is an eigenvector of $\SCMNh_{\IndFreqbin,\LatestFrame}$ corresponding to its zero eigenvalue that satisfies $\NoiseSV_{\IndFreqbin}^{\Hermite} \DiagVec_{\IndFreqbin} = 1$, and $\SCMNb_{\IndFreqbin,\LatestFrame} \in \mathbb{C}^{\NumMic \times \NumMic}$ is a matrix defined as 
\begin{align}
    \label{eq:O-RCSCME:def_SCMNb}
    \SCMNb_{\IndFreqbin,\LatestFrame} = \bigl( \Identity_{\NumMic} - \DiagVec_{\IndFreqbin} \NoiseSV_{\IndFreqbin}^{\Hermite} \bigr) \bigl( \SCMNh_{\IndFreqbin,\LatestFrame} \bigr)^{+} \bigl( \Identity_{\NumMic} - \NoiseSV_{\IndFreqbin} \DiagVec_{\IndFreqbin}^{\Hermite} \bigr).
\end{align} 
Note that from (\ref{eq:O-RCSCME:transform_model_noise_SCM_rank-M-1}) and (\ref{eq:O-RCSCME:model_noise_image_mat}), $\SCMNh_{\IndFreqbin,\LatestFrame} \DemixVec_{\IndFreqbin\IndTargetSrc} = \bm{0}$ holds, and thus, $\DiagVec_{\IndFreqbin}$ is equivalent to $\DemixVec_{\IndFreqbin\IndTargetSrc}$. 
Here, $\Cost{\RCSCME,\LatestFrame} = \AuxCost{\RCSCME,\LatestFrame}$ holds if and only if the following conditions hold:
\begin{align}
    \label{eq:O-RCSCME:AuxMatT_condition}
    \AuxMatT_{\IndFreqbin\IndFrame} &= \VarT_{\IndFreqbin\IndFrame} \SV_{\IndFreqbin} \bigl(\SV_{\IndFreqbin}\bigr)^{\Hermite} \bigl( \SCMX_{\IndFreqbin\IndFrame} \bigr)^{-1} \frac{\bm{\ObsSignal}_{\IndFreqbin\IndFrame} \bm{\ObsSignal}_{\IndFreqbin\IndFrame}^{\Hermite}}{\| \bm{\ObsSignal}_{\IndFreqbin\IndFrame} \|^{2}},
    \\
    \label{eq:O-RCSCME:AuxMatN_condition}
    \AuxMatN_{\IndFreqbin\IndFrame} &= \VarN_{\IndFreqbin\IndFrame} \SCMN_{\IndFreqbin} \bigl( \SCMX_{\IndFreqbin\IndFrame} \bigr)^{-1} \frac{\bm{\ObsSignal}_{\IndFreqbin\IndFrame} \bm{\ObsSignal}_{\IndFreqbin\IndFrame}^{\Hermite}}{\| \bm{\ObsSignal}_{\IndFreqbin\IndFrame} \|^{2}},
    \\
    \label{eq:O-RCSCME:AuxMatSCMX_condition}
    \AuxMatSCMX_{\IndFreqbin\IndFrame} &= \SCMX_{\IndFreqbin\IndFrame},
    \\
    \label{eq:O-RCSCME:AuxScaVarT_condition}
    \AuxScaVarT_{\IndFreqbin\IndFrame} &= \VarT_{\IndFreqbin\IndFrame}.
\end{align}
This auxiliary function has the same form as that in \cite{Kubo2020TASLP}, except for a frame-dependent weight parameter.
Thus, the update rule based on the ME algorithm can be derived in a similar way to that in \cite{Kubo2020TASLP} as follows:
\begin{align}
    \label{eq:O-RCSCME:naive_update_VarT}
    \VarT_{\IndFreqbin\IndFrame} &\leftarrow \frac{1}{\VarT_{\IndFreqbin\IndFrame}} \left( \frac{
        \bigl|  \bm{\ObsSignal}_{\IndFreqbin\IndFrame}^{\Hermite} \AuxMatT_{\IndFreqbin\IndFrame} \SV_{\IndFreqbin} \bigr|^{2} + \bar{\ScaleIG} 
    }{
        \bigl( \SV_{\IndFreqbin} \bigr)^{\Hermite} \AuxMatSCMX_{\IndFreqbin\IndFrame}^{-1} \SV_{\IndFreqbin} + \frac{\bar{\ShapeIG}+1}{\AuxScaVarT_{\IndFreqbin\IndFrame}}
    }\right),
    \\
    \label{eq:O-RCSCME:naive_update_VarN}
    \VarN_{\IndFreqbin\IndFrame} &\leftarrow \frac{1}{\VarN_{\IndFreqbin\IndFrame}} \left(\frac{
        \bm{\ObsSignal}_{\IndFreqbin\IndFrame}^{\Hermite} \bigl( \AuxMatN_{\IndFreqbin\IndFrame} \bigr)^{\Hermite} \SCMNb_{\IndFreqbin,\LatestFrame} \AuxMatN_{\IndFreqbin\IndFrame} \bm{\ObsSignal}_{\IndFreqbin\IndFrame} + \frac{| \DiagVec_{\IndFreqbin}^{\Hermite} \AuxMatN_{\IndFreqbin\IndFrame} \bm{\ObsSignal}_{\IndFreqbin\IndFrame} |^{2}}{\WeightSCMN_{\IndFreqbin,\LatestFrame}}
    }{
        \WeightSCMN_{\IndFreqbin,\LatestFrame} \NoiseSV_{\IndFreqbin}^{\Hermite} \AuxMatSCMX_{\IndFreqbin\IndFrame}^{-1} \NoiseSV_{\IndFreqbin} + \trace \bigl( \AuxMatSCMX_{\IndFreqbin\IndFrame}^{-1} \SCMNh_{\IndFreqbin,\LatestFrame} \bigr)
    }\right),
    \\
    \label{eq:O-RCSCME:naive_update_WeightSCMN}
    \WeightSCMN_{\IndFreqbin,\LatestFrame} &\leftarrow \frac{1}{\WeightSCMN_{\IndFreqbin,\LatestFrame}} \left(\frac{
        \sum_{\IndFrame=1}^{\LatestFrame} \WeightORCSCME_{\IndFrame,\LatestFrame} \frac{| \DiagVec_{\IndFreqbin}^{\Hermite} \AuxMatN_{\IndFreqbin\IndFrame} \bm{\ObsSignal}_{\IndFreqbin\IndFrame} |^{2}}{\VarN_{\IndFreqbin\IndFrame}} 
    }{
        \sum_{\IndFrame=1}^{\LatestFrame} \WeightORCSCME_{\IndFrame,\LatestFrame} \VarN_{\IndFreqbin\IndFrame} \NoiseSV_{\IndFreqbin}^{\Hermite} \AuxMatSCMX_{\IndFreqbin\IndFrame}^{-1} \NoiseSV_{\IndFreqbin}
    }\right).
\end{align}
Here, since the auxiliary function is minimized with respect to the auxiliary variables $\AuxMatT_{\IndFreqbin\IndFrame}$, $\AuxMatN_{\IndFreqbin\IndFrame}$, $\AuxMatSCMX_{\IndFreqbin\IndFrame}$, and $\AuxScaVarT_{\IndFreqbin\IndFrame}$ when the equalities in (\ref{eq:O-RCSCME:AuxMatT_condition})--(\ref{eq:O-RCSCME:AuxScaVarT_condition}) hold, $\AuxMatT_{\IndFreqbin\IndFrame}$, $\AuxMatN_{\IndFreqbin\IndFrame}$, $\AuxMatSCMX_{\IndFreqbin\IndFrame}$, and $\AuxScaVarT_{\IndFreqbin\IndFrame}$ are updated using (\ref{eq:O-RCSCME:AuxMatT_condition})--(\ref{eq:O-RCSCME:AuxScaVarT_condition}), respectively, after each update of $\VarT_{\IndFreqbin\IndFrame}$, $\VarN_{\IndFreqbin\IndFrame}$, and $\WeightSCMN_{\IndFreqbin,\LatestFrame}$.
It is guaranteed that the update of each variable monotonically nonincreases the framewise cost function $\Cost{\RCSCME,\LatestFrame}$.
Note that if we set $\WeightORCSCME_{\IndFrame,\LatestFrame} = 1 / \LatestFrame \ (\forall \IndFrame)$, the entire update rule of online RCSCME coincides with that of offline RCSCME. 
In addition, if (\ref{eq:O-RCSCME:condition_online_to_blockwise}) holds, the entire update rule coincides with that of RCSCME in B-RCSCME using a $\NumFrame^{(\mathrm{R})}$-frame-long batch, as described in Section~\ref{sec:related_methods}-\ref{ssec:blockwise-RCSCME}.

\subsubsection{Online Update Rule}
\label{sssec:O-RCSCME:online_update_rule}

In the update rules derived in Section~\ref{sec:proposed_method}-\ref{ssec:online_RCSCME}-\ref{sssec:O-RCSCME:naive_update_rule}, the auxiliary variables need to be recalculated each time the objective variables $\VarT_{\IndFreqbin\IndFrame}$, $\VarN_{\IndFreqbin\IndFrame}$, and $\WeightSCMN_{\IndFreqbin,\LatestFrame}$ are updated.
To reduce the computational cost while maintaining high speech extraction performance, we approximate some intermediate parameters that involve summation over time frames in the auxiliary function using estimates of the time-varying parameters corresponding to $\IndFrame = 1, ..., \LatestFrame - \ApproxFrameORCSCME\ (\ApproxFrameORCSCME \geq 1)$ at the $(\LatestFrame - 1)$th-frame estimation\footnote{
    It is also possible for online NSR-ILRMA to update the time-varying parameters over multiple past frames.
    However, in that case, we would not be able to use the Sherman--Morrison formula to calculate the inverse matrices efficiently, which would increase the computational cost of the iterative updates.
    Therefore, to ensure real-time capability, we fix the time-varying parameters for all frames except the latest one in online NSR-ILRMA.
}.
We then minimize the approximate auxiliary function.

To derive an efficient update rule, we approximate $\VarT_{\IndFreqbin\IndFrame}$, $\VarN_{\IndFreqbin\IndFrame}$, $\AuxMatT_{\IndFreqbin\IndFrame}$, $\AuxMatN_{\IndFreqbin\IndFrame}$, $\AuxMatSCMX_{\IndFreqbin\IndFrame}$, and $\AuxScaVarT_{\IndFreqbin\IndFrame}$ for $\IndFrame = 1, ..., \LatestFrame - \ApproxFrameORCSCME$ in (\ref{eq:O-RCSCME:auxiliary_cost}) using their estimates obtained at the ($\LatestFrame-1$)th-frame estimation, $\EstVarT_{\IndFreqbin\IndFrame,\LatestFrame-1}$, $\EstVarN_{\IndFreqbin\IndFrame,\LatestFrame-1}$, $\EstAuxMatT_{\IndFreqbin\IndFrame,\LatestFrame-1}$, $\EstAuxMatN_{\IndFreqbin\IndFrame,\LatestFrame-1}$, $\EstAuxMatSCMX_{\IndFreqbin\IndFrame,\LatestFrame-1}$, and $\EstAuxScaVarT_{\IndFreqbin\IndFrame,\LatestFrame-1}$, respectively.
Then, the approximate auxiliary function with respect to the objective variables $\VarT_{\IndFreqbin\IndFrame}$ and $\VarN_{\IndFreqbin\IndFrame}$ for $\IndFrame = \LatestFrame - \ApproxFrameORCSCME + 1, ..., \LatestFrame$ and $\WeightSCMN_{\IndFreqbin,\LatestFrame}$ is expressed as follows:
\begin{align}
    \label{eq:O-RCSCME:approx_aux_function}
    &\Cost{\RCSCME,\LatestFrame}
    \nonumber\\
    &\approx
    \sum_{\IndFreqbin} \Biggl[
        \sum_{\IndFrame = \LatestFrame - \ApproxFrameORCSCME + 1}^{\LatestFrame} \WeightORCSCME_{\IndFrame,\LatestFrame} \Biggl(
            \frac{\bigl| \bm{\ObsSignal}_{\IndFreqbin\IndFrame}^{\Hermite} \AuxMatT_{\IndFreqbin\IndFrame} \SV_{\IndFreqbin} \bigr|^{2} + \bar{\ScaleIG}}{\VarT_{\IndFreqbin\IndFrame}} 
    \nonumber\\
            &\phantom{\approx}+ \frac{\bm{\ObsSignal}_{\IndFreqbin\IndFrame}^{\Hermite} \bigl( \AuxMatN_{\IndFreqbin\IndFrame} \bigr)^{\Hermite} \SCMNb_{\IndFreqbin,\LatestFrame} \AuxMatN_{\IndFreqbin\IndFrame} \bm{\ObsSignal}_{\IndFreqbin\IndFrame}}{\VarN_{\IndFreqbin\IndFrame}} + \frac{\bigl| \DiagVec_{\IndFreqbin}^{\Hermite} \AuxMatN_{\IndFreqbin\IndFrame} \bm{\ObsSignal}_{\IndFreqbin\IndFrame} \bigr|^{2}}{\VarN_{\IndFreqbin\IndFrame} \WeightSCMN_{\IndFreqbin,\LatestFrame}} 
    \nonumber\\
            &\phantom{\approx}+ \VarT_{\IndFreqbin\IndFrame} \biggl( \bigl( \SV_{\IndFreqbin} \bigr)^{\Hermite} \AuxMatSCMX_{\IndFreqbin\IndFrame}^{-1} \SV_{\IndFreqbin} + \frac{\bar{\ShapeIG} + 1}{\AuxScaVarT_{\IndFreqbin\IndFrame}} \biggr)
    \nonumber\\
            &\phantom{\approx}+ \VarN_{\IndFreqbin\IndFrame} \Bigl( \WeightSCMN_{\IndFreqbin,\LatestFrame} \NoiseSV_{\IndFreqbin}^{\Hermite} \AuxMatSCMX_{\IndFreqbin\IndFrame}^{-1} \NoiseSV_{\IndFreqbin} + \trace \bigl( \AuxMatSCMX_{\IndFreqbin\IndFrame}^{-1} \SCMNh_{\IndFreqbin,\LatestFrame} \bigr) \Bigr)
        \Biggr)
    \nonumber\\
    &\phantom{\approx}+ \frac{\EstNumerWeightSCMN_{\IndFreqbin,\LatestFrame}}{\WeightSCMN_{\IndFreqbin,\LatestFrame}} + \EstDenomWeightSCMN_{\IndFreqbin,\LatestFrame} \WeightSCMN_{\IndFreqbin,\LatestFrame}
    \Biggr] + \const,
\end{align}
where $\const$ denotes the term independent of $\VarT_{\IndFreqbin\IndFrame}\ (\forall \IndFreqbin, \IndFrame = \LatestFrame - \ApproxFrameORCSCME + 1, ..., \LatestFrame)$, $\VarN_{\IndFreqbin\IndFrame}\ (\forall \IndFreqbin, \IndFrame = \LatestFrame - \ApproxFrameORCSCME + 1, ..., \LatestFrame)$, and $\WeightSCMN_{\IndFreqbin,\LatestFrame}\ (\forall \IndFreqbin)$.
Here, $\EstNumerWeightSCMN_{\IndFreqbin,\LatestFrame}$ and $\EstDenomWeightSCMN_{\IndFreqbin,\LatestFrame}$ are nonnegative intermediate variables that aggregate the approximated time-varying parameters for $\IndFrame = 1, ..., \LatestFrame - \ApproxFrameORCSCME$, and they are defined as
\begin{align}
    \label{eq:O-RCSCME:define_Est_numer_WeightSCMN}
    \EstNumerWeightSCMN_{\IndFreqbin,\LatestFrame} &= \sum_{\IndFrame = 1}^{\LatestFrame - \ApproxFrameORCSCME} \WeightORCSCME_{\IndFrame,\LatestFrame} \frac{
        \bigl| \DiagVec_{\IndFreqbin}^{\Hermite} \EstAuxMatN_{\IndFreqbin\IndFrame,\LatestFrame-1} \bm{\ObsSignal}_{\IndFreqbin\IndFrame} \bigr|^{2}
    }{
        \EstVarN_{\IndFreqbin\IndFrame,\LatestFrame-1}
    },
    \\
    \label{eq:O-RCSCME:define_Est_denom_WeightSCMN}
    \EstDenomWeightSCMN_{\IndFreqbin,\LatestFrame} &= \sum_{\IndFrame = 1}^{\LatestFrame - \ApproxFrameORCSCME} \WeightORCSCME_{\IndFrame,\LatestFrame} \EstVarN_{\IndFreqbin\IndFrame,\LatestFrame-1} \NoiseSV_{\IndFreqbin}^{\Hermite} \EstAuxMatSCMX_{\IndFreqbin\IndFrame,\LatestFrame-1}^{-1} \NoiseSV_{\IndFreqbin},
\end{align}
where $\EstNumerWeightSCMN_{\IndFreqbin,\LatestFrame}$ and $\EstDenomWeightSCMN_{\IndFreqbin,\LatestFrame}$ are set to $0$ when $\LatestFrame \leq \ApproxFrameORCSCME$.
When focusing on each objective variable, the approximate auxiliary function becomes a summation of a reciprocal term, a linear term, and a constant term as in \cite{Kubo2020TASLP}.
Thus, the update rule based on the ME algorithm can be derived in a similar way to that in \cite{Kubo2020TASLP} as follows:
\begin{align}
    \label{eq:O-RCSCME:update_approx_aux_VarT}
    &\VarT_{\IndFreqbin\IndFrame} \leftarrow \frac{1}{\VarT_{\IndFreqbin\IndFrame}} \left(
        \frac{
            \bigl| \bm{\ObsSignal}_{\IndFreqbin\IndFrame}^{\Hermite} \AuxMatT_{\IndFreqbin\IndFrame} \SV_{\IndFreqbin} \bigr|^{2} + \bar{\ScaleIG}
        }{
            \bigl( \SV_{\IndFreqbin} \bigr)^{\Hermite} \AuxMatSCMX_{\IndFreqbin\IndFrame}^{-1} \SV_{\IndFreqbin} + \frac{\bar{\ShapeIG} + 1}{\AuxScaVarT_{\IndFreqbin\IndFrame}}
        }
    \right),
    \\
    \label{eq:O-RCSCME:update_approx_aux_VarN}
    &\VarN_{\IndFreqbin\IndFrame} \leftarrow \frac{1}{\VarN_{\IndFreqbin\IndFrame}} \left(
        \frac{
            \bm{\ObsSignal}_{\IndFreqbin\IndFrame}^{\Hermite} \bigl( \AuxMatN_{\IndFreqbin\IndFrame} \bigr)^{\Hermite} \SCMNb_{\IndFreqbin,\LatestFrame} \AuxMatN_{\IndFreqbin\IndFrame} \bm{\ObsSignal}_{\IndFreqbin\IndFrame} + \frac{\bigl| \DiagVec_{\IndFreqbin}^{\Hermite} \AuxMatN_{\IndFreqbin\IndFrame} \bm{\ObsSignal}_{\IndFreqbin\IndFrame} \bigr|^{2}}{\WeightSCMN_{\IndFreqbin,\LatestFrame}}
        }{
            \WeightSCMN_{\IndFreqbin,\LatestFrame} \NoiseSV_{\IndFreqbin}^{\Hermite} \AuxMatSCMX_{\IndFreqbin\IndFrame}^{-1} \NoiseSV_{\IndFreqbin} + \trace \Bigl( \AuxMatSCMX_{\IndFreqbin\IndFrame}^{-1} \SCMNh_{\IndFreqbin,\LatestFrame} \Bigr)
        }
    \right),
    \\
    &\NumerWeightSCMN_{\IndFreqbin,\LatestFrame} = \EstNumerWeightSCMN_{\IndFreqbin,\LatestFrame} + \sum_{\IndFrame = \LatestFrame - \ApproxFrameORCSCME + 1}^{\LatestFrame} \WeightORCSCME_{\IndFrame,\LatestFrame} \frac{\bigl| \DiagVec_{\IndFreqbin}^{\Hermite} \AuxMatN_{\IndFreqbin\IndFrame} \bm{\ObsSignal}_{\IndFreqbin\IndFrame} \bigr|^{2}}{\VarN_{\IndFreqbin\IndFrame}},
    \\
    &\DenomWeightSCMN_{\IndFreqbin,\LatestFrame} = \EstDenomWeightSCMN_{\IndFreqbin,\LatestFrame} + \sum_{\IndFrame = \LatestFrame - \ApproxFrameORCSCME + 1}^{\LatestFrame} \WeightORCSCME_{\IndFrame,\LatestFrame} \VarN_{\IndFreqbin\IndFrame} \NoiseSV_{\IndFreqbin}^{\Hermite} \AuxMatSCMX_{\IndFreqbin\IndFrame}^{-1} \NoiseSV_{\IndFreqbin},
    \\
    \label{eq:O-RCSCME:update_approx_aux_WeightSCMN}
    &\WeightSCMN_{\IndFreqbin,\LatestFrame} \leftarrow \frac{\NumerWeightSCMN_{\IndFreqbin,\LatestFrame}}{\DenomWeightSCMN_{\IndFreqbin,\LatestFrame} \WeightSCMN_{\IndFreqbin,\LatestFrame}},
\end{align}
where $\NumerWeightSCMN_{\IndFreqbin}$ and $\DenomWeightSCMN_{\IndFreqbin}$ are nonnegative intermediate variables.
For initialization, $\VarT_{\IndFreqbin\LatestFrame}$ and $\VarN_{\IndFreqbin\LatestFrame}$ are set to
\begin{align}
    \label{eq:O-RCSCME:initialization_VarT}
    \VarT_{\IndFreqbin\LatestFrame} &= \abs{\DemixVec_{\IndFreqbin\IndTargetSrc}^{\Hermite} \bm{\ObsSignal}_{\IndFreqbin\LatestFrame}}^{2},
    \\
    \label{eq:O-RCSCME:initialization_VarN}
    \VarN_{\IndFreqbin\LatestFrame} &= \frac{1}{\NumMic} \bm{\ObsSignal}_{\IndFreqbin\LatestFrame}^{\Hermite} \NoiseImageMat_{\IndFreqbin}^{\Hermite} \bigl( \SCMNh_{\IndFreqbin,\LatestFrame} \bigr)^{+} \NoiseImageMat_{\IndFreqbin} \bm{\ObsSignal}_{\IndFreqbin\LatestFrame},
\end{align}
and $\VarT_{\IndFreqbin\IndFrame}$ and $\VarN_{\IndFreqbin\IndFrame}$ for $\IndFrame = \LatestFrame - \ApproxFrameORCSCME + 1, ..., \LatestFrame-1$ are set to their estimates obtained at the $(\LatestFrame-1)$th-frame estimation.
$\WeightSCMN_{\IndFreqbin,1}$ is initialized as a sufficiently small parameter $\epsilon_{\WeightSCMN}$ for stability.
Furthermore, we transform (\ref{eq:O-RCSCME:define_Est_numer_WeightSCMN}) and (\ref{eq:O-RCSCME:define_Est_denom_WeightSCMN}) as 
\begin{align}
    \label{eq:O-RCSCME:transform_Est_numer_WeightSCMN}
    \EstNumerWeightSCMN_{\IndFreqbin,\LatestFrame} &= \WeightORCSCME_{\LatestFrame - \ApproxFrameORCSCME, \LatestFrame} \frac{
        \bigl| \DiagVec_{\IndFreqbin}^{\Hermite} \EstAuxMatN_{\IndFreqbin(\LatestFrame - \ApproxFrameORCSCME), \LatestFrame-1} \bm{\ObsSignal}_{\IndFreqbin(\LatestFrame-\ApproxFrameORCSCME)} \bigr|^{2}
    }{
        \EstVarN_{\IndFreqbin(\LatestFrame-\ApproxFrameORCSCME),\LatestFrame-1}
    }
    \nonumber\\
    &\phantom{=} + \sum_{\IndFrame = 1}^{(\LatestFrame - 1) - \ApproxFrameORCSCME} \frac{\WeightORCSCME_{\IndFrame,\LatestFrame}}{\WeightORCSCME_{\IndFrame,\LatestFrame-1}} \WeightORCSCME_{\IndFrame,\LatestFrame-1} \frac{
        \bigl| \DiagVec_{\IndFreqbin}^{\Hermite} \EstAuxMatN_{\IndFreqbin\IndFrame,\LatestFrame-1} \bm{\ObsSignal}_{\IndFreqbin\IndFrame} \bigr|^{2}
    }{\EstVarN_{\IndFreqbin\IndFrame,\LatestFrame-1}},
    \\
    \label{eq:O-RCSCME:transform_Est_denom_WeightSCMN}
    \EstDenomWeightSCMN_{\IndFreqbin,\LatestFrame} &= \WeightORCSCME_{\LatestFrame - \ApproxFrameORCSCME, \LatestFrame} \EstVarN_{\IndFreqbin(\LatestFrame - \ApproxFrameORCSCME),\LatestFrame-1} \NoiseSV_{\IndFreqbin}^{\Hermite} \EstAuxMatSCMX_{\IndFreqbin(\LatestFrame-\ApproxFrameORCSCME),\LatestFrame-1}^{-1} \NoiseSV_{\IndFreqbin}
    \nonumber\\
    &\phantom{=} + \sum_{\IndFrame = 1}^{(\LatestFrame-1) - \ApproxFrameORCSCME} \frac{\WeightORCSCME_{\IndFrame,\LatestFrame}}{\WeightORCSCME_{\IndFrame,\LatestFrame-1}} \WeightORCSCME_{\IndFrame,\LatestFrame-1} \EstVarN_{\IndFreqbin\IndFrame,\LatestFrame-1} \NoiseSV_{\IndFreqbin}^{\Hermite} \EstAuxMatSCMX_{\IndFreqbin\IndFrame,\LatestFrame-1}^{-1} \NoiseSV_{\IndFreqbin}.
\end{align}
Here, since $\EstVarN_{\IndFreqbin\IndFrame,\LatestFrame-1}$, $\EstAuxMatN_{\IndFreqbin\IndFrame,\LatestFrame-1}$, and $\EstAuxMatSCMX_{\IndFreqbin\IndFrame,\LatestFrame-1}$ for $\IndFrame = 1, ..., (\LatestFrame - 1) - \ApproxFrameORCSCME$ are fixed to $\EstVarN_{\IndFreqbin\IndFrame,\LatestFrame-2}$, $\EstAuxMatN_{\IndFreqbin\IndFrame,\LatestFrame-2}$, and $\EstAuxMatSCMX_{\IndFreqbin\IndFrame,\LatestFrame-2}$ and are not updated at the $(\LatestFrame-1)$th-frame estimation, $\EstVarN_{\IndFreqbin\IndFrame,\LatestFrame-1} = \EstVarN_{\IndFreqbin\IndFrame,\LatestFrame-2}$, $\EstAuxMatN_{\IndFreqbin\IndFrame,\LatestFrame-1} = \EstAuxMatN_{\IndFreqbin\IndFrame,\LatestFrame-2}$, and $\EstAuxMatSCMX_{\IndFreqbin\IndFrame,\LatestFrame-1} = \EstAuxMatSCMX_{\IndFreqbin\IndFrame,\LatestFrame-2}$ hold for $\IndFrame = 1, ..., (\LatestFrame-1)-\ApproxFrameORCSCME$.
Therefore, similarly to the discussion in Section~\ref{sec:proposed_method}-\ref{ssec:online_NSR-ILRMA}-\ref{sssec:O-NSR-ILRMA:online_update_rule}, if $\WeightORCSCME_{\IndFrame,\LatestFrame} / \WeightORCSCME_{\IndFrame,\LatestFrame-1}$ is independent of $\IndFrame$, the second term of the right-hand sides of (\ref{eq:O-RCSCME:define_Est_numer_WeightSCMN}) and (\ref{eq:O-RCSCME:define_Est_denom_WeightSCMN}) can be expressed using the estimates of $\EstNumerWeightSCMN_{\IndFreqbin,\LatestFrame-1}$ and $\EstDenomWeightSCMN_{\IndFreqbin,\LatestFrame-1}$, respectively.
To satisfy this condition, we set $\WeightORCSCME_{\IndFrame,\LatestFrame}$ as follows, in the same manner as $\WeightOILRMA_{\IndFrame,\LatestFrame}$:
\begin{align}
    \label{eq:O-RCSCME:WeightORCSCME}
    \WeightORCSCME_{\IndFrame,\LatestFrame} = \frac{1 - \ForgetRCSCME}{1 - \ForgetRCSCME^{\LatestFrame}} \ForgetRCSCME^{\LatestFrame - \IndFrame},
\end{align}
where $\ForgetRCSCME \in (0, 1)$ denotes the forgetting factor for online RCSCME.
As a result, the update of $\EstNumerWeightSCMN_{\IndFreqbin,\LatestFrame}$ and $\EstDenomWeightSCMN_{\IndFreqbin,\LatestFrame}$ can be expressed as 
\begin{align}
    \label{eq:O-RCSCME:update_Est_numer_WeightSCMN}
    \EstNumerWeightSCMN_{\IndFreqbin,\LatestFrame} &= \frac{1 - \ForgetRCSCME}{1 - \ForgetRCSCME^{\LatestFrame}} \ForgetRCSCME^{\ApproxFrameORCSCME} \frac{
        \bigl| \DiagVec_{\IndFreqbin}^{\Hermite} \EstAuxMatN_{\IndFreqbin(\LatestFrame - \ApproxFrameORCSCME), \LatestFrame-1} \bm{\ObsSignal}_{\IndFreqbin(\LatestFrame-\ApproxFrameORCSCME)} \bigr|^{2}
    }{
        \EstVarN_{\IndFreqbin(\LatestFrame-\ApproxFrameORCSCME),\LatestFrame-1}
    }
    \nonumber\\
    &\phantom{=} + \frac{\ForgetRCSCME - \ForgetRCSCME^{\LatestFrame}}{1 - \ForgetRCSCME^{\LatestFrame}} \EstNumerWeightSCMN_{\IndFreqbin,\LatestFrame-1},
    \\
    \label{eq:O-RCSCME:update_Est_denom_WeightSCMN}
    \EstDenomWeightSCMN_{\IndFreqbin,\LatestFrame} &= \frac{1 - \ForgetRCSCME}{1 - \ForgetRCSCME^{\LatestFrame}} \ForgetRCSCME^{\ApproxFrameORCSCME} \EstVarN_{\IndFreqbin(\LatestFrame - \ApproxFrameORCSCME),\LatestFrame-1} \NoiseSV_{\IndFreqbin}^{\Hermite} \EstAuxMatSCMX_{\IndFreqbin(\LatestFrame-\ApproxFrameORCSCME),\LatestFrame-1}^{-1} \NoiseSV_{\IndFreqbin}
    \nonumber\\
    &\phantom{=} + \frac{\ForgetRCSCME - \ForgetRCSCME^{\LatestFrame}}{1 - \ForgetRCSCME^{\LatestFrame}} \EstDenomWeightSCMN_{\IndFreqbin,\LatestFrame-1}.
\end{align}
Finally, by substituting (\ref{eq:O-RCSCME:AuxMatT_condition})--(\ref{eq:O-RCSCME:AuxScaVarT_condition}) and (\ref{eq:O-RCSCME:WeightORCSCME}) into (\ref{eq:O-RCSCME:update_approx_aux_VarT})--(\ref{eq:O-RCSCME:update_approx_aux_WeightSCMN}), (\ref{eq:O-RCSCME:update_Est_numer_WeightSCMN}), and (\ref{eq:O-RCSCME:update_Est_denom_WeightSCMN}), we can derive the entire update rule as follows:
\begin{align}
    \label{eq:O-RCSCME:online_update_VarT}
    \VarT_{\IndFreqbin\IndFrame} &\leftarrow \VarT_{\IndFreqbin\IndFrame} \left(\frac{
        \bigl| \bm{\ObsSignal}_{\IndFreqbin\IndFrame}^{\Hermite} \bigl( \SCMX_{\IndFreqbin\IndFrame} \bigr)^{-1} \SV_{\IndFreqbin} \bigr|^{2} + \frac{\bar{\ScaleIG}}{( \VarT_{\IndFreqbin\IndFrame} )^{2}}
    }{
        \bigl( \SV_{\IndFreqbin} \bigr)^{\Hermite} \bigl( \SCMX_{\IndFreqbin\IndFrame} \bigr)^{-1} \SV_{\IndFreqbin} + \frac{\bar{\ShapeIG} + 1}{\VarT_{\IndFreqbin\IndFrame}}
    }\right),
    \\
    \label{eq:O-RCSCME:online_update_VarN}
    \VarN_{\IndFreqbin\IndFrame} &\leftarrow \VarN_{\IndFreqbin\IndFrame} \left(\frac{
        \bm{\ObsSignal}_{\IndFreqbin\IndFrame}^{\Hermite} \bigl( \SCMX_{\IndFreqbin\IndFrame} \bigr)^{-1} \SCMN_{\IndFreqbin} \bigl( \SCMX_{\IndFreqbin\IndFrame} \bigr)^{-1} \bm{\ObsSignal}_{\IndFreqbin\IndFrame}
    }{
        \trace \Bigl( \bigl( \SCMX_{\IndFreqbin\IndFrame} \bigr)^{-1} \SCMN_{\IndFreqbin} \Bigr)
    }\right),
    \\
    \label{eq:O-RCSCME:online_update_numer_weightSCMN}
    \NumerWeightSCMN_{\IndFreqbin,\LatestFrame} &\leftarrow \EstNumerWeightSCMN_{\IndFreqbin,\LatestFrame} 
    \nonumber\\
    &+ \WeightSCMN_{\IndFreqbin,\LatestFrame}^{2} \frac{1 - \ForgetRCSCME}{1 - \ForgetRCSCME^{\LatestFrame}} \sum_{\IndFrame = \LatestFrame - \ApproxFrameORCSCME + 1}^{\LatestFrame} \ForgetRCSCME^{\LatestFrame - \IndFrame} \VarN_{\IndFreqbin\IndFrame} \bigl| \NoiseSV_{\IndFreqbin}^{\Hermite} \bigl( \SCMX_{\IndFreqbin\IndFrame} \bigr)^{-1} \bm{\ObsSignal}_{\IndFreqbin\IndFrame} \bigr|^{2},
    \\
    \label{eq:O-RCSCME:online_update_denom_weightSCMN}
    \DenomWeightSCMN_{\IndFreqbin,\LatestFrame} &\leftarrow \EstDenomWeightSCMN_{\IndFreqbin,\LatestFrame}
    \nonumber\\
    &+ \frac{1 - \ForgetRCSCME}{1 - \ForgetRCSCME^{\LatestFrame}} \sum_{\IndFrame = \LatestFrame - \ApproxFrameORCSCME + 1}^{\LatestFrame} \ForgetRCSCME^{\LatestFrame - \IndFrame} \VarN_{\IndFreqbin\IndFrame} \NoiseSV_{\IndFreqbin}^{\Hermite} \bigl( \SCMX_{\IndFreqbin\IndFrame} \bigr)^{-1} \NoiseSV_{\IndFreqbin},
    \\
    \label{eq:O-RCSCME:online_update_weightSCMN}
    \WeightSCMN_{\IndFreqbin,\LatestFrame} &\leftarrow \frac{\NumerWeightSCMN_{\IndFreqbin,\LatestFrame}}{\DenomWeightSCMN_{\IndFreqbin,\LatestFrame} \WeightSCMN_{\IndFreqbin,\LatestFrame}}.
\end{align}
Note that $\SCMN_{\IndFreqbin}$ is updated after the update of $\WeightSCMN_{\IndFreqbin,\LatestFrame}$, and $\SCMX_{\IndFreqbin\IndFrame}$ is updated after the update of $\VarN_{\IndFreqbin\IndFrame}$ and $\WeightSCMN_{\IndFreqbin,\LatestFrame}$.
$\EstNumerWeightSCMN_{\IndFreqbin,\LatestFrame}$ and $\EstDenomWeightSCMN_{\IndFreqbin,\LatestFrame}$ are updated before the iterations in the $\LatestFrame$th frame as 
\begin{align}
    \label{eq:O-RCSCME:online_update_Est_numer_weightSCMN}
    \EstNumerWeightSCMN_{\IndFreqbin,\LatestFrame} &\leftarrow \frac{1 - \ForgetRCSCME}{1 - \ForgetRCSCME^{\LatestFrame}} \ForgetRCSCME^{\ApproxFrameORCSCME} \EstVarN_{\IndFreqbin(\LatestFrame-\ApproxFrameORCSCME),\LatestFrame-1}
    \nonumber\\
    &\phantom{\leftarrow} \cdot \abs{\NoiseSV_{\IndFreqbin}^{\Hermite} \bigl( \EstSCMX_{\IndFreqbin(\LatestFrame-\ApproxFrameORCSCME),\LatestFrame-1}\bigr)^{-1} \bm{\ObsSignal}_{\IndFreqbin(\LatestFrame-\ApproxFrameORCSCME)}}^{2} + \frac{\ForgetRCSCME - \ForgetRCSCME^{\LatestFrame}}{1 - \ForgetRCSCME^{\LatestFrame}} \EstNumerWeightSCMN_{\IndFreqbin,\LatestFrame-1},
    \\
    \label{eq:O-RCSCME:online_update_Est_denom_weightSCMN}
    \EstDenomWeightSCMN_{\IndFreqbin,\LatestFrame} &\leftarrow \frac{1 - \ForgetRCSCME}{1 - \ForgetRCSCME^{\LatestFrame}} \ForgetRCSCME^{\ApproxFrameORCSCME} \EstVarN_{\IndFreqbin(\LatestFrame-\ApproxFrameORCSCME),\LatestFrame-1} \NoiseSV_{\IndFreqbin}^{\Hermite} \bigl( \EstSCMX_{\IndFreqbin(\LatestFrame-\ApproxFrameORCSCME),\LatestFrame-1} \bigr)^{-1} \NoiseSV_{\IndFreqbin}
    \nonumber\\
    &\phantom{\leftarrow} + \frac{\ForgetRCSCME - \ForgetRCSCME^{\LatestFrame}}{1 - \ForgetRCSCME^{\LatestFrame}} \EstDenomWeightSCMN_{\IndFreqbin,\LatestFrame-1},
\end{align}
where $\EstSCMX_{\IndFreqbin(\LatestFrame-\ApproxFrameORCSCME),\LatestFrame-1}$ denotes the estimate of $\SCMX_{\IndFreqbin(\LatestFrame-\ApproxFrameORCSCME)}$ obtained at the $(\LatestFrame-1)$th-frame estimation.

\subsubsection{Techniques for Acceleration}
\label{sssec:O-RCSCME:techniques_for_acceleration}

The update rules for online RCSCME (\ref{eq:O-RCSCME:online_update_VarT})--(\ref{eq:O-RCSCME:online_update_Est_denom_weightSCMN}) require numerous matrix operations including the computation of $( \SCMX_{\IndFreqbin\IndFrame} )^{-1}$, which hinders real-time implementation.
In \cite{Kubo2020TASLP}, by analytically transforming the matrix operations, such as $(\SCMX_{\IndFreqbin\IndFrame})^{-1}$ and $( \SCMN_{\IndFreqbin} )^{-1}$, efficient update rules consisting of scalar operations are derived.
Following this, we apply this acceleration technique into (\ref{eq:O-RCSCME:online_update_VarT})--(\ref{eq:O-RCSCME:online_update_Est_denom_weightSCMN}).
First, we define the following intermediate variables:
\begin{align}
    \label{eq:O-RCSCME:define_ScalarSecond_aRa}
    \ScalarSecond_{\IndFreqbin}^{\upaRa} &\coloneqq \bigl( \SV_{\IndFreqbin} \bigr)^{\Hermite} \SCMNb_{\IndFreqbin,\LatestFrame} \SV_{\IndFreqbin},
    \\
    \label{eq:O-RCSCME:define_ScalarSecond_aRx}
    \ScalarSecond_{\IndFreqbin\IndFrame}^{\upaRx} &\coloneqq \bigl( \SV_{\IndFreqbin} \bigr)^{\Hermite} \SCMNb_{\IndFreqbin,\LatestFrame} \bm{\ObsSignal}_{\IndFreqbin\IndFrame},
    \\
    \label{eq:O-RCSCME:define_ScalarSecond_xRx}
    \ScalarSecond_{\IndFreqbin\IndFrame}^{\upxRx} &\coloneqq \bm{\ObsSignal}_{\IndFreqbin\IndFrame}^{\Hermite} \SCMNb_{\IndFreqbin,\LatestFrame} \bm{\ObsSignal}_{\IndFreqbin\IndFrame},
    \\
    \label{eq:O-RCSCME:define_ScalarSecond_ua}
    \ScalarSecond_{\IndFreqbin}^{\upua} &\coloneqq \DiagVec_{\IndFreqbin}^{\Hermite} \SV_{\IndFreqbin},
    \\
    \label{eq:O-RCSCME:define_ScalarSecond_ux}
    \ScalarSecond_{\IndFreqbin\IndFrame}^{\upux} &\coloneqq \DiagVec_{\IndFreqbin}^{\Hermite} \bm{\ObsSignal}_{\IndFreqbin\IndFrame},
    \\
    \label{eq:O-RCSCME:define_ScalarFirst_aRa}
    \ScalarFirst_{\IndFreqbin}^{\upaRa} &\coloneqq \ScalarSecond_{\IndFreqbin}^{\upaRa} + \frac{\abs{\ScalarSecond_{\IndFreqbin}^{\upua}}^{2}}{\WeightSCMN_{\IndFreqbin,\LatestFrame}},
    \\
    \label{eq:O-RCSCME:define_ScalarFirst_aRx}
    \ScalarFirst_{\IndFreqbin\IndFrame}^{\upaRx} &\coloneqq \ScalarSecond_{\IndFreqbin\IndFrame}^{\upaRx} + \frac{\bigl( \ScalarSecond_{\IndFreqbin}^{\upua} \bigr)^{*} \ScalarSecond_{\IndFreqbin\IndFrame}^{\upux}}{\WeightSCMN_{\IndFreqbin,\LatestFrame}},
    \\
    \label{eq:O-RCSCME:define_ScalarFirst_xRx}
    \ScalarFirst_{\IndFreqbin\IndFrame}^{\upxRx} &\coloneqq \ScalarSecond_{\IndFreqbin\IndFrame}^{\upxRx} + \frac{\abs{\ScalarSecond_{\IndFreqbin\IndFrame}^{\upux}}^{2}}{\WeightSCMN_{\IndFreqbin,\LatestFrame}},
    \\
    \label{eq:O-RCSCME:define_ScalarFirst_bRa}
    \ScalarFirst_{\IndFreqbin}^{\upbRa} &\coloneqq \frac{\ScalarSecond_{\IndFreqbin}^{\upua}}{\WeightSCMN_{\IndFreqbin,\LatestFrame}}
    \\
    \label{eq:O-RCSCME:define_ScalarFirst_bRx}
    \ScalarFirst_{\IndFreqbin\IndFrame}^{\upbRx} &\coloneqq \frac{\ScalarSecond_{\IndFreqbin\IndFrame}^{\upux}}{\WeightSCMN_{\IndFreqbin,\LatestFrame}},
    \\
    \label{eq:O-RCSCME:define_ScalarFirst_bRb}
    \ScalarFirst_{\IndFreqbin}^{\upbRb} &\coloneqq \frac{1}{\WeightSCMN_{\IndFreqbin,\LatestFrame}},
    \\
    \label{eq:O-RCSCME:define_ScalarComb}
    \ScalarComb_{\IndFreqbin\IndFrame} &\coloneqq \frac{\VarT_{\IndFreqbin\IndFrame}}{\VarN_{\IndFreqbin\IndFrame} + \VarT_{\IndFreqbin\IndFrame} \ScalarFirst_{\IndFreqbin}^{\upaRa}},
\end{align}
where $^{*}$ denotes the conjugate of a complex value.
Then, using these scalar terms, we can transform (\ref{eq:O-RCSCME:online_update_VarT})--(\ref{eq:O-RCSCME:online_update_denom_weightSCMN}), (\ref{eq:O-RCSCME:online_update_Est_numer_weightSCMN}), and (\ref{eq:O-RCSCME:online_update_Est_denom_weightSCMN}) as follows:
\begin{align}
    \label{eq:O-RCSCME:accelerated_online_update_VarT}
    \VarT_{\IndFreqbin\IndFrame} &\leftarrow \VarT_{\IndFreqbin\IndFrame} \left(\frac{
        \abs{\frac{\ScalarFirst_{\IndFreqbin\IndFrame}^{\upaRx}}{\VarN_{\IndFreqbin\IndFrame} + \VarT_{\IndFreqbin\IndFrame} \ScalarFirst_{\IndFreqbin}^{\upaRa}}}^{2} + \frac{\bar{\ScaleIG}}{( \VarT_{\IndFreqbin\IndFrame} )^{2}}
    }{
        \frac{\ScalarFirst_{\IndFreqbin}^{\upaRa}}{\VarN_{\IndFreqbin\IndFrame} + \VarT_{\IndFreqbin\IndFrame} \ScalarFirst_{\IndFreqbin}^{\upaRa}} + \frac{\bar{\ShapeIG} + 1}{\VarT_{\IndFreqbin\IndFrame}}
    }\right),
    \\
    \label{eq:O-RCSCME:accelerated_online_update_VarN}
    \VarN_{\IndFreqbin\IndFrame} &\leftarrow  
    \frac{
        \ScalarFirst_{\IndFreqbin\IndFrame}^{\upxRx} \hspace{-0.1em} - 2 \ScalarComb_{\IndFreqbin\IndFrame} \abs{\ScalarFirst_{\IndFreqbin\IndFrame}^{\upaRx}}^{2} + \ScalarComb_{\IndFreqbin\IndFrame}^{2} \ScalarFirst_{\IndFreqbin}^{\upaRa} \abs{\ScalarFirst_{\IndFreqbin\IndFrame}^{\upaRx}}^{2}
    }{
        \NumMic - \ScalarComb_{\IndFreqbin\IndFrame} \ScalarFirst_{\IndFreqbin}^{\upaRa}
    },
    \\
    \label{eq:O-RCSCME:accelerated_online_update_numer_WeightSCMN}
    \NumerWeightSCMN_{\IndFreqbin,\LatestFrame} &\leftarrow \EstNumerWeightSCMN_{\IndFreqbin,\LatestFrame} + \WeightSCMN_{\IndFreqbin,\LatestFrame}^{2} \frac{1 - \ForgetRCSCME}{1 - \ForgetRCSCME^{\LatestFrame}}
    \nonumber\\
    &\phantom{\leftarrow} \cdot \sum_{\IndFrame = \LatestFrame - \ApproxFrameORCSCME + 1}^{\LatestFrame} \ForgetRCSCME^{\LatestFrame - \IndFrame} \frac{\abs{\ScalarFirst_{\IndFreqbin\IndFrame}^{\upbRx} - \ScalarComb_{\IndFreqbin\IndFrame} \ScalarFirst_{\IndFreqbin}^{\upbRa} \ScalarFirst_{\IndFreqbin\IndFrame}^{\upaRx}}^{2}}{\VarN_{\IndFreqbin\IndFrame}},
    \\
    \label{eq:O-RCSCME:accelerated_online_update_denom_WeightSCMN}
    \DenomWeightSCMN_{\IndFreqbin,\LatestFrame} &\leftarrow \EstDenomWeightSCMN_{\IndFreqbin,\LatestFrame} + \frac{1 - \ForgetRCSCME}{1 - \ForgetRCSCME^{\LatestFrame}} 
    \nonumber\\
    &\phantom{\leftarrow} \cdot \sum_{\IndFrame = \LatestFrame - \ApproxFrameORCSCME + 1}^{\LatestFrame} \ForgetRCSCME^{\LatestFrame - \IndFrame} \Bigl( \ScalarFirst_{\IndFreqbin}^{\upbRb} - \ScalarComb_{\IndFreqbin\IndFrame} \abs{\ScalarFirst_{\IndFreqbin}^{\upaRb}}^{2} \Bigr),
    \\
    \label{eq:O-RCSCME:accelerated_online_update_Est_numer_WeightSCMN}
    \EstNumerWeightSCMN_{\IndFreqbin,\LatestFrame} &\leftarrow \frac{1 - \ForgetRCSCME}{1 - \ForgetRCSCME^{\LatestFrame}} \ForgetRCSCME^{\ApproxFrameORCSCME} \bigl( \EstVarN_{\IndFreqbin(\LatestFrame-\ApproxFrameORCSCME),\LatestFrame-1}\bigr)^{-1}
    \nonumber\\
    &\phantom{\leftarrow} \cdot \abs{\Est{\ScalarFirst}_{\IndFreqbin(\LatestFrame-\ApproxFrameORCSCME),\LatestFrame-1}^{\upbRx} - \Est{\ScalarComb}_{\IndFreqbin(\LatestFrame-\ApproxFrameORCSCME),\LatestFrame-1} \Est{\ScalarFirst}_{\IndFreqbin,\LatestFrame-1}^{\upbRa} \Est{\ScalarFirst}_{\IndFreqbin(\LatestFrame-\ApproxFrameORCSCME),\LatestFrame-1}^{\upaRx}}^{2}
    \nonumber\\
    &\phantom{=} + \frac{\ForgetRCSCME - \ForgetRCSCME^{\LatestFrame}}{1 - \ForgetRCSCME^{\LatestFrame}} \EstNumerWeightSCMN_{\IndFreqbin,\LatestFrame-1},
    \\
    \label{eq:O-RCSCME:accelerated_online_update_Est_denom_WeightSCMN}
    \EstDenomWeightSCMN_{\IndFreqbin,\LatestFrame} &\leftarrow \frac{1 - \ForgetRCSCME}{1 - \ForgetRCSCME^{\LatestFrame}} \ForgetRCSCME^{\ApproxFrameORCSCME} \Bigl( \Est{\ScalarFirst}_{\IndFreqbin,\LatestFrame-1}^{\upbRb} - \Est{\ScalarComb}_{\IndFreqbin(\LatestFrame-\ApproxFrameORCSCME),\LatestFrame-1} \abs{\Est{\ScalarFirst}_{\IndFreqbin,\LatestFrame-1}^{\upbRa}}^{2} \Bigr)
    \nonumber\\
    &\phantom{\leftarrow} + \frac{\ForgetRCSCME - \ForgetRCSCME^{\LatestFrame}}{1 - \ForgetRCSCME^{\LatestFrame}} \EstDenomWeightSCMN_{\IndFreqbin,\LatestFrame-1},
\end{align}
where $\Est{\ScalarFirst}_{\IndFreqbin\IndFrame,\LatestFrame-1}^{\upaRx}$, $\Est{\ScalarFirst}_{\IndFreqbin,\LatestFrame-1}^{\upbRa}$, $\Est{\ScalarFirst}_{\IndFreqbin\IndFrame,\LatestFrame-1}^{\upbRx}$, $\Est{\ScalarFirst}_{\IndFreqbin,\LatestFrame-1}^{\upbRb}$, and $\Est{\ScalarComb}_{\IndFreqbin\IndFrame,\LatestFrame-1}$ are the estimates of $\ScalarFirst_{\IndFreqbin\IndFrame}^{\upaRx}$, $\ScalarFirst_{\IndFreqbin}^{\upbRa}$, $\ScalarFirst_{\IndFreqbin\IndFrame}^{\upbRx}$, $\ScalarFirst_{\IndFreqbin}^{\upbRb}$, and $\ScalarComb_{\IndFreqbin\IndFrame}$ obtained at the $(\LatestFrame-1)$th-frame estimation, respectively.
Note that $\ScalarComb_{\IndFreqbin\IndFrame}$ is updated at the initialization and after the update of $\VarN_{\IndFreqbin\IndFrame}$ and $\WeightSCMN_{\IndFreqbin,\LatestFrame}$, and $\ScalarFirst_{\IndFreqbin}^{\upaRa}$, $\ScalarFirst_{\IndFreqbin\IndFrame}^{\upaRx}$, $\ScalarFirst_{\IndFreqbin\IndFrame}^{\upxRx}$, $\ScalarFirst_{\IndFreqbin}^{\upbRa}$, $\ScalarFirst_{\IndFreqbin\IndFrame}^{\upbRx}$, and $\ScalarFirst_{\IndFreqbin}^{\upbRb}$ are updated at the initialization and after the update of $\WeightSCMN_{\IndFreqbin,\LatestFrame}$. 
Since $\SCMNb_{\IndFreqbin,\LatestFrame}$ is constant during the iterative updates at the $\LatestFrame$th-frame estimation, all computations involving matrix operations in (\ref{eq:O-RCSCME:define_ScalarSecond_aRa})--(\ref{eq:O-RCSCME:define_ScalarSecond_ux}) can be performed collectively at the initialization of the $\LatestFrame$th-frame estimation.
Consequently, similar to \cite{Kubo2020TASLP}, online RCSCME can perform iterative updates using only scalar operations, resulting in a significant acceleration.

Next, we consider the computation of $\SCMNb_{\IndFreqbin,\LatestFrame}$ required at the initialization of each frame.
The computation of $\SCMNb_{\IndFreqbin,\LatestFrame}$ in (\ref{eq:O-RCSCME:def_SCMNb}) includes the computation of the MP inverse $(\SCMNh_{\IndFreqbin,\LatestFrame})^{+}$, which becomes a bottleneck for real-time implementation.
To address this problem, we consider calculating the MP inverse efficiently.
First, we define $\breve{\DemixMat}_{\IndFreqbin} \in \mathbb{C}^{(\NumSrc - 1) \times \NumMic}$ and $\breve{\MixMat}_{\IndFreqbin} \in \mathbb{C}^{\NumMic \times (\NumSrc-1)}$ as 
\begin{align}
    \label{eq:O-RCSCME:def_W_breve}
    \breve{\DemixMat}_{\IndFreqbin} &\coloneqq \bigl( \DemixVec_{\IndFreqbin1}, ..., \DemixVec_{\IndFreqbin(\IndTargetSrc-1)}, \DemixVec_{\IndFreqbin(\IndTargetSrc+1)}, ..., \DemixVec_{\IndFreqbin\NumSrc} \bigr)^{\Hermite},
    \\
    \label{eq:O-RCSCME:def_A_breve}
    \breve{\MixMat}_{\IndFreqbin} &\coloneqq \bigl( \MixVec_{\IndFreqbin1}, ..., \MixVec_{\IndFreqbin(\IndTargetSrc-1)}, \MixVec_{\IndFreqbin(\IndTargetSrc+1)}, ..., \MixVec_{\IndFreqbin\NumSrc}\bigr).
\end{align}
Note that $\MixVec_{\IndFreqbin\IndTargetSrc} = \SV_{\IndFreqbin}$ holds, and thus, we use $\SV_{\IndFreqbin}$ instead of $\MixVec_{\IndFreqbin\IndTargetSrc}$.
Here, for the products of $\breve{\DemixMat}_{\IndFreqbin}$ and $\breve{\MixMat}_{\IndFreqbin}$,
\begin{align}
    \label{eq:O-RCSCME:matrix_product_of_W_breve_and_A_breve}
    \breve{\DemixMat}_{\IndFreqbin} \breve{\MixMat}_{\IndFreqbin} &= \Identity_{\NumSrc-1},
    \\
    \label{eq:O-RCSCME:matrix_product_of_A_breve_and_W_breve}
    \breve{\MixMat}_{\IndFreqbin} \breve{\DemixMat}_{\IndFreqbin} 
    &= \NoiseImageMat_{\IndFreqbin}
\end{align}
hold.
Then, (\ref{eq:O-RCSCME:model_noise_SCM_rank-M-1}) can be expressed as 
\begin{align}
    \label{eq:O-RCSCME:transform_SCMNh_with_W_breve_and_A_breve}
    \SCMNh_{\IndFreqbin,\LatestFrame} = \breve{\MixMat}_{\IndFreqbin} \bigl( \breve{\DemixMat}_{\IndFreqbin} \XMat_{\IndFreqbin,\LatestFrame} \breve{\DemixMat}_{\IndFreqbin}^{\Hermite} \bigr) \breve{\MixMat}_{\IndFreqbin}^{\Hermite}.
\end{align}
Here, $\breve{\DemixMat}_{\IndFreqbin} \XMat_{\IndFreqbin,\LatestFrame} \breve{\DemixMat}_{\IndFreqbin}^{\Hermite} \in \mathbb{C}^{(\NumSrc-1) \times (\NumSrc-1)}$ is regular, and the following matrix $\MPinv_{\IndFreqbin,\LatestFrame}$ satisfies all the conditions for the MP inverse of $\SCMNh_{\IndFreqbin,\LatestFrame}$\footnote{
    That is, $\MPinv_{\IndFreqbin,\LatestFrame} \SCMNh_{\IndFreqbin,\LatestFrame} \MPinv_{\IndFreqbin,\LatestFrame} = \MPinv_{\IndFreqbin,\LatestFrame}$, $\SCMNh_{\IndFreqbin,\LatestFrame} \MPinv_{\IndFreqbin,\LatestFrame} \SCMNh_{\IndFreqbin,\LatestFrame} = \SCMNh_{\IndFreqbin,\LatestFrame}$, $(\SCMNh_{\IndFreqbin,\LatestFrame} \MPinv_{\IndFreqbin,\LatestFrame})^{\Hermite} = \SCMNh_{\IndFreqbin,\LatestFrame} \MPinv_{\IndFreqbin,\LatestFrame}$, and $(\MPinv_{\IndFreqbin,\LatestFrame} \SCMNh_{\IndFreqbin,\LatestFrame})^{\Hermite} = \MPinv_{\IndFreqbin,\LatestFrame} \SCMNh_{\IndFreqbin,\LatestFrame}$ hold.
}: 
\begin{align}
    \label{eq:O-RCSCME:def_MPinv_of_SCMNh}
    \MPinv_{\IndFreqbin,\LatestFrame} &\coloneqq \biggl( \Identity_{\NumSrc} - \frac{\DemixVec_{\IndFreqbin\IndTargetSrc} \DemixVec_{\IndFreqbin\IndTargetSrc}^{\Hermite}}{\| \DemixVec_{\IndFreqbin\IndTargetSrc} \|^{2}} \biggr) \breve{\DemixMat}_{\IndFreqbin}^{\Hermite} \bigl( \breve{\DemixMat}_{\IndFreqbin} \XMat_{\IndFreqbin,\LatestFrame} \breve{\DemixMat}_{\IndFreqbin}^{\Hermite} \bigr)^{-1}
    \nonumber\\
    &\phantom{\coloneqq = } \cdot \breve{\DemixMat}_{\IndFreqbin} \biggl( \Identity_{\NumSrc} - \frac{\DemixVec_{\IndFreqbin\IndTargetSrc} \DemixVec_{\IndFreqbin\IndTargetSrc}^{\Hermite}}{\| \DemixVec_{\IndFreqbin\IndTargetSrc} \|^{2}} \biggr).
\end{align}
Furthermore, the inverse of $\breve{\DemixMat}_{\IndFreqbin} \XMat_{\IndFreqbin,\LatestFrame} \breve{\DemixMat}_{\IndFreqbin}^{\Hermite}$ can be expressed as follows:
\begin{align}
    \label{eq:O-RCSCME:W_breve_X_W_breve_inv}
    &\bigl( \breve{\DemixMat}_{\IndFreqbin} \XMat_{\IndFreqbin,\LatestFrame} \breve{\DemixMat}_{\IndFreqbin}^{\Hermite} \bigr)^{-1}
    \nonumber\\
    &= \breve{\MixMat}_{\IndFreqbin}^{\Hermite} \XMatInv_{\IndFreqbin,\LatestFrame} \breve{\MixMat}_{\IndFreqbin}
     - \frac{
        \breve{\MixMat}_{\IndFreqbin}^{\Hermite} \XMatInv_{\IndFreqbin,\LatestFrame} \SV_{\IndFreqbin} \bigl( \SV_{\IndFreqbin} \bigr)^{\Hermite} \XMatInv_{\IndFreqbin,\LatestFrame} \breve{\MixMat}_{\IndFreqbin}
    }{
        \bigl( \SV_{\IndFreqbin} \bigr)^{\Hermite} \XMatInv_{\IndFreqbin,\LatestFrame} \SV_{\IndFreqbin}
    },
\end{align}
where $\XMatInv_{\IndFreqbin,\LatestFrame} \coloneqq \bigl( \XMat_{\IndFreqbin,\LatestFrame} \bigr)^{-1}$.
Since $\NoiseImageMat_{\IndFreqbin} (\Identity_{\NumSrc} - \DemixVec_{\IndFreqbin\IndTargetSrc} \DemixVec_{\IndFreqbin\IndTargetSrc}^{\Hermite} / \| \DemixVec_{\IndFreqbin\IndTargetSrc} \|^{2}) = \Identity_{\NumSrc} - \DemixVec_{\IndFreqbin\IndTargetSrc} \DemixVec_{\IndFreqbin\IndTargetSrc}^{\Hermite} / \| \DemixVec_{\IndFreqbin\IndTargetSrc} \|^{2}$ holds, by substituting (\ref{eq:O-RCSCME:W_breve_X_W_breve_inv}) and (\ref{eq:O-RCSCME:matrix_product_of_A_breve_and_W_breve}) into (\ref{eq:O-RCSCME:def_MPinv_of_SCMNh}), we can transform $\MPinv_{\IndFreqbin,\LatestFrame}$ as
\begin{align}
    \label{eq:O-RCSCME:def_intermediate_mat_for_MPinv}
    \bm{\Upsilon}_{\IndFreqbin,\LatestFrame} &\coloneqq \XMatInv_{\IndFreqbin,\LatestFrame} - \frac{\XMatInv_{\IndFreqbin,\LatestFrame} \SV_{\IndFreqbin} \bigl( \SV_{\IndFreqbin} \bigr)^{\Hermite} \XMatInv_{\IndFreqbin,\LatestFrame}}{\bigl( \SV_{\IndFreqbin} \bigr)^{\Hermite} \XMatInv_{\IndFreqbin,\LatestFrame} \SV_{\IndFreqbin}},
    \\
    \label{eq:O-RCSCME:transform_MPinv_of_SCMh}
    \MPinv_{\IndFreqbin,\LatestFrame}
    &= \Biggl( \Identity_{\NumSrc} - \frac{\DemixVec_{\IndFreqbin\IndTargetSrc} \DemixVec_{\IndFreqbin\IndTargetSrc}^{\Hermite}}{\| \DemixVec_{\IndFreqbin\IndTargetSrc} \|^{2}} \Biggr) \bm{\Upsilon}_{\IndFreqbin,\LatestFrame} \Biggl( \Identity_{\NumSrc} - \frac{\DemixVec_{\IndFreqbin\IndTargetSrc} \DemixVec_{\IndFreqbin\IndTargetSrc}^{\Hermite}}{\| \DemixVec_{\IndFreqbin\IndTargetSrc} \|^{2}} \Biggr),
\end{align}
where $\bm{\Upsilon}_{\IndFreqbin,\LatestFrame}$ is an intermediate variable.
Next, we focus on the update of $\XMatInv_{\IndFreqbin,\LatestFrame}$.
For the update of $\XMat_{\IndFreqbin,\LatestFrame}$, we can transform (\ref{eq:O-RCSCME:model_XMat}) using (\ref{eq:O-RCSCME:WeightORCSCME}) as 
\begin{align}
    \label{eq:O-RCSCME:transform_XMat}
    \XMat_{\IndFreqbin,\LatestFrame} &= \frac{1 - \ForgetRCSCME}{1 - \ForgetRCSCME^{\LatestFrame}} \bm{\ObsSignal}_{\IndFreqbin\LatestFrame} \bm{\ObsSignal}_{\IndFreqbin\LatestFrame}^{\Hermite}
    \nonumber\\
    &\phantom{=} + \frac{\ForgetRCSCME - \ForgetRCSCME^{\LatestFrame}}{1 - \ForgetRCSCME^{\LatestFrame}} \left( \sum_{\IndFrame=1}^{\LatestFrame-1} \frac{1 - \ForgetRCSCME}{1 - \ForgetRCSCME^{\LatestFrame-1}} \ForgetRCSCME^{(\LatestFrame-1)-\IndFrame} \bm{\ObsSignal}_{\IndFreqbin\IndFrame} \bm{\ObsSignal}_{\IndFreqbin\IndFrame}^{\Hermite} \right)
    \nonumber\\
    &= \frac{1 - \ForgetRCSCME}{1 - \ForgetRCSCME^{\LatestFrame}} \bm{\ObsSignal}_{\IndFreqbin\LatestFrame} \bm{\ObsSignal}_{\IndFreqbin,\LatestFrame}^{\Hermite} + \frac{\ForgetRCSCME - \ForgetRCSCME^{\LatestFrame}}{1 - \ForgetRCSCME^{\LatestFrame}} \XMat_{\IndFreqbin,\LatestFrame-1}.
\end{align}
Since $\XMat_{\IndFreqbin,\LatestFrame}$ is updated by summing the full-rank matrix $\XMat_{\IndFreqbin,\LatestFrame-1}$ and the rank-1 matrix, $\XMatInv_{\IndFreqbin,\LatestFrame}$ can be updated using the Sherman--Morrison formula as follows:
\begin{align}
    \label{eq:O-RCSCME:transform_XMatInv_using_Sherman-Morrison-formula}
    \XMatInv_{\IndFreqbin,\LatestFrame} = \frac{1 - \ForgetRCSCME^{\LatestFrame}}{\ForgetRCSCME - \ForgetRCSCME^{\LatestFrame}} \Biggl( \XMatInv_{\IndFreqbin,\LatestFrame-1} - \frac{\XMatInv_{\IndFreqbin,\LatestFrame-1} \bm{\ObsSignal}_{\IndFreqbin\LatestFrame} \bm{\ObsSignal}_{\IndFreqbin\LatestFrame}^{\Hermite} \XMatInv_{\IndFreqbin,\LatestFrame-1}}{\frac{\ForgetRCSCME - \ForgetRCSCME^{\LatestFrame}}{1 - \ForgetRCSCME} + \bm{\ObsSignal}_{\IndFreqbin\LatestFrame}^{\Hermite} \XMatInv_{\IndFreqbin,\LatestFrame-1} \bm{\ObsSignal}_{\IndFreqbin\LatestFrame}} \Biggr),
\end{align}
where $\XMatInv_{\IndFreqbin,1}$ is set to $\bigl( \epsilon_{\mathrm{emp}} \Identity_{\NumSrc} + \bm{\ObsSignal}_{\IndFreqbin1} \bm{\ObsSignal}_{\IndFreqbin1}^{\Hermite} \bigr)^{-1} = \epsilon_{\mathrm{emp}}^{-1} \Identity_{\NumSrc} - \bm{\ObsSignal}_{\IndFreqbin1} \bm{\ObsSignal}_{\IndFreqbin1}^{\Hermite} / \bigl( \epsilon_{\mathrm{emp}}^{2} + \epsilon_{\mathrm{emp}} \bm{\ObsSignal}_{\IndFreqbin1}^{\Hermite} \bm{\ObsSignal}_{\IndFreqbin1} \bigr)$ with a sufficiently small parameter $\epsilon_{\mathrm{emp}}$ for stability.
By substituting (\ref{eq:O-RCSCME:transform_XMatInv_using_Sherman-Morrison-formula}) into (\ref{eq:O-RCSCME:transform_MPinv_of_SCMh}), we obtain the efficient update rule of $\MPinv_{\IndFreqbin,\LatestFrame}$ as follows:
\begin{align}
    \label{eq:O-RCSCME:efficient_update_of_MPinv_XMatInv}
    &\XMatInv_{\IndFreqbin,\LatestFrame} = \frac{1 - \ForgetRCSCME}{\ForgetRCSCME - \ForgetRCSCME^{\LatestFrame}} \Biggl( \XMatInv_{\IndFreqbin,\LatestFrame-1} - \frac{\XMatInv_{\IndFreqbin,\LatestFrame-1} \bm{\ObsSignal}_{\IndFreqbin\LatestFrame} \bm{\ObsSignal}_{\IndFreqbin\LatestFrame}^{\Hermite} \XMatInv_{\IndFreqbin,\LatestFrame-1}}{\frac{\ForgetRCSCME - \ForgetRCSCME^{\LatestFrame}}{1 - \ForgetRCSCME} + \bm{\ObsSignal}_{\IndFreqbin\LatestFrame}^{\Hermite} \XMatInv_{\IndFreqbin,\LatestFrame-1} \bm{\ObsSignal}_{\IndFreqbin\LatestFrame} } \Biggr),
    \\
    \label{eq:O-RCSCME:efficient_update_of_Intermediate_Variable}
    &\bm{\Upsilon}_{\IndFreqbin,\LatestFrame} = \XMatInv_{\IndFreqbin,\LatestFrame} - \frac{\XMatInv_{\IndFreqbin,\LatestFrame} \SV_{\IndFreqbin} \bigl(\SV_{\IndFreqbin}\bigr)^{\Hermite} \XMatInv_{\IndFreqbin,\LatestFrame}}{\bigl( \SV_{\IndFreqbin} \bigr)^{\Hermite} \XMatInv_{\IndFreqbin,\LatestFrame} \SV_{\IndFreqbin}}
    \\
    \label{eq:O-RCSCME:efficient_update_of_MPinv_MPinv}
    &\MPinv_{\IndFreqbin,\LatestFrame} = \Biggl( \Identity_{\NumSrc} - \frac{\DemixVec_{\IndFreqbin\IndTargetSrc} \DemixVec_{\IndFreqbin\IndTargetSrc}^{\Hermite}}{\| \DemixVec_{\IndFreqbin\IndTargetSrc} \|^{2}} \Biggr) \bm{\Upsilon}_{\IndFreqbin,\LatestFrame} \Biggl( \Identity_{\NumSrc} - \frac{\DemixVec_{\IndFreqbin\IndTargetSrc} \DemixVec_{\IndFreqbin\IndTargetSrc}^{\Hermite}}{\| \DemixVec_{\IndFreqbin\IndTargetSrc} \|^{2}} \Biggr).
\end{align}
Using $\MPinv_{\IndFreqbin,\LatestFrame}$, we calculate $\SCMNb_{\IndFreqbin,\LatestFrame}$ as 
\begin{align}
    \label{eq:O-RCSCME:update_of_SCMNb_using_efficient_MPinv}
    \SCMNb_{\IndFreqbin,\LatestFrame} = \bigl( \Identity_{\NumSrc} - \DiagVec_{\IndFreqbin} \NoiseSV_{\IndFreqbin}^{\Hermite} \bigr) \MPinv_{\IndFreqbin,\LatestFrame} \bigl( \Identity_{\NumSrc} - \NoiseSV_{\IndFreqbin} \DiagVec_{\IndFreqbin}^{\Hermite} \bigr).
\end{align}
In addition, the initialization of $\VarN_{\IndFreqbin\LatestFrame}$, (\ref{eq:O-RCSCME:initialization_VarN}), can be rewritten as 
\begin{align}
    \label{eq:O-RCSCME:initialization_VarN_with_MPinv}
    \VarN_{\IndFreqbin\LatestFrame} = \frac{1}{\NumMic} \NoiseSrcImage_{\IndFreqbin\LatestFrame}^{\Hermite} \MPinv_{\IndFreqbin,\LatestFrame} \NoiseSrcImage_{\IndFreqbin\LatestFrame}.
\end{align}

In summary, the entire framewise update algorithm for online RCSCME consists of (\ref{eq:O-RCSCME:efficient_update_of_MPinv_XMatInv})--(\ref{eq:O-RCSCME:update_of_SCMNb_using_efficient_MPinv}), (\ref{eq:O-RCSCME:define_ScalarSecond_aRa})--(\ref{eq:O-RCSCME:define_ScalarSecond_ux}), (\ref{eq:O-RCSCME:accelerated_online_update_Est_numer_WeightSCMN}), and (\ref{eq:O-RCSCME:accelerated_online_update_Est_denom_WeightSCMN}) for the initialization and (\ref{eq:O-RCSCME:define_ScalarFirst_aRa})--(\ref{eq:O-RCSCME:accelerated_online_update_denom_WeightSCMN}) and (\ref{eq:O-RCSCME:online_update_weightSCMN}) for the iterative update.

After the parameter estimation at the $\LatestFrame$th frame, the source image of the directional target speech signal at the $\LatestFrame$th frame is extracted using a multichannel Wiener filter as follows:
\begin{align}
    \label{eq:O-RCSCME:multichannel_Wiener_filtering}
    \TargetSrcImage_{\IndFreqbin\LatestFrame} &= \VarT_{\IndFreqbin\LatestFrame} \SV_{\IndFreqbin} \bigl( \SV_{\IndFreqbin} \bigr)^{\Hermite} \bigl( \SCMX_{\IndFreqbin\LatestFrame} \bigr)^{-1} \bm{\ObsSignal}_{\IndFreqbin\LatestFrame}
    \nonumber\\
    &= \ScalarComb_{\IndFreqbin\LatestFrame} \ScalarFirst_{\IndFreqbin\LatestFrame}^{\upaRx} \SV_{\IndFreqbin}.
\end{align}

\section{EXPERIMENTS}
\label{sec:experiments}

In this section, we demonstrate the effectiveness of the proposed method through several experiments using simulated diffuse noise environments and a real-world recording.
First, in Section~\ref{sec:experiments}-\ref{ssec:simulation_with_stationary}, we simulate a scenario where the target speaker is stationary and evaluate the real-time speech extraction performance and processing time.
Then, in Section~\ref{sec:experiments}-\ref{ssec:simulation_with_moving}, we simulate a scenario where the target speaker moves and confirm the robustness of the proposed method to a moving speaker compared with conventional methods.
We also evaluate the effectiveness of the proposed online NSR-ILRMA alone in Section~\ref{sec:experiments}-\ref{ssec:ablation_ilrma_module} and conduct ablation study on combinations of the real-time extensions of the RCSCME-based methods (i.e., NSR-ILRMA and RCSCME) based on blockwise batch and online algorithms in Section~\ref{sec:experiments}-\ref{ssec:ablation_combination}.
Finally, in Section~\ref{sec:experiments}-\ref{ssec:real-world_recording}, we use real-world-recorded signals and confirm that the proposed method outperforms conventional methods even under a practical condition.

\subsection{EXPERIMENTS WITH SIMULATED STATIONARY SPEAKER}
\label{ssec:simulation_with_stationary}

\begin{figure}[tbp]
    \centering
    \includegraphics[width=\columnwidth]{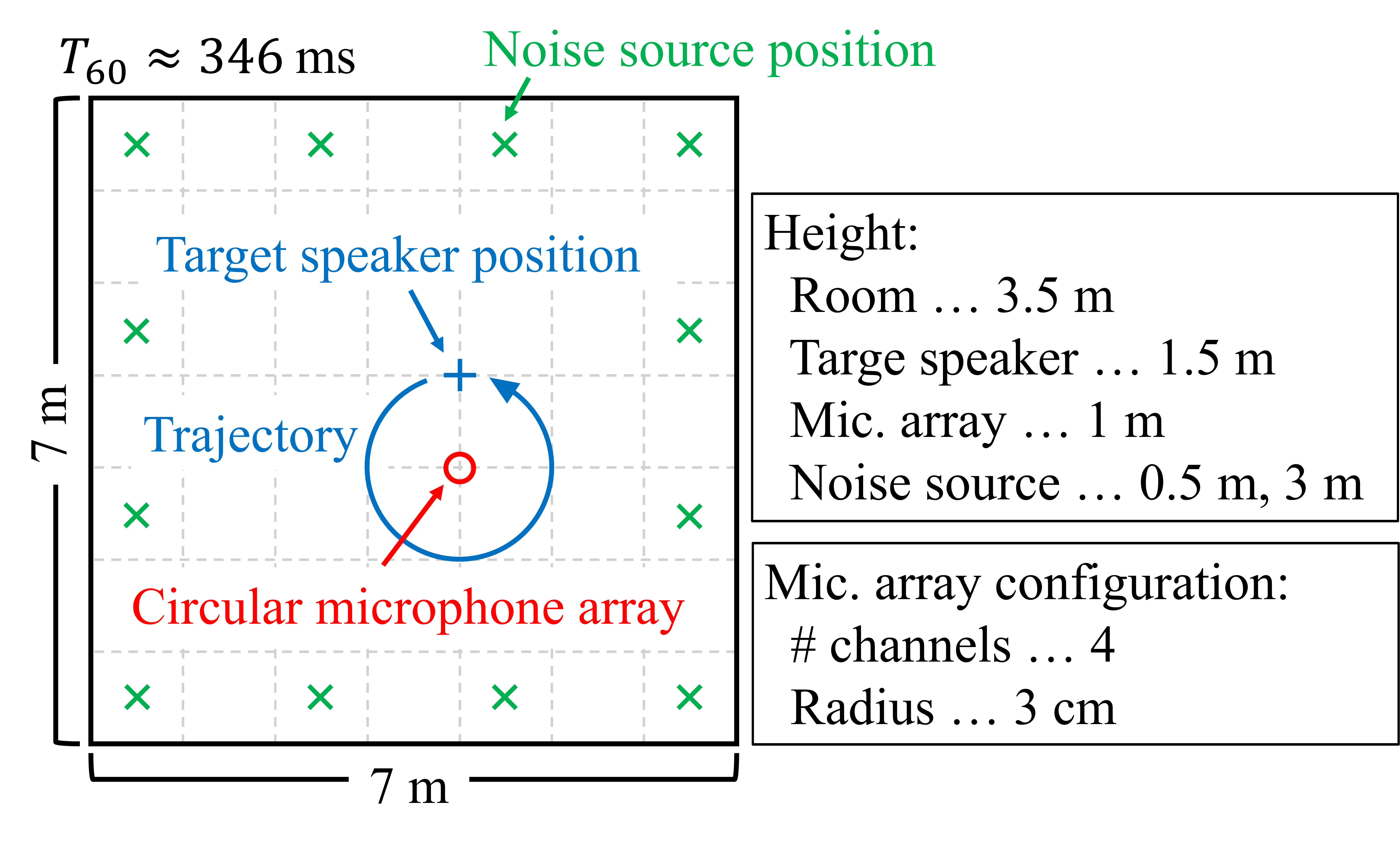}
    \vspace{-1em}
    \caption{
        Room layout for simulating impulse responses.
        Noise sources were placed at green `$\times$' positions.
        Position of each noise source was perturbed by adding positional noise generated from three-dimensional standard normal distribution and scaled by $0.01$.
        Directivity of each noise source was hyper cardioid and was set to point outwards from center of room.
        Circular microphone array consisting of four omnidirectional microphones with radius of 3~cm was placed at red `$\bigcirc$' position.
        In experiment described in Section~\ref{sec:experiments}-\ref{ssec:simulation_with_stationary}, target speaker is stationary at blue `$+$' position.
        In experiment described in Section~\ref{sec:experiments}-\ref{ssec:simulation_with_moving}, target speaker moves along blue arrow trajectory at constant speed from 10 to 20~s and from 30 to 40~s after start.
        At all other times, target speaker is stationary at blue `$+$' position.
    }
    \label{fig:room_layout}
\end{figure}

In this section, as the simplest case, we simulate a scenario where the target speaker is stationary and evaluate the real-time speech extraction performance and processing time.

For the dry sources of the target speech signal, we used speech signals of 100 speakers from the JVS dataset~\cite{Takamichi2020AST} and created a dry source by concatenating their speech signals to a total length of 20~s.
For the dry sources of the diffuse noise, we prepared six types of noise (\textit{station}, \textit{cafe}, \textit{traffic}, \textit{cafeteria}, \textit{restaurant}, and \textit{public square}) and created 24 different 20-s-long noise signals from the DEMAND dataset~\cite{Thiemann2013DEMAND} for each noise type by using different starting points.
We then convolved all dry sources with room impulse responses generated using the image method implemented in Pyroomacoustics~\cite{Scheibler2018ICASSP}.
Fig.~\ref{fig:room_layout} shows the room layout for simulating the impulse responses.
The reverberation time $T_{60}$ was approximately 346~ms.
Finally, we applied a low-cut filter with a cut-off frequency of 50~Hz~\cite{Allen1979JASA} to the simulated source images, and mixed them so that the input SNR becomes 0~dB at a reference microphone.
The sampling rate was 16~kHz, and the STFT was performed using a 64-ms-long Hann window with a shift length of 32~ms.

We compared five methods: O-IVA using IP to update the demixing matrix (\textit{O-IVA-IP})~\cite{Taniguchi2014HSCMA}, O-IVA using ISS to update the demixing matrix (\textit{O-IVA-ISS})~\cite{Nakashima2022APSIPA}, online spatially regularized IVE (\textit{O-SR-IVE})~\cite{Ueda2024IEEE-TASLP}, \textit{B-RCSCME}~\cite{Ishikawa2025Access}, and the proposed \textit{O-RCSCME}.
For O-IVA-IP and O-IVA-ISS, the forgetting factor $\ForgetIVA$ and the stability parameter $\epsilon_{\IVA}$ were set to 0.95 and $10^{-3}$, respectively.
The number of iterations per frame was set to 1.
Since both O-IVA-IP and O-IVA-ISS are BSS methods, it is necessary to determine the channel index corresponding to the target speech.
To address this, following the strategy in \cite{Ishikawa2025Access}, we performed the kurtosis-based channel selection method~\cite{Fujihara2008IWAENC} every 16~frames (corresponding to 512~ms).
For O-SR-IVE, the forgetting factor was set to 0.95, and the weight parameters for unit response regularization and scale regularization were set to 10 and $10^{-4}$, respectively.
These weight parameters were determined by grid search over $\{10^{-4}, 10^{-3}, ..., 10^{3}\}$, and the combination yielding the best performance was used.
The number of iterations per frame was set to 1. 
For B-RCSCME, the ILRMA part used NSR-ILRMA and was performed at a minimum interval of 512~ms using the most recent 5-s-long observed signals, and the RCSCME part was performed every 32~ms (the shift length of the STFT) using the most recent 1-s-long observed signals.
The weight of the regularizer for NSR-ILRMA was set to 1.
The prior steering vector of the target speech, $\Prior{\MixVec}_{\IndFreqbin,\LatestFrame}$, was calculated as in \cite{Ishikawa2025Access} under the following conditions:
the height of the virtual source was the same as that of the microphone array;
the horizontal distance and relative horizontal angle between the virtual source and the microphone array were set to be the same as those between the target speaker and the microphone array in the simulation;
a free field and attenuation in accordance with the distance were assumed.
The channel index corresponding to the target speech, $\IndTargetSrc$, was set to $\NumSrc$.
The demixing matrix $\DemixMat_{\IndFreqbin}$ was initialized to the inverse of the matrix $\Prior{\MixMat}_{\IndFreqbin,\LatestFrame}$ whose $\IndTargetSrc$th (i.e., $\NumSrc$th) column was $\Prior{\MixVec}_{\IndFreqbin,\LatestFrame}$ and others were the orthonormal bases of the orthogonal complementary space of $\Prior{\MixVec}_{\IndFreqbin,\LatestFrame}$ every time the ILRMA part is executed.
All other conditions were the same as those in \cite{Ishikawa2025Access}.
For the proposed O-RCSCME, the number of iterations per frame for online NSR-ILRMA and online RCSCME was set to 1 and 2, respectively.
The forgetting factors for online NSR-ILRMA and online RCSCME, $\ForgetILRMA$ and $\ForgetRCSCME$, respectively, were set to 0.95.
The number of NMF bases in online NSR-ILRMA was set to 10.
To stabilize the update of the NMF variables, we only updated the NMF variables for $\IndBasis = 1, ..., \min(\lceil \LatestFrame / 2 \rceil, \NumBasis)$, and the reset threshold for the basis variable, $\epsilon_{\mathrm{basis}}$, was set to 0.4.
The frame-independent weight parameter of the regularizer, $\bar{\RegWeight}_{\IndFreqbin\IndSrc}$, was set to 1.
The frame-independent shape and scale parameters for the inverse gamma distribution, $\bar{\ShapeIG}$ and $\bar{\ScaleIG}$, were set to 1.6 and $10^{-16}$, respectively. 
The number of past frames used for updating the time-varying parameters, $\ApproxFrameORCSCME$, was set to 30.
The stability parameters $\epsilon_{\ILRMA}$, $\epsilon_{\WeightSCMN}$, and $\epsilon_{\mathrm{emp}}$ were set to $10^{-3}$.
Here, the hyperparameter $\bar{\ScaleIG}$ was determined to a sufficiently small value to induce sparsity, and $\ApproxFrameORCSCME$ was determined to be approximately equal to the batch size of the RCSCME part in B-RCSCME.
The hyperparameters $\bar{\RegWeight}_{\IndFreqbin\IndSrc}$ and $\bar{\ShapeIG}$, as well as those for the stabilization techniques of online NSR-ILRMA, were experimentally determined in terms of speech extraction performance.
The basis variable $\Basis_{\IndFreqbin\IndBasis\IndSrc}$ was initialized with a uniform random value in the range of $[10^{-10}, 1]$ before the 1st-frame estimation.
The activation variable $\Activation_{\IndBasis\LatestFrame\IndSrc}$ was initialized with a uniform random value in the range of $[10^{-10}, 1]$ for $\LatestFrame = 1$ and $\Est{\Activation}_{\IndBasis(\LatestFrame-1)\IndSrc,\LatestFrame-1}$ for $\LatestFrame > 1$.
The demixing matrix $\DemixMat_{\IndFreqbin}$ was initialized to the inverse of $\Prior{\MixMat}_{\IndFreqbin,1}$ when $\LatestFrame = 1$.
These hyperparameters were determined experimentally in terms of speech extraction performance.
These methods were implemented in Python, and the computation was performed on a PC equipped with Intel Core i9-13900KF CPU and 128~GB RAM.
All computations were performed on the CPU.
The data type of all variables was set to double-precision floating points.

The evaluation measures were the source-to-distortion ratio (SDR) and source-to-interferences ratio (SIR)~\cite{Vincent2006TASLP} improvements and the maximum processing time per frame.
To evaluate the real-time speech extraction performance, we first calculated the segmentwise SDR and SIR improvements using the signals between $(\iota - 1)$~s and $(\iota + 1)$~s $(\iota = 1, ..., 19)$.
We used the average SDR and SIR improvements across all 100 target speech signals for each noise condition.
We also compared the maximum processing times per frame for O-IVA-IP, O-IVA-ISS, O-SR-IVE, the RCSCME part of B-RCSCME, and O-RCSCME using a total of 375000 frames across all noise conditions.

\begin{figure*}[tbp]
    \centering
    \includegraphics[width=0.99\linewidth]{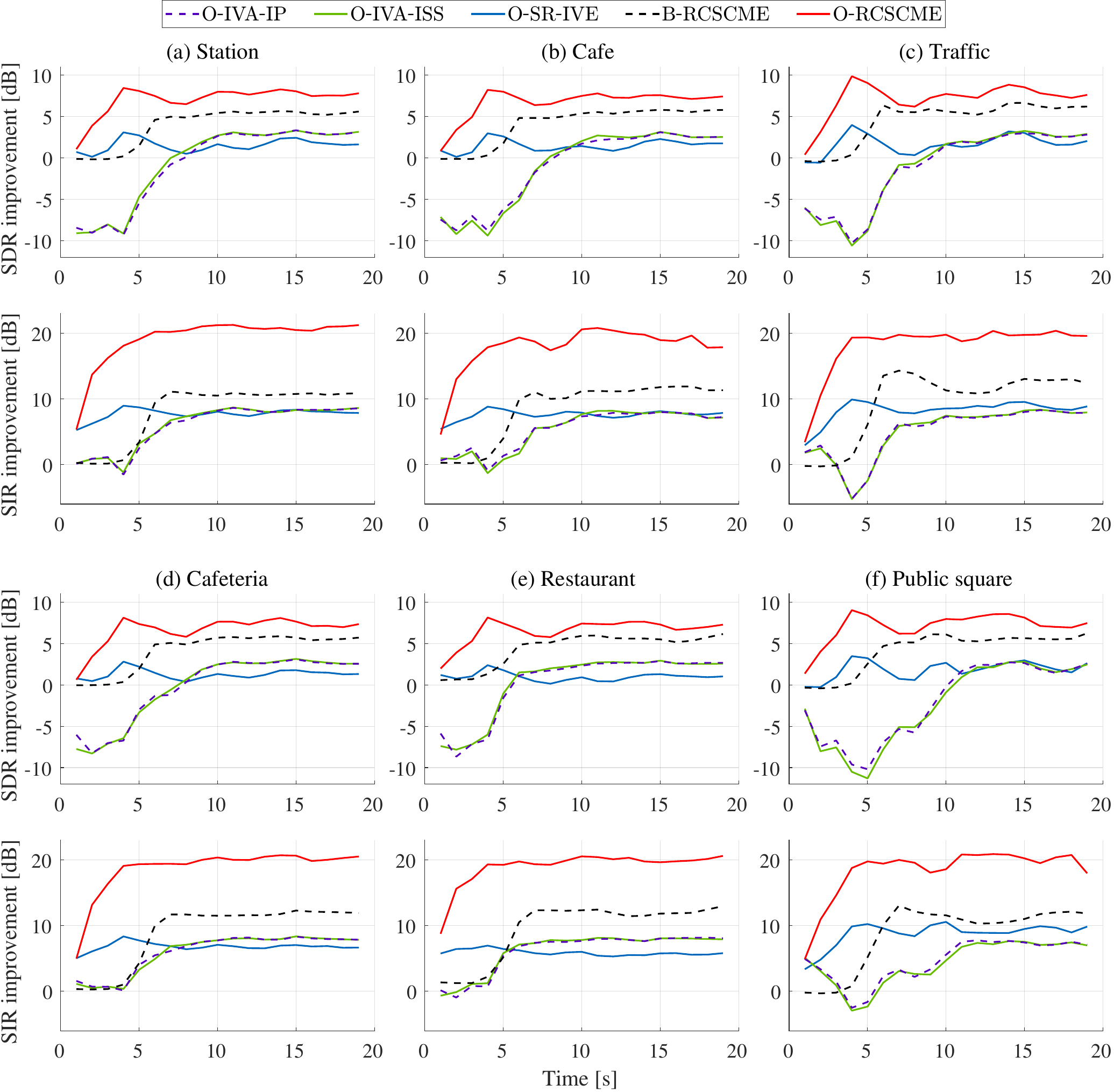}  
    \caption{
        Average SDR and SIR improvements for each method in experiment with simulated stationary speaker.
        For each noise condition, upper and lower panels represent SDR and SIR improvements, respectively.
        The noise conditions are (a) station, (b) cafe, (c) traffic, (d) cafeteria, (e) restaurant, and (f) public square.
    }
    \label{fig:SDR-SIR_experiment_1}
\end{figure*}

Fig.~\ref{fig:SDR-SIR_experiment_1} shows the average SDR and SIR improvements for each method under each noise condition.
We confirmed that the proposed method outperforms the conventional methods under all noise conditions.
Since O-IVA-IP, O-IVA-ISS, and O-SR-IVE are methods based on linear demixing filters, these methods cannot completely separate the diffuse noise and the target speech signals in principle, which may result in lower speech extraction performance.
In the early stages of the execution, O-RCSCME achieves high performance quickly, whereas B-RCSCME, which is based on the blockwise batch algorithm, does not perform well until a full batch of observed signals had been obtained.

\begin{figure}[tbp]
    \centering
    \includegraphics[width=0.99\columnwidth]{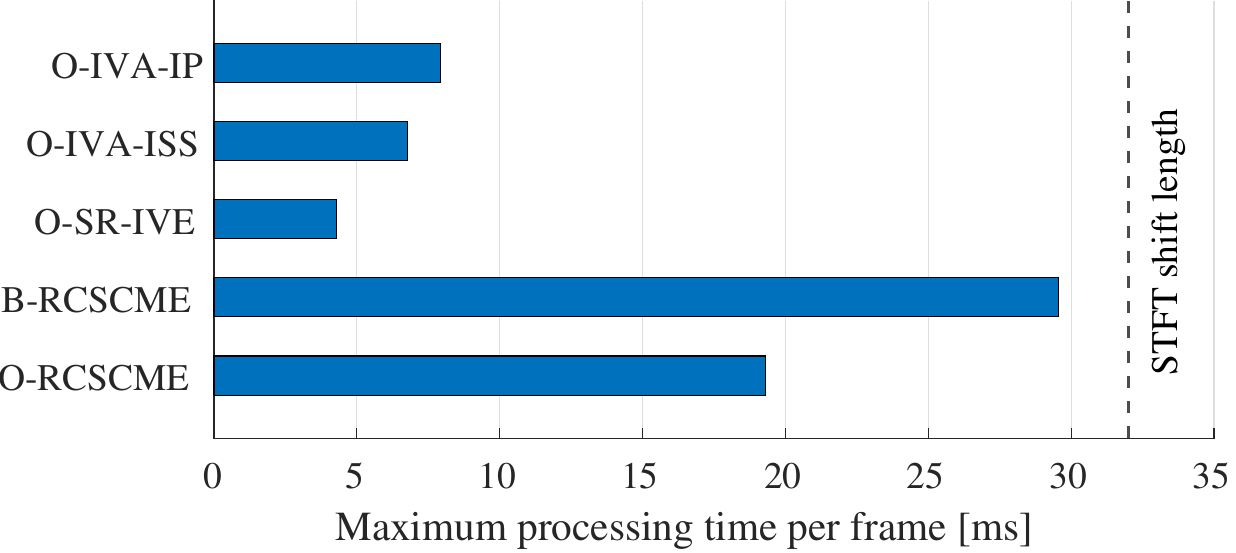}
    \vspace{-1em}
    \caption{
        Maximum processing time per frame for all methods.
        Black dashed line indicates real-time threshold (i.e., 32~ms STFT shift length).
    }
    \label{fig:max_proctime}
\end{figure}

Next, Fig.~\ref{fig:max_proctime} shows the maximum processing time per frame for each method.
All maximum processing times were less than the shift length of the STFT (32~ms), indicating that all methods could function in real time.
Furthermore, although the maximum processing time of O-RCSCME was longer than those of O-IVA-IP, O-IVA-ISS, and O-SR-IVE, there was still a sufficient margin compared with the STFT shift length, and thus, O-RCSCME is expected to perform in real time even with limited computational resources or an increased number of microphones.

From these results, we confirm that the proposed method outperforms the conventional methods in terms of real-time speech extraction performance while maintaining a sufficiently short processing time.

\subsection{EXPERIMENTS WITH SIMULATED MOVING SPEAKER}
\label{ssec:simulation_with_moving}

\begin{figure*}[tbp]
    \centering
    \includegraphics[width=0.99\linewidth]{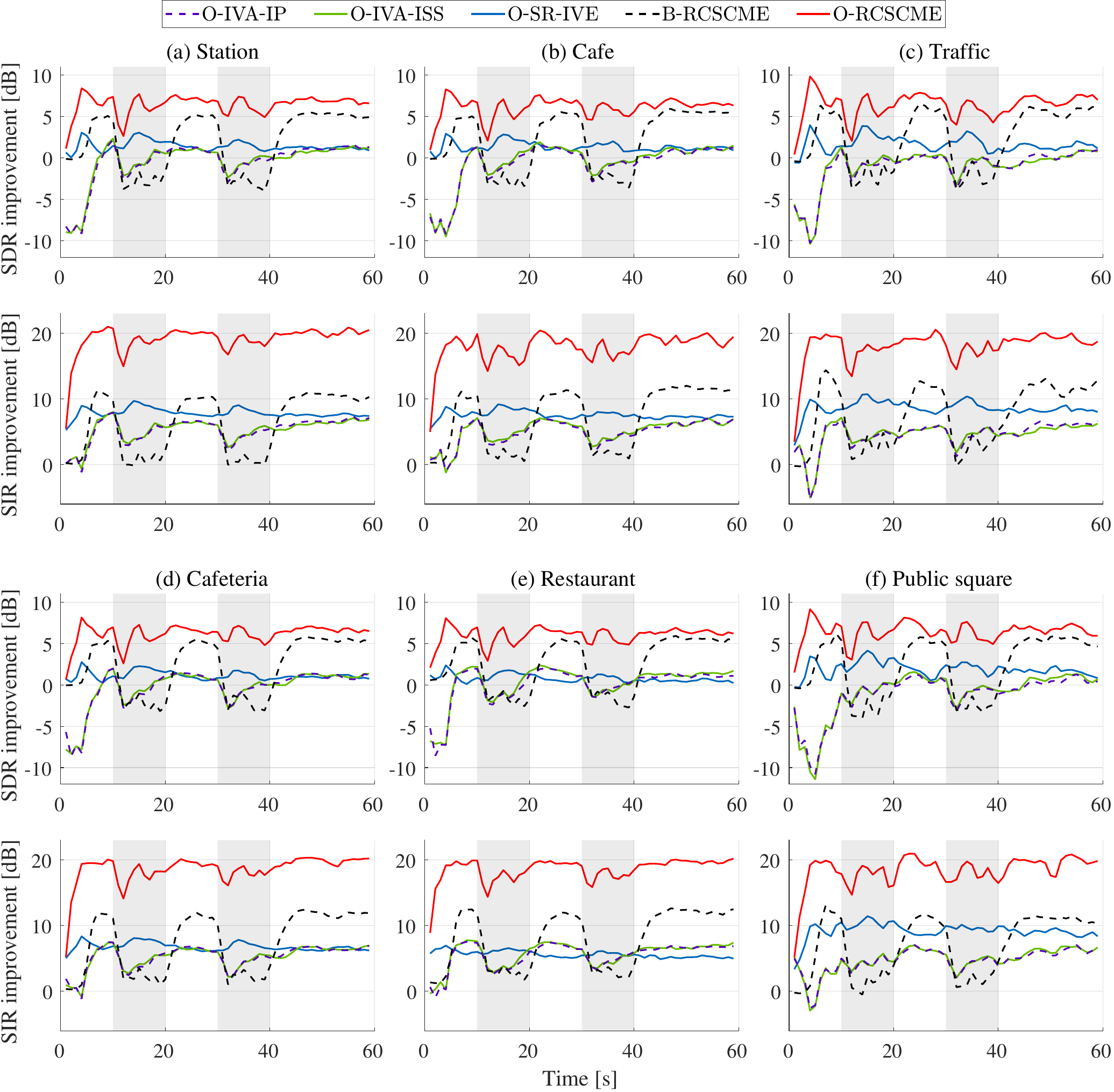}
    \caption{
        Average SDR and SIR improvements for each method in experiment with simulated moving speaker.
        For each noise condition, upper and lower panels represent SDR and SIR improvements, respectively.
        The noise conditions are (a) station, (b) cafe, (c) traffic, (d) cafeteria, (e) restaurant, and (f) public square.
        Gray shaded areas represent time interval in which target speaker moved around microphone array.
    }
    \label{fig:SDR-SIR_experiment_2}
\end{figure*}

In this section, we evaluate the real-time speech extraction performance in a more dynamic scenario than that in Section~\ref{sec:experiments}-\ref{ssec:simulation_with_stationary} by simulating a situation where the target speaker moves.

We created the dry sources for the target speech and the diffuse noise signals in the same manner as in Section~\ref{sec:experiments}-\ref{ssec:simulation_with_stationary}, except that the total length of each signal was changed to 60~s.
The room layout used to simulate the impulse responses is shown in Fig.~\ref{fig:room_layout}.
The source image of the target speech signal was created in the following steps.
First, we set the target speaker to move around the microphone array at a constant speed along the arrow trajectory during the 10--20~s and 30--40~s periods, and to remain stationary at the `$+$' position at all other times.
Next, at each sampling period (i.e., 1/16000~s), we generated the room impulse response corresponding to the target speaker's current position using the image method implemented in Pyroomacoustics~\cite{Scheibler2018ICASSP}.
Finally, we created the source image of the target speech signal by convolving the dry source with the room impulse response at each time step and summing the results.
The source image of the diffuse noise was created in the same way as in Section~\ref{sec:experiments}-\ref{ssec:simulation_with_stationary}.
The low-cut filter and mixing condition were the same as those in Section~\ref{sec:experiments}-\ref{ssec:simulation_with_stationary}.
The compared methods were the same as those in Section~\ref{sec:experiments}-\ref{ssec:simulation_with_stationary}, that is, O-IVA-IP, O-IVA-ISS, O-SR-IVE, B-RCSCME, and O-RCSCME.
The prior steering vector of the target speech for each frame, $\Prior{\MixVec}_{\IndFreqbin,\LatestFrame}$, was calculated in the same manner as in Section~\ref{sec:experiments}-\ref{ssec:simulation_with_stationary}, except that the position of the virtual source was set by adding a positional noise, generated from a three-dimensional standard normal distribution and scaled by 0.01, to the true position of the target speaker at each frame.
For B-RCSCME, the ILRMA part used $\Prior{\MixVec}_{\IndFreqbin,\LatestFrame}$ at the end of its processing block for the regularizer.
The other parameter settings were the same as those in Section~\ref{sec:experiments}-\ref{ssec:simulation_with_stationary}.
The evaluation measures were the SDR and SIR improvements calculated in the same manner as in Section~\ref{sec:experiments}-\ref{ssec:simulation_with_stationary}.

Fig.~\ref{fig:SDR-SIR_experiment_2} shows the average SDR and SIR improvements for each method under each noise condition.
Focusing on the intervals where the target speaker was moving (gray shaded areas in Fig.~\ref{fig:SDR-SIR_experiment_2}), we found that the performance of B-RCSCME degraded significantly compared with the stationary intervals.
B-RCSCME is based on the blockwise batch algorithm and executed NSR-ILRMA using a 5-s-long batch of observed signals in this experimental setting.
However, in this scenario where the target speaker moves 360$^{\circ}$ around the microphone array within 10~s, the relative horizontal angle between the target speaker and the microphone array can change by up to 180$^{\circ}$ within a single batch used in the ILRMA part.
Therefore, it is difficult to perform source separation using a single time-invariant demixing matrix, resulting in the degradation of speech extraction performance.
On the other hand, although the proposed method shows some instantaneous performance dips, it maintains the high speech extraction performance even while the target speaker is moving.
We consider that this robustness stems from the capability of O-RCSCME to estimate the demixing matrix by placing greater weights on the recent observed signals at each frame.
O-IVA-IP, O-IVA-ISS, and O-SR-IVE also estimate the demixing matrix every frame and show robust speech extraction performance while the target speaker is moving.
However, since they cannot completely exclude diffuse noise from the extracted target speech signal, their speech extraction performance is very low, similar to the result in Section~\ref{sec:experiments}-\ref{ssec:simulation_with_stationary}.
From this result, we confirmed that the proposed method achieves higher speech extraction performance than the conventional methods even in situations where the target speaker moves.

\subsection{EVALUATION OF ONLINE NSR-ILRMA}
\label{ssec:ablation_ilrma_module}
In this section, we evaluate the real-time speech extraction performance of the proposed online NSR-ILRMA alone and confirm the effectiveness of the proposed stabilization techniques.

For the observed signals, we used those described in Sections~\ref{sec:experiments}-\ref{ssec:simulation_with_stationary} and \ref{sec:experiments}-\ref{ssec:simulation_with_moving} for the stationary and moving target speaker conditions, respectively.
We compared four methods: \textit{O-IVA-ISS}~\cite{Nakashima2022APSIPA}, the blockwise-batch-algorithm-based NSR-ILRMA (\textit{B-NSR-ILRMA}), the proposed online NSR-ILRMA without the stabilization techniques described in Section~\ref{sec:proposed_method}-\ref{ssec:online_NSR-ILRMA}-\ref{sssec:O-NSR-ILRMA:technique_for_stability} (\textit{O-NSR-ILRMA w/o stabilization}), and the proposed online NSR-ILRMA with the stabilization techniques (\textit{O-NSR-ILRMA}).
For O-IVA-ISS, we used the same settings as in Section~\ref{sec:experiments}-\ref{ssec:simulation_with_stationary}.
For B-NSR-ILRMA, we employed a variant of B-RCSCME in which the RCSCME part is replaced with simple linear separation using the latest estimates of the demixing matrix and the target channel index, and the same settings as in Section~\ref{sec:experiments}-\ref{ssec:simulation_with_stationary} were used.
For O-NSR-ILRMA w/o stabilization and O-NSR-ILRMA, we used the same settings as the online NSR-ILRMA part of O-RCSCME in Section~\ref{sec:experiments}-\ref{ssec:simulation_with_stationary}.
The prior steering vector for the target speech was calculated in the same manner as in Sections~\ref{sec:experiments}-\ref{ssec:simulation_with_stationary} and \ref{sec:experiments}-\ref{ssec:simulation_with_moving} for the stationary and moving target speaker conditions, respectively.
In addition, for the experiments with the stationary target speaker, we employed a reference method, denoted as \textit{Potential}, representing the upper-bound performance of O-NSR-ILRMA, where all observed signals are used and a sufficient number of iterations are performed without considering real-time constraints.
For the evaluation metrics, the SDR and SIR improvements were calculated in the same manner as in Section~\ref{sec:experiments}-\ref{ssec:simulation_with_stationary} and averaged over all noise conditions.

\begin{figure}[tbp]
    \centering
    \includegraphics[width=0.99\linewidth]{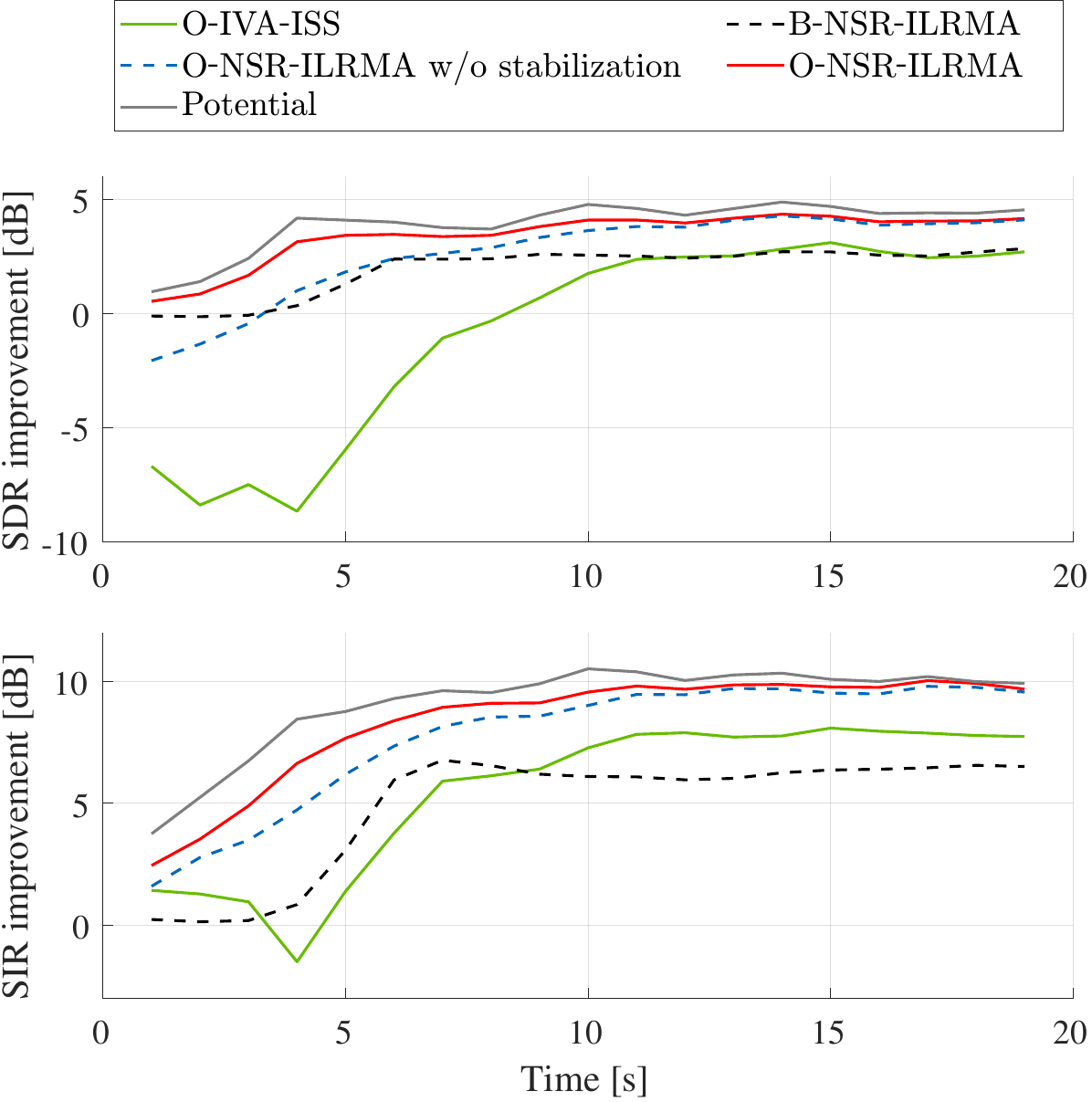}
    \vspace{-1em}
    \caption{
        Average SDR (top panel) and SIR (bottom panel) improvements for each method in experiment evaluating online NSR-ILRMA alone with simulated stationary speaker.
    }
    \label{fig:SDR-SIR_ablation_ilrma_stat}
\end{figure}

\begin{figure}[tbp]
    \centering
    \includegraphics[width=0.99\linewidth]{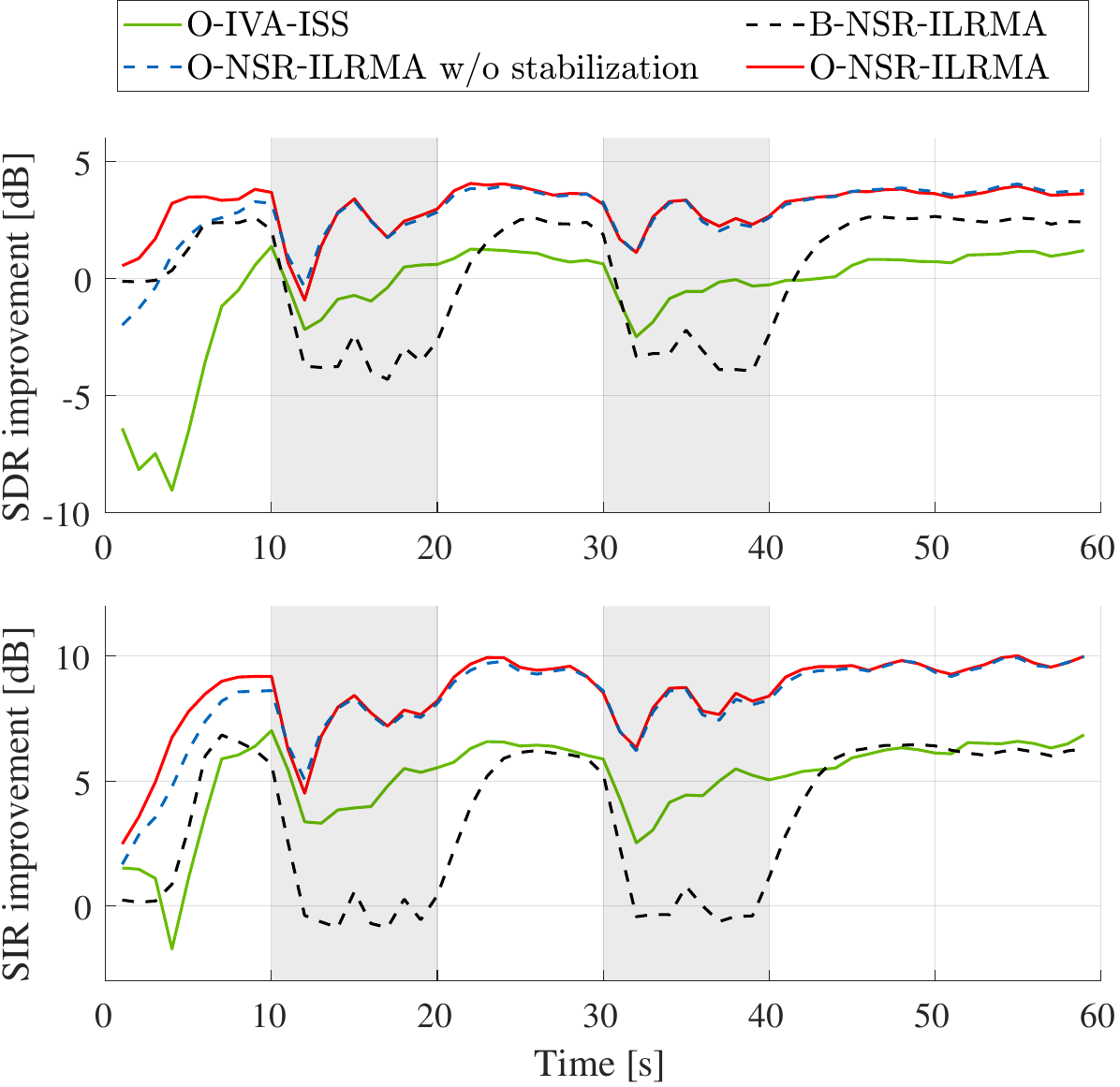}
    \vspace{-1em}
    \caption{
        Average SDR (top panel) and SIR (bottom panel) improvements for each method in experiment evaluating online NSR-ILRMA alone with simulated moving speaker.
        Gray shaded areas represent time interval in which target speaker moved around microphone array.
    }
    \label{fig:SDR-SIR_ablation_ilrma_mov}
\end{figure}

Figs.~\ref{fig:SDR-SIR_ablation_ilrma_stat} and \ref{fig:SDR-SIR_ablation_ilrma_mov} show the average SDR and SIR improvements for each method under stationary and moving target speaker conditions, respectively.
The proposed method outperforms the conventional methods and maintains higher speech extraction performance even when the target speaker is moving.
In particular, O-NSR-ILRMA achieves higher performance than B-NSR-ILRMA, which suggests that O-NSR-ILRMA can track changes in the background noise and estimate the appropriate demixing filters for the current frame.
Compared with O-NSR-ILRMA w/o stabilization, O-NSR-ILRMA achieves higher performance in the early stage, which demonstrates the effectiveness of the proposed stabilization techniques.
Furthermore, compared with Potential, the proposed method exhibits less than 1~dB degradation in SDR while enabling real-time operation, indicating that it remains sufficiently practical.

\subsection{ABLATION STUDY FOR COMBINATION OF BLOCKWISE BATCH AND ONLINE ALGORITHMS}
\label{ssec:ablation_combination}
Since the RCSCME-based method consists of NSR-ILRMA and RCSCME, we can consider combinations by introducing the proposed online NSR-ILRMA and online RCSCME into B-RCSCME.
In this section, taking B-RCSCME as a baseline, we conduct an ablation study to demonstrate the effectiveness of each component of the proposed method.

For the observed signals, we used those described in Sections~\ref{sec:experiments}-\ref{ssec:simulation_with_stationary} and \ref{sec:experiments}-\ref{ssec:simulation_with_moving} for the stationary and moving target speaker conditions, respectively.
We compared four combinations of the real-time extensions of NSR-ILRMA and RCSCME: \textit{Batch-Batch}, \textit{Batch-Online}, \textit{Online-Batch}, and \textit{Online-Online}.
Table~2 summarizes the labels and corresponding combinations of the real-time extensions of NSR-ILRMA and RCSCME.
Each component of all methods was configured using the same settings as the corresponding component of B-RCSCME or O-RCSCME in Section~\ref{sec:experiments}-\ref{ssec:simulation_with_stationary}.
The prior steering vector for the target speech was calculated in the same manner as in Sections~\ref{sec:experiments}-\ref{ssec:simulation_with_stationary} and \ref{sec:experiments}-\ref{ssec:simulation_with_moving} for the stationary and moving target speaker conditions, respectively.
The evaluation measures were the SDR and SIR improvements and the maximum processing time per frame.
The SDR and SIR improvements were calculated in the same manner as in Section~\ref{sec:experiments}-\ref{ssec:simulation_with_stationary} and averaged over all noise conditions.
We also compared the maximum processing time per frame using a total of 375000 frames across all noise conditions with a stationary target speaker.
Note that since Batch-Batch and Batch-Online perform blockwise-batch-algorithm-based NSR-ILRMA in parallel, we used the maximum processing time of blockwise-batch-algorithm-based and online RCSCME for Batch-Batch and Batch-Online, respectively.

\begin{table}[tbp]
    \label{table:ablation_label}
    \caption{
        Labels for combinations of real-time extensions of NSR-ILRMA and RCSCME based on blockwise batch and online algorithms
    }
    \centering
    \small
    \setlength{\tabcolsep}{4pt}
    \begin{tabular}{cccc}
            & \multicolumn{1}{c!{\vrule width 1pt}}{}  & \multicolumn{2}{c}{RCSCME}    \\
            & \multicolumn{1}{c!{\vrule width 1pt}}{}  & \multicolumn{1}{c|}{Blockwise batch} & Online \\ 
        \noalign{\hrule height 1pt}
        \multirow{2}{*}{NSR-ILRMA} & \multicolumn{1}{c!{\vrule width 1pt}}{Blockwise batch} & \multicolumn{1}{c|}{\textit{Batch-Batch}} & \textit{Batch-Online} \\
        \cline{2-4}
            & \multicolumn{1}{c!{\vrule width 1pt}}{Online} & \multicolumn{1}{c|}{\textit{Online-Batch}} & \textit{Online-Online}
    \end{tabular}
\end{table}

\begin{figure}[tbp]
    \centering
    \includegraphics[width=0.99\linewidth]{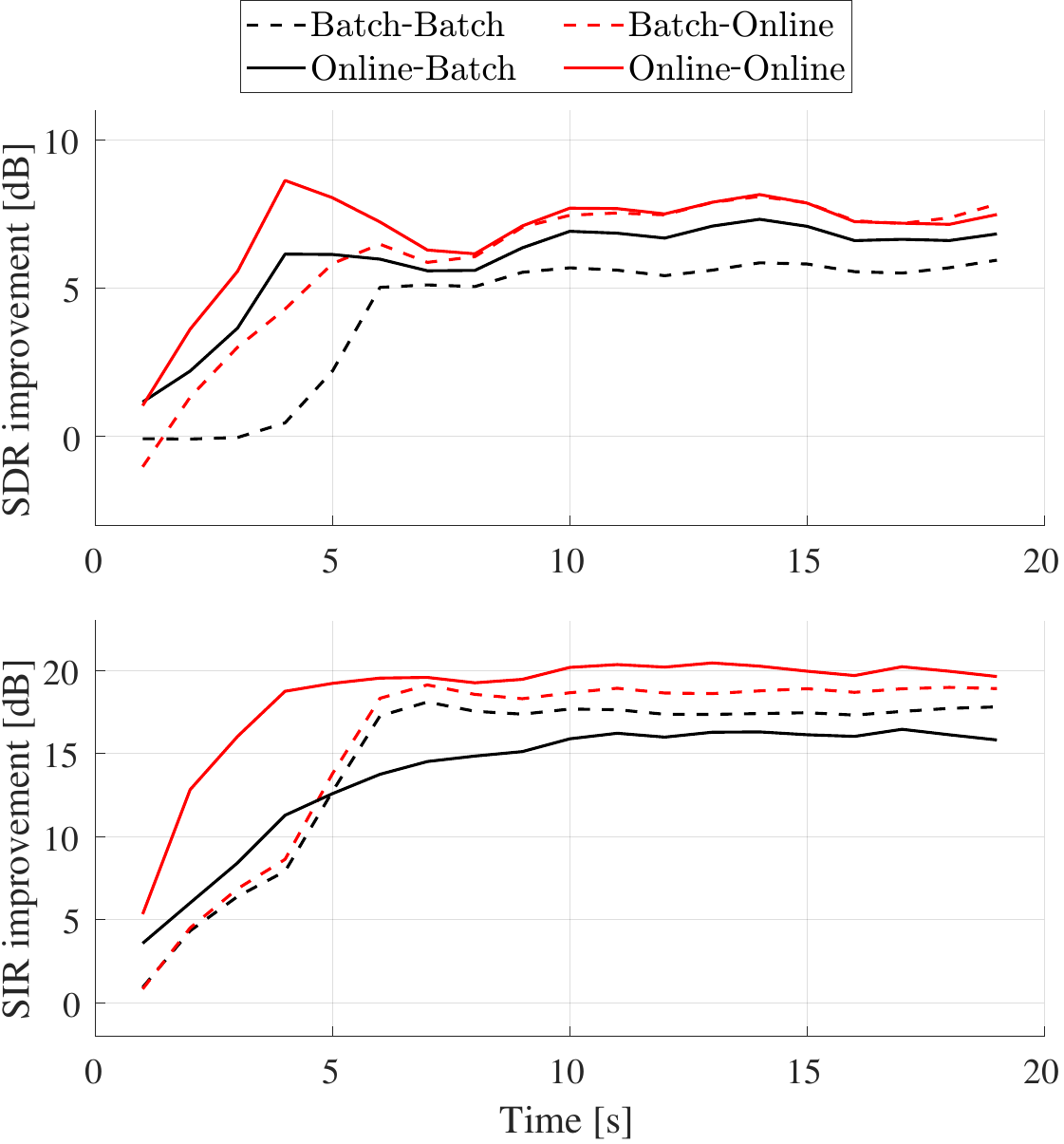}
    \vspace{-1em}
    \caption{
        Average SDR (top panel) and SIR (bottom panel) improvements for each method in experiment evaluating combinations of blockwise batch and online algorithms with simulated stationary speaker.
    }
    \label{fig:SDR-SIR_ablation_combination_stat}
\end{figure}

\begin{figure}[tbp]
    \centering
    \includegraphics[width=0.99\linewidth]{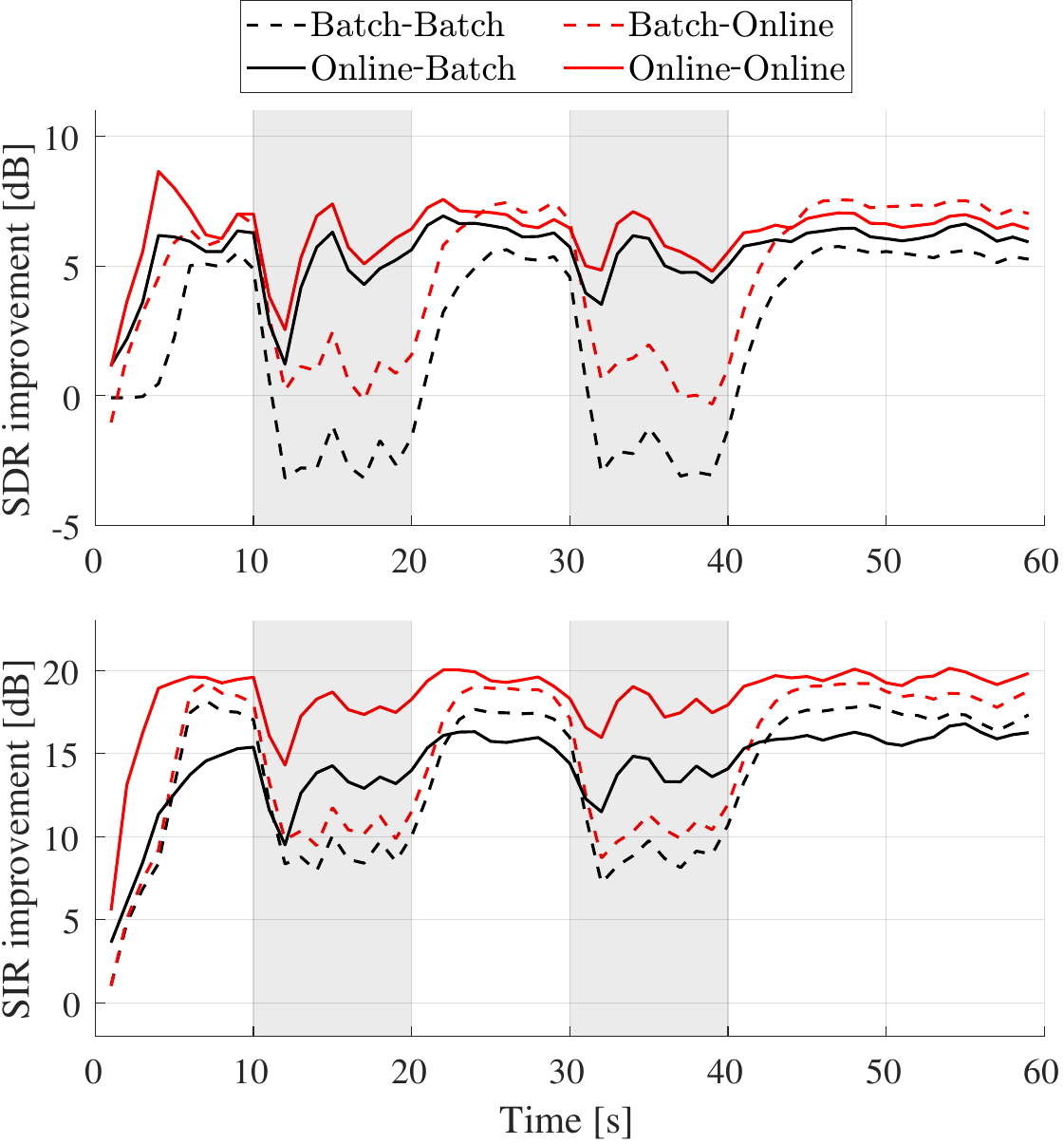} 
    \vspace{-1em}
    \caption{
        Average SDR (top panel) and SIR (bottom panel) improvements for each method in experiment evaluating combinations of blockwise batch and online algorithms with simulated moving speaker.
        Gray shaded areas represent time interval in which target speaker moved around microphone array.
    }
    \label{fig:SDR-SIR_ablation_combination_mov}
\end{figure}

Figs.~\ref{fig:SDR-SIR_ablation_combination_stat} and \ref{fig:SDR-SIR_ablation_combination_mov} show the average SDR and SIR improvements for each method under stationary and moving target speaker conditions, respectively.
As shown in Figs.~\ref{fig:SDR-SIR_ablation_ilrma_stat} and \ref{fig:SDR-SIR_ablation_combination_stat}, methods with the proposed online RCSCME achieve higher speech extraction performance than those without it under the stationary target speaker condition. 
The same tendency can be seen under the moving speaker condition as shown in Figs.~\ref{fig:SDR-SIR_ablation_ilrma_mov} and \ref{fig:SDR-SIR_ablation_combination_mov}. 
When comparing combinations that use the same RCSCME variants (i.e., Batch-Batch with Online-Batch, and Batch-Online with Online-Online), those using online NSR-ILRMA achieve higher speech extraction performance than those using blockwise-batch-algorithm-based NSR-ILRMA under both the stationary and moving target speaker condition.
These results also demonstrate the effectiveness of the proposed online NSR-ILRMA even when it was combined with the real-time extensions of RCSCME.
When comparing combinations that use the same NSR-ILRMA variants (i.e., Batch-Batch with Batch-Online, and Online-Batch with Online-Online), those using online RCSCME achieve higher speech extraction performance than those using blockwise-batch-algorithm-based RCSCME.
This result may suggest that since RCSCME estimates time-varying filters, overfitting the parameters to observations near recent time frames enables accurate modeling of local time-frequency structures and improves real-time speech extraction performance.
It is worth noting that although the SDR improvement of blockwise-batch-algorithm-based NSR-ILRMA is about 2~dB lower than that of online NSR-ILRMA in the later stage (see Fig.~\ref{fig:SDR-SIR_ablation_ilrma_stat}, B-NSR-ILRMA vs. O-NSR-ILRMA), this performance gap is reduced when comparing methods that use the proposed online RCSCME (see Fig.~\ref{fig:SDR-SIR_ablation_combination_stat}, Batch-Online vs. Online-Online).
This result suggests that online RCSCME is robust to errors in spatial filter estimation, further reinforcing its effectiveness.

\begin{figure}[tbp]
    \centering
    \includegraphics[width=0.99\linewidth]{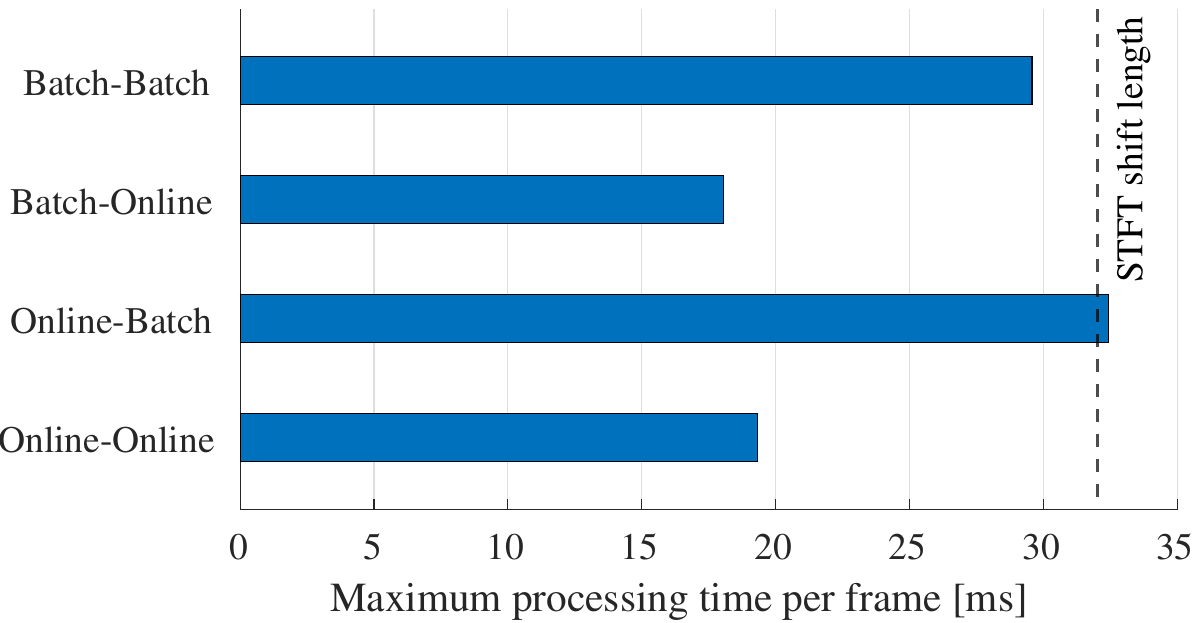}
    \vspace{-1em}
    \caption{
        Maximum processing time per frame for all combinations of real-time extensions of RCSCME-based methods based on blockwise batch and online algorithms. 
        Black dashed line indicates real-time threshold (i.e., 32~ms STFT shift length).
    }
    \label{fig:proctime_ablation_combination}
\end{figure}

Next, Fig.~\ref{fig:proctime_ablation_combination} shows the maximum processing time per frame for each method.
When comparing methods using the same NSR-ILRMA variant, the maximum processing time of methods using online RCSCME is approximately 10~ms shorter than methods using blockwise-batch-algorithm-based RCSCME.
This result indicates that the proposed online RCSCME significantly contributes to reducing the computational cost.
Note that although the maximum processing time of Online-Batch slightly exceeds the STFT shift length (i.e., 32~ms), the output signal is generated after the completion of processing even when the processing time exceeds the shift length for a fair comparison of speech extraction performance.

\subsection{EXPERIMENT WITH REAL-WORLD RECORDING}
\label{ssec:real-world_recording}

\begin{figure}[tbp]
    \centering
    \includegraphics[width=0.99\columnwidth]{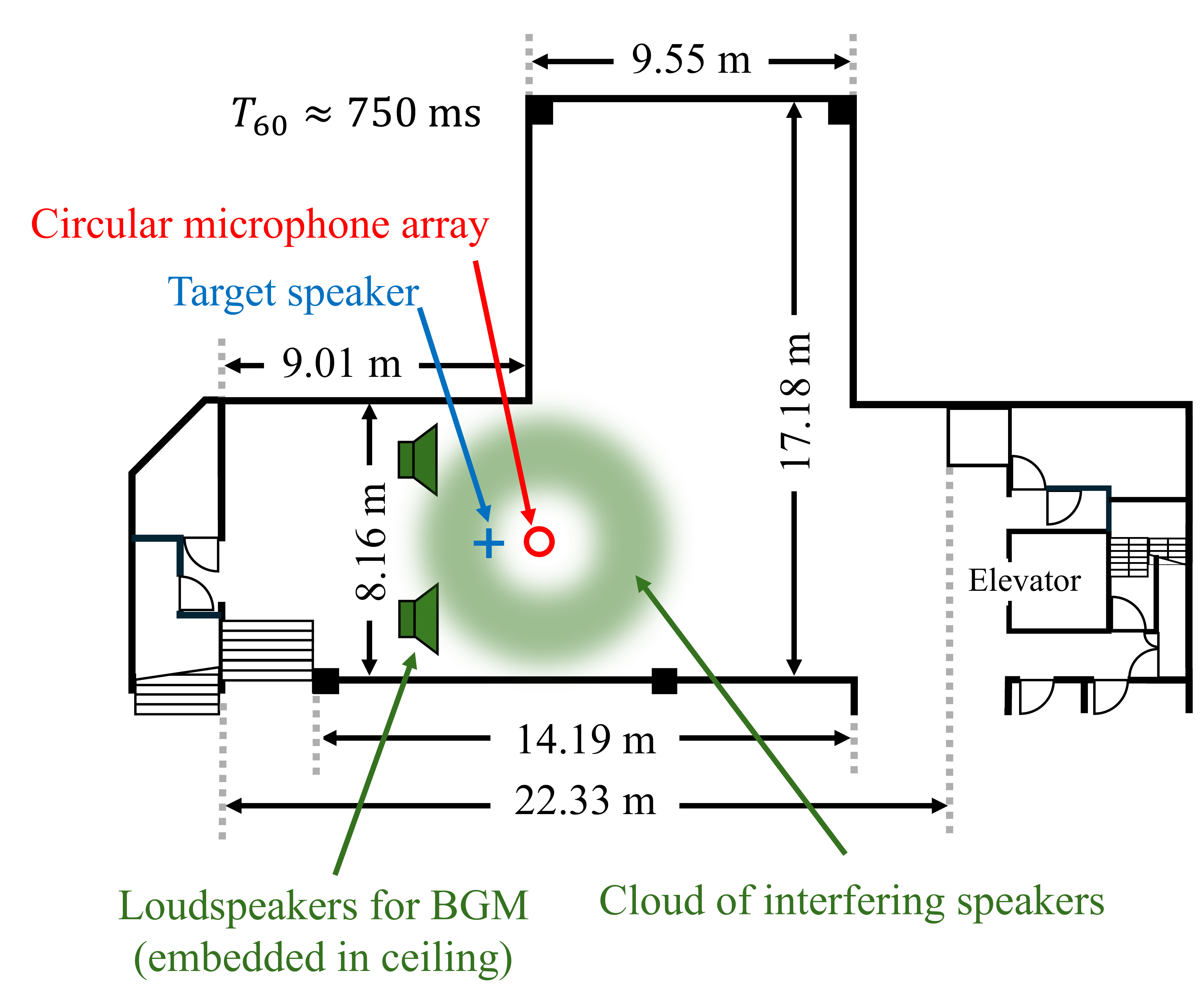}
    \vspace{-1em}
    \caption{
        Room layout for real-world recording at the Ito International Research Center, The University of Tokyo.
        Green shaded area represents cloud of interfereing speakers located 2--4~m aroung microphone array.
        Two loudspeakers for playing music were embedded in ceiling.
        Circular microphone array consisting of four omnidirectional microphones with a radius of 3.25~cm was placed at red `$\bigcirc$' position.
        Target speaker is stationary at blue `$+$' position.
    }
    \label{fig:room_layout_for_real_rec}
\end{figure}

In this section, we evaluate the real-time speech extraction performance under a practical condition using a real-world recording and confirm that the proposed method is also effective in a real-world environment.

The diffuse noise and the impulse responses for the target speech were recorded at the Ito International Research Center, The University of Tokyo.
Fig.~\ref{fig:room_layout_for_real_rec} shows the room layout for real-world recording.
A circular microphone array with four omnidirectional microphones and a radius of 3.25~cm was placed at a height of 1~m from the floor.
During the diffuse noise recording, 10 participants sat 2--4~m from the microphone array and either talked to other participants around them or read a pre-assigned text.
Simultaneously, music was played from the loudspeakers embedded in the ceiling.
The impulse responses for the target speech were recorded under the following conditions: the height of the target speaker was 1.1~m, the horizontal distance between the microphone array and the target speaker was 1~m, and the reverberation time $T_{60}$ was approximately 750~ms.
Note that the target speaker was stationary.
The dry source for the target speech signal was the same as that used in Section~\ref{sec:experiments}-\ref{ssec:simulation_with_stationary}, and the length of each speech signal was 20~s.
The input SNR was set to 0~dB at a reference microphone.
The sampling rate was 16~kHz, and the STFT was performed using a 64-ms-long Hann window with a shift length of 32~ms.
The compared methods were the same as those in Section~\ref{sec:experiments}-\ref{ssec:simulation_with_stationary}, that is, O-IVA-IP, O-IVA-ISS, O-SR-IVE, B-RCSCME, and O-RCSCME.
The shape parameter of the inverse gamma distribution in the RCSCME part of B-RCSCME, $\ShapeIG$, and the frame-independent shape parameter in O-RCSCME, $\bar{\ShapeIG}$, were both set to 0.1.
These hyperparameters were experimentally determined on the basis of the speech extraction performance.
The other parameter settings were the same as those in Section~\ref{sec:experiments}-\ref{ssec:simulation_with_stationary}.
The evaluation measures were the SDR and SIR improvements calculated in the same manner as in Section~\ref{sec:experiments}-\ref{ssec:simulation_with_stationary}.

\begin{figure}[tbp]
    \centering
    \includegraphics[width=0.99\columnwidth]{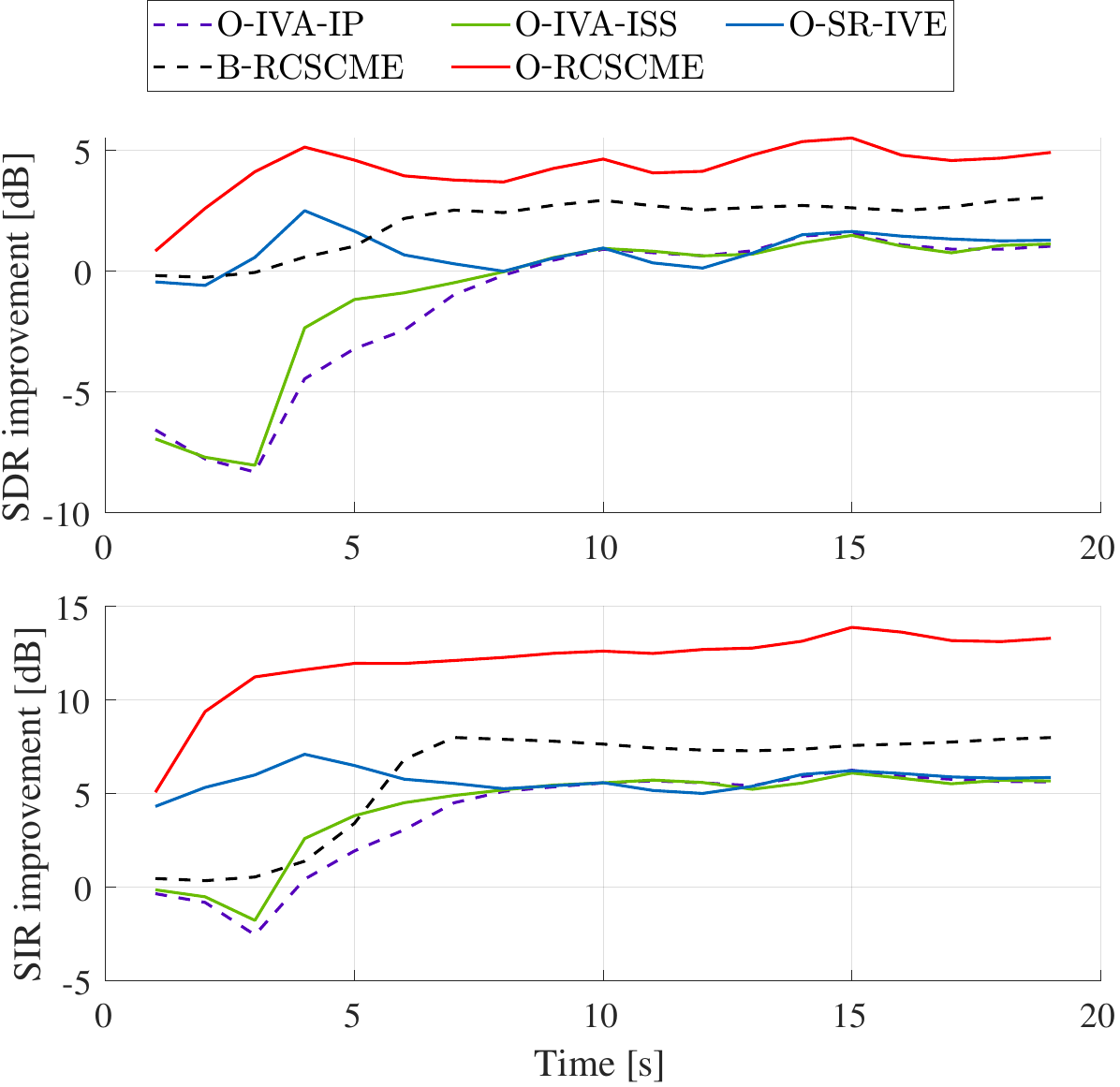}
    \vspace{-1em}
    \caption{
        Average SDR (top panel) and SIR (bottom panel) improvements for each method in experiment using real-world recording.
    }
    \label{fig:SDR-SIR_experiment_3}
\end{figure}

Fig.~\ref{fig:SDR-SIR_experiment_3} shows the average SDR and SIR improvements for each method.
Consistent with the simulation results in Sections~\ref{sec:experiments}-\ref{ssec:simulation_with_stationary} and \ref{sec:experiments}-\ref{ssec:simulation_with_moving}, we confirmed that the proposed O-RCSCME outperformed the conventional methods.
Note that, compared with the experiment in Section~\ref{sec:experiments}-\ref{ssec:simulation_with_stationary}, the performance of all methods decreased.
This is likely because the reverberation time in the real-world recording was more than double that in the simulation, making the speech extraction task more difficult.

\section{CONCLUSION}
\label{sec:conclusion}

In this paper, to achieve real-time speech extraction under diffuse noise condition that operate robustly even when the target speaker moves, we proposed online algorithms for NSR-ILRMA and RCSCME.
We derived the proposed online algorithms in three steps.
First, we formulated framewise cost functions on the basis of MWLE by assigning smaller weights to the likelihoods of past frames, which allowed us to emphasize recent frames of the observed signals.
Second, we derived the naive update rules for the framewise cost functions on the basis of the MM and ME algorithms.
In the last step, to reduce the computational cost of the naive update rules, we approximated some time-varying parameters using their estimates and derived the online update rules by minimizing the approximate cost or auxiliary functions.
Furthermore, we introduced stabilization techniques for online NSR-ILRMA and further acceleration techniques for online RCSCME.
In the experiments, we demonstrated that the proposed method outperforms the conventional methods in terms of real-time speech extraction performance and can operate robustly in dynamic scenarios where the target speaker moves.
The experiment using real-world recordings also confirmed the effectiveness of the proposed method in practical situations.

Finally, we discuss the limitation of the proposed method and directions for future work.
As a limitation, the proposed method is applicable only to single-target speech extraction because it is based on offline RCSCME designed for such scenarios.
However, practical applications often require simultaneous extraction of multiple speakers.
Therefore, extending the proposed online RCSCME to multi-speaker extraction is an important direction for future work.
Furthermore, since our approach to deriving online update rules consists of formulating a framewise cost function on the basis of MWLE and approximating time-varying parameters, it can be applied to various offline BSS and MSE methods based on MLE (e.g., MNMF).
Therefore, the proposed method is expected to contribute to the development of new online extensions of such offline methods.

\section*{APPENDIX A: MM AND ME ALGORITHMS}
\label{appendix:MM_and_ME_algorithms}

The MM and ME algorithms are optimization schemes for cost functions that are difficult to minimize directly~\cite{Hunter2000JCGS,Fevotte2009NC}.
They consider an upper-bound function (often referred to as an \textit{auxiliary function}) and iteratively update parameters using this function.

Let us consider the minimization problem of a cost function $\Cost{}(\Theta)$ with respect to a set of objective variables $\Theta$.
In the MM and ME algorithms, we design the auxiliary function $\bar{\Cost{}}(\Theta, \Omega)$ that satisfies
\begin{align}
    \label{eq:MM_ME:AuxFunc_def_1}
    \Cost{}(\Theta) &\leq \bar{\Cost{}}(\Theta, \Omega), &(\forall \Theta, \forall \Omega)
    \\
    \label{eq:MM_ME:AuxFunc_def_2}
    \Cost{}(\Theta) &= \min_{\Omega} \bar{\Cost{}}(\Theta, \Omega),  &(\forall \Theta)
\end{align}
where $\Omega$ denotes a set of auxiliary variables.
Then, the following two steps are iteratively performed.
First, we update the auxiliary variables using the latest objective variables $\hat{\Theta}$ as 
\begin{align}
    \label{eq:MM_ME:update_AuxVar}
    \hat{\Omega} \leftarrow \arg \min_{\Omega} \bar{\Cost{}}(\hat{\Theta}, \Omega).
\end{align}
Considering (\ref{eq:MM_ME:AuxFunc_def_2}), we can use $\hat{\Omega}$ that satisfies $\bar{\Cost{}}(\hat{\Theta}, \hat{\Omega}) = \Cost{}(\hat{\Theta})$ for update instead of (\ref{eq:MM_ME:update_AuxVar}).
Next, we update the objective variables using the latest auxiliary variables $\hat{\Omega}$ as 
\begin{align}
    \label{eq:MM_ME:update_ObjVar_MM}
    \Theta \leftarrow \arg \min_{\hat{\Theta}} \bar{\Cost{}}(\hat{\Theta}, \hat{\Omega})
\end{align}
in the MM algorithm and as
\begin{align}
    \label{eq:MM_ME:update_ObjVar_ME}
    \Theta \leftarrow \breve{\Theta}\ \mathrm{s.t.}\ \bar{\Cost{}}(\breve{\Theta}, \hat{\Omega}) = \bar{\Cost{}}(\hat{\Theta}, \hat{\Omega}), \breve{\Theta} \ne \hat{\Theta} 
\end{align}
in the ME algorithm.
These algorithms guarantee the monotonic nonincrease in the original cost function $\Cost{}(\Theta)$.
In particular, it has been experimentally found that the ME algorithm tends to converge faster than the MM algorithm because the changes in parameters are likely to be larger at each iteration~\cite{Fevotte2009NC}.

\section*{APPENDIX B: UPDATE RULES OF O-IVA USING IP AND ISS}
\label{appendix:update_rule_for_o_iva}

In O-IVA~\cite{Taniguchi2014HSCMA,Nakashima2022APSIPA}, the demixing matrix $\DemixMat_{\IndFreqbin}$ is updated iteratively using IP or ISS.
The IP-based update rule proposed in \cite{Taniguchi2014HSCMA} is given as follows:
\begin{align}
    \label{eq:appendix:O-IVA-IP:w1}
    \DemixVec_{\IndFreqbin\IndSrc} &\leftarrow \bigl( \DemixMat_{\IndFreqbin} \Est{\CovIVA}_{\IndFreqbin\IndFrame,\LatestFrame} \bigr)^{-1} \UnitVec_{\IndSrc},
    \\
    \label{eq:appendix:O-IVA-IP:w2}
    \DemixVec_{\IndFreqbin\IndSrc} &\leftarrow \DemixVec_{\IndFreqbin\IndSrc} / \sqrt{\DemixVec_{\IndFreqbin\IndSrc}^{\Hermite} \Est{\CovIVA}_{\IndFreqbin\IndSrc,\LatestFrame} \DemixVec_{\IndFreqbin\IndSrc}},
\end{align}
where $\Est{\CovIVA}_{\IndFreqbin\IndSrc,\LatestFrame}$ is updated by (\ref{eq:O-IVA_update_CovMat}).
Here, since $\Est{\CovIVA}_{\IndFreqbin\IndSrc,\LatestFrame}$ is updated by summing the full-rank and rank-1 matrices in (\ref{eq:O-IVA_update_CovMat}), the following fast algorithm based on the Sherman--Morrison formula has been also proposed in \cite{Taniguchi2014HSCMA}:
\begin{align}
    \label{eq:appendix:O-IVA-IP2:InvCovIVA}
    \Est{\InvCovIVA}_{\IndFreqbin\IndSrc,\LatestFrame} &\leftarrow \frac{1}{\ForgetIVA} \Biggl( \Est{\InvCovIVA}_{\IndFreqbin\IndSrc,\LatestFrame-1} - \frac{\Est{\InvCovIVA}_{\IndFreqbin\IndSrc,\LatestFrame-1} \bm{\ObsSignal}_{\IndFreqbin\LatestFrame} \bm{\ObsSignal}_{\IndFreqbin\LatestFrame}^{\Hermite} \Est{\InvCovIVA}_{\IndFreqbin\IndSrc,\LatestFrame-1}}{2 \AuxVarIVA_{\LatestFrame\IndSrc} \frac{\ForgetIVA}{1 - \ForgetIVA}  +  \bm{\ObsSignal}_{\IndFreqbin\LatestFrame}^{\Hermite} \Est{\InvCovIVA}_{\IndFreqbin\IndSrc,\LatestFrame-1} \bm{\ObsSignal}_{\IndFreqbin\LatestFrame}} \Biggr),
    \\
    \label{eq:appendix:O-IVA-IP2:diffDemixVecIVA}
    \diffDemixVecIVA_{\IndFreqbin\IndSrc} &\leftarrow \Est{\InvCovIVA}_{\IndFreqbin\IndSrc,\LatestFrame} \MixVec_{\IndFreqbin\IndSrc} / \sqrt{\MixVec_{\IndFreqbin\IndSrc}^{\Hermite} \Est{\InvCovIVA}_{\IndFreqbin\IndSrc,\LatestFrame} \MixVec_{\IndFreqbin\IndSrc}},
    \\
    \label{eq:appendix:O-IVA-IP2:MixMat}
    \MixMat_{\IndFreqbin} &\leftarrow \Biggl( \Identity_{\NumSrc} - \frac{\MixVec_{\IndFreqbin\IndSrc} (\diffDemixVecIVA_{\IndFreqbin\IndSrc} - \DemixVec_{\IndFreqbin\IndSrc})^{\Hermite}}{\diffDemixVecIVA_{\IndFreqbin\IndSrc}^{\Hermite} \MixVec_{\IndFreqbin\IndSrc}} \Biggr) \MixMat_{\IndFreqbin},
    \\
    \label{eq:appendix:O-IVA-IP2:DemixVec}
    \DemixVec_{\IndFreqbin\IndSrc} &\leftarrow \diffDemixVecIVA_{\IndFreqbin\IndSrc},
\end{align}
where $\Est{\InvCovIVA}_{\IndFreqbin\IndSrc,\LatestFrame} := \Est{\CovIVA}_{\IndFreqbin\IndSrc,\LatestFrame}^{-1}$, and $\Est{\InvCovIVA}_{\IndFreqbin\IndSrc,0}$ is initialized as $\epsilon_{\IVA}^{-1} \Identity_{\NumSrc}$.
On the other hand, the ISS-based update rule proposed in \cite{Nakashima2022APSIPA} is expressed as
\begin{align}
    \label{eq:appendix:O-IVA-ISS:Var}
    \VarISS_{\IndFreqbin\IndSrc\IndSrc'} &\leftarrow \begin{cases}
        \frac{\DemixVec_{\IndFreqbin\IndSrc'}^{\Hermite} \Est{\CovIVA}_{\IndFreqbin\IndSrc',\LatestFrame} \DemixVec_{\IndFreqbin\IndSrc}}{\DemixVec_{\IndFreqbin\IndSrc}^{\Hermite} \Est{\CovIVA}_{\IndFreqbin\IndSrc',\LatestFrame} \DemixVec_{\IndFreqbin\IndSrc}}, &\text{(if $\IndSrc' \ne \IndSrc$)}
        \\
        1 - \bigl( \DemixVec_{\IndFreqbin\IndSrc}^{\Hermite} \Est{\CovIVA}_{\IndFreqbin\IndSrc,\LatestFrame} \DemixVec_{\IndFreqbin\IndSrc} \bigr)^{-\frac{1}{2}}, &\text{(if $\IndSrc' = \IndSrc$)}
    \end{cases}
    \\
    \label{eq:appendix:O-IVA-ISS:DemixMat}
    \DemixMat_{\IndFreqbin} &\leftarrow \DemixMat_{\IndFreqbin} - ( \VarISS_{\IndFreqbin\IndSrc1}, ..., \VarISS_{\IndFreqbin\IndSrc\NumSrc} )^{\Transpose} \DemixVec_{\IndFreqbin\IndSrc}^{\Hermite}.
\end{align}

In offline IVA, it is guaranteed that both the update rules based on IP and ISS monotonically nonincrease the cost function.
\cite{Scheibler2020ICASSP} has demonstrated that when the number of microphones $\NumMic$ is small (e.g., $\NumMic = 4$), the processing time is almost identical for offline IVA using IP and ISS, while IVA using IP achieves slightly better separation performance. 
In contrast, when the number of microphones is sufficiently large, ISS becomes computationally faster than IP.
This difference is attributed to the fact that ISS avoids the inversion of an $\NumMic$-dimensional matrix required in IP by replacing it with alternative matrix operations.


\itemsep 30pt
\bibliographystyle{IEEEtrans.bst}
\bibliography{reference}



\vfill\pagebreak

\end{document}